%% file: zzz_trend_detection.tex
\documentclass[twocolumn]{aastex62}
\usepackage{float}
\usepackage{textcomp}
\usepackage{graphicx}
\usepackage[caption=false]{subfig}
\usepackage{enumitem}
\usepackage{comment}
\usepackage{makecell}
\usepackage{environ}
\usepackage{multirow}
\usepackage{orcidlink}
\usepackage{lineno}
\accepted{\today}

\shortauthors{Peñil et al.}

\begin{document}

\title{Systematic Search for Long-Term Trends in \textit{Fermi}-LAT Jetted Active Galactic Nuclei}

\correspondingauthor{P. Peñil, A. Dom\'inguez, S. Buson, M. Ajello}
\email{ppenil@clemson.edu, alberto.d@ucm.es}
\email{sara.buson@uni-wuerzburg.de, majello@clemson.edu}

\author{P. Pe\~nil\orcidlink{0000-0003-3741-9764}}
\affil{Department of Physics and Astronomy, Clemson University,	Kinard Lab of Physics, Clemson, SC 29634-0978, USA}

\author{A. Dom\'inguez\orcidlink{0000-0002-3433-4610}}
\affil{IPARCOS and Department of EMFTEL, Universidad Complutense de Madrid, E-28040 Madrid, Spain}

\author{S. Buson\orcidlink{0000-0002-3308-324X}}
\affil{Julius-Maximilians-Universit\"at, 97070, W\"urzburg, Germany}

\author{M. Ajello\orcidlink{0000-0002-6584-1703}}
\affil{Department of Physics and Astronomy, Clemson University,	Kinard Lab of Physics, Clemson, SC 29634-0978, USA}

\author{S. Adhikari\orcidlink{0009-0006-1029-1026}}
\affil{Department of Physics and Astronomy, Clemson University,	Kinard Lab of Physics, Clemson, SC 29634-0978, USA}

\author{A. Rico\orcidlink{0000-0001-5233-7180}}
\affil{Department of Physics and Astronomy, Clemson University,	Kinard Lab of Physics, Clemson, SC 29634-0978, USA}



\begin{abstract}
Jetted Active Galactic Nuclei (AGN) exhibit variability across a wide range of time scales. Traditionally, this variability can often be modeled well as a stochastic process. However, in certain cases, jetted AGN variability displays regular patterns, enabling us to conduct investigations aimed at understanding its origins. Additionally, a novel type of variability has emerged in jetted AGN lightcurves, specifically, the observation of a long-term trend characterized by a linear increase of the flux with time in blazars such as PG 1553+113, which is among the objects most likely to display periodic behavior. In this paper, we present the results of a systematic search for long-term trends, spanning $\approx$10\, years, utilizing 12 years of Fermi-LAT observations. The study is focused on detecting the presence of linear or quadratic long-term trends in a sample of 3308 jetted AGN. Our analysis has identified 40 jetted AGN that exhibit long-term trends, each with distinct properties, which we also characterize in this study. These long-term trends may originate from the dynamics of a supermassive black hole binary system, or they could be the result of intrinsic phenomena within the jet itself. Our findings can help in addressing questions pertaining to the astrophysical origins of variability and periodicity within jetted AGN.  
\end{abstract}


\section{Introduction} \label{sec:intro}
Jetted Active Galactic Nuclei (AGN) emit collimated and highly relativistic jets, which, in certain instances, are oriented towards our line of sight \citep[e.g.,][]{wiita_lecture}. The emission from jetted AGN exhibits significant variability across the entire electromagnetic spectrum \citep[e.g.,][]{urry_multiwavelengh}, spanning various time scales \citep[e.g.,][]{urry_variability}. This variability is typically attributed to stochastic processes \citep[e.g.,][]{ruan_stocastic}. However, recent research has identified quasi-periodic oscillations (QPOs) in a few select blazars \citep[e.g.,][]{ackermann_pg1553, penil_2020}. The physical models for these QPOs fall into two distinct categories: phenomena associated with the accretion disk or the jet in a supermassive black hole (SMBH) system \citep[e.g.,][]{gracias_modulation_disk, camenzind_jet}, and processes generated by an SMBH binary \citep[e.g.,][]{sandrinelli_redfit, sobacchi_binary}.

Recently, a new type of variability has been identified in blazars. Specifically, an increase or decrease in flux over time (trend) has been observed in different sources, exhibiting linear or quadratic structures that extend over years and potentially even decades (long-term). These long-term trends have been detected in the radio flux of objects such as BL Lac \citep[][]{marscher_radio_trend_bllac} as well as 3C 111 and 3C 454.3 \citep[][]{marscher_radio_trend_3c111}. More recently, this long-term variability has been noted in the $\gamma$-ray emissions of blazars such as 3C 84 and 1ES 1215+303 \citep[][respectively]{rani_3c_84, 1ES1215+303_trend}. Additionally, \citet{penil_2022_multiwave} identified similar long-term trends in the multiwavelength emissions of PG 1553+113. These extended trends have been linked to effects within the jet as magnetic reconnections \citep[][]{giannios_trend_magnetic} or result from interactions between the SMBH binary and its surrounding material \cite[e.g.,][]{farris_trend_binary}. Additionally, these observed trends could arise from stochastic phenomena without any relation to a specific physical origin \citep[e.g.,][]{kaaz_accretion_disk_trends}. However, detailed investigations of this long-term variability have been limited, with only a few studies focusing on select objects. To address this gap, we conducted a systematic search for long-term trends in $\gamma$-ray emissions of a sample of 3308 jetted AGN using data obtained in the first 12 years of observations from the \textit{Fermi} Large Area Telescope (LAT).

The structure of the paper is outlined as follows. In Section \ref{sec:sample}, we introduce the jetted AGN sample that forms the basis of our study. Section \ref{sec:methodology} delves into the systematic search for long-term trends, providing an in-depth overview of our analysis methodology. Our study's findings and a comprehensive discussion are presented in Section \ref{sec:results}. Section \ref{sec:summary} provides a summary of our key findings and conclusions.

\section{Jetted AGN sample} \label{sec:sample}

\subsection{Source Selection}
In this study, we analyze a sample comprising 3308 jetted AGN observed by the \textit{Fermi}-LAT (where $\approx$97\% of these sources are blazars), all of which are part of the 4FGL-DR2 catalog \citep[][]{4fgl_dr2}. To facilitate our analysis, we employ $\gamma$-ray lightcurves (LCs) binned at 28-day intervals. This choice strikes a balance between computational feasibility and sensitivity to long-term variations. These LCs are generated by extracting source fluxes at E$\geq$ 100~MeV. This energy threshold is selected to enhance photon statistics and minimize the presence of upper limits in the LCs compared to previous analyses, such as the one conducted by \citet{penil_2020}.

\subsection{{\it Fermi}-LAT Data Analysis}\label{sec:fermianalysis}
The {\it Fermi}-LAT data reduction is performed by using the Python package \texttt{fermipy}\footnote{We use the version 0.19.0} \citep{Wood:2017yyb} and applied to each jetted AGN included in our sample. The specific procedure is explained below. We select the photons belonging to the \texttt{Pass 8 SOURCE} class \citep[][]{atwood_source_class, bruel_pass8}, in a region of interest (ROI) of 15$^\circ$ $\times$ 15$^\circ$ square, centered at the target. The ROI model includes all 4FGL-DR2 catalog sources \citep[][]{4fgl_dr2} located within 20$^{\circ}$ from the ROI center, as well as the Galactic and isotropic diffuse emission \footnote{\url{https://fermi.gsfc.nasa.gov/ssc/data/access/lat/BackgroundModels.html}} (\texttt{gll\_iem\_v07.fits} and \texttt{iso\_P8R3\_SOURCE\_V2.txt}). We apply a zenith angle cut of $\theta < 90^{\circ}$ to minimize the contamination from $\gamma$ rays produced in the Earth’s upper atmosphere. To avoid potential spurious effects in the periodicity analysis, we remove the time periods coinciding with solar flares and $\gamma$-ray bursts detected by the LAT. Finally, the standard data quality cuts ($ \rm DATA\_QUAL > 0) \&\& (LAT\_CONFIG == 1$) are applied. 

We use the \texttt{P8R3\_SOURCE\_V2} instrument response functions to carry out a binned analysis in the 0.1-800 GeV energy range using ten bins per decade in energy and 0.1$^{\circ}$ spatial bins. Considering a full-time range of 2008 Aug 04 15:43:36 UTC to 2020 Dec 10 00:01:26 UTC, we perform a maximum likelihood analysis. Each source is modeled using the spectral shapes and parameters reported in 4FGL as starting values for the fit. We first perform a fit of the ROI by means of the \textit{fermipy} method ``optimize'' to ensure that all spectral parameters are close to their global likelihood maxima. This is done by iteratively optimizing the components of the ROI model in sequential steps, starting from the largest components.\footnote{\url{https://fermipy.readthedocs.io/en/latest/fitting.html}}  
The initial results are evaluated for potential newly-detected sources with an iterative procedure by a test statistic (TS) map since our data span a different integration time to 4FGL. The TS is defined as $2\log(L/L_0)$, where \textit{$L_0$} is the likelihood without the source and \textit{$L$} is the likelihood of the model with a point source at a given position. TS=25 corresponds to a statistical significance of $\gtrsim4.0\sigma$~\citep[according to the prescription adopted in][]{mattox1996, 4fgl_catalogue}. We produce a TS map with a putative point source at each map pixel and evaluate its significance over the current best-fit model. The test source is modeled with a power-law spectrum where only the normalization can vary in the fit process, whereas the photon index is fixed at 2. We search for significant peaks (TS$>$25) in the TS map, with a minimum separation of 0.5$^{\circ}$ from existing sources in the model. We add a new point source to the model at the position of the most significant peak found. Then, the ROI is fitted again, and a new TS map is produced. This process is iterated until no more significant excesses are found, generally leading to adding two point-sources. 

\begin{figure*}[!ht]
	\centering
	\includegraphics[scale=0.2295]{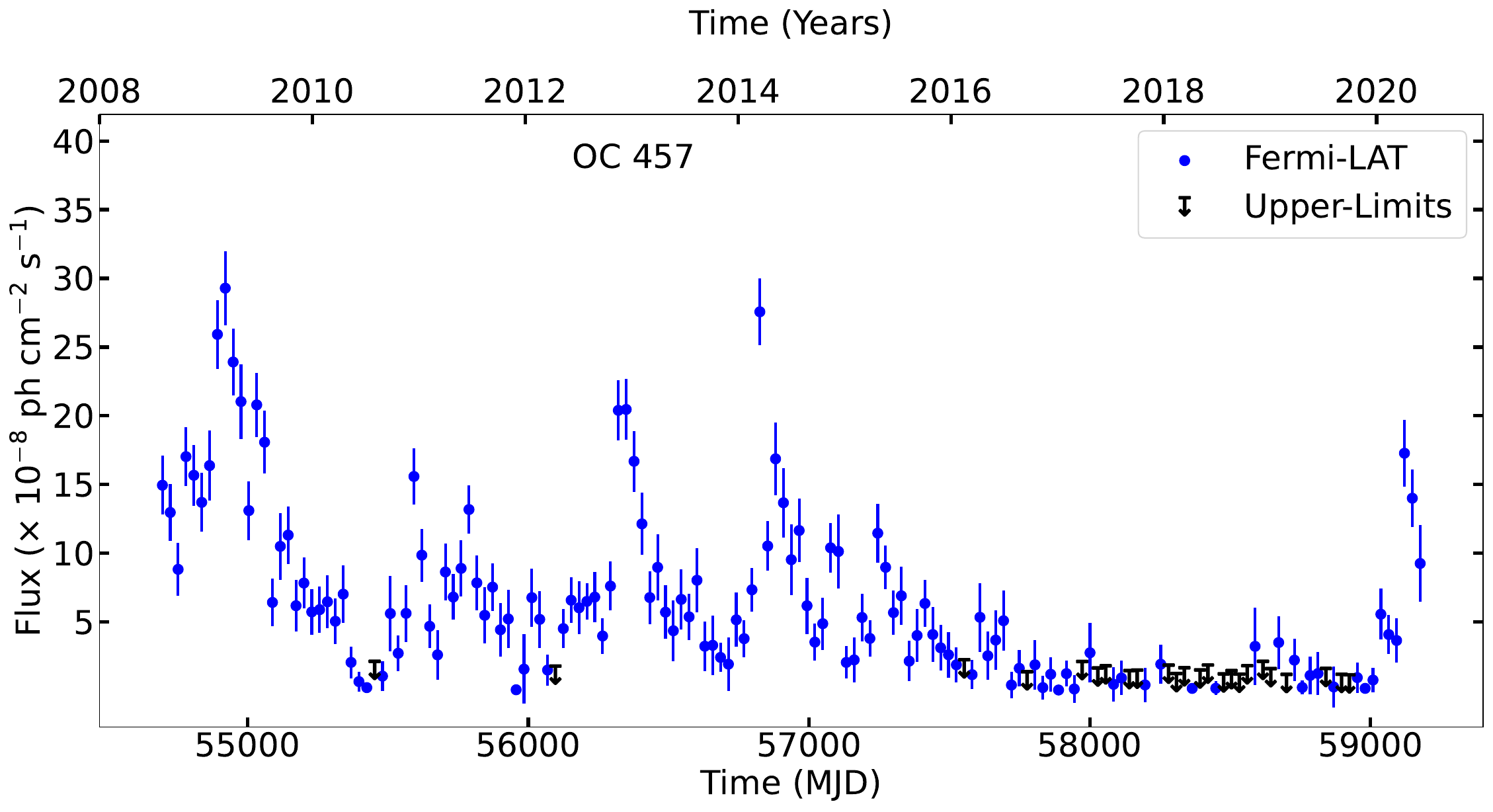}
        \includegraphics[scale=0.2295]{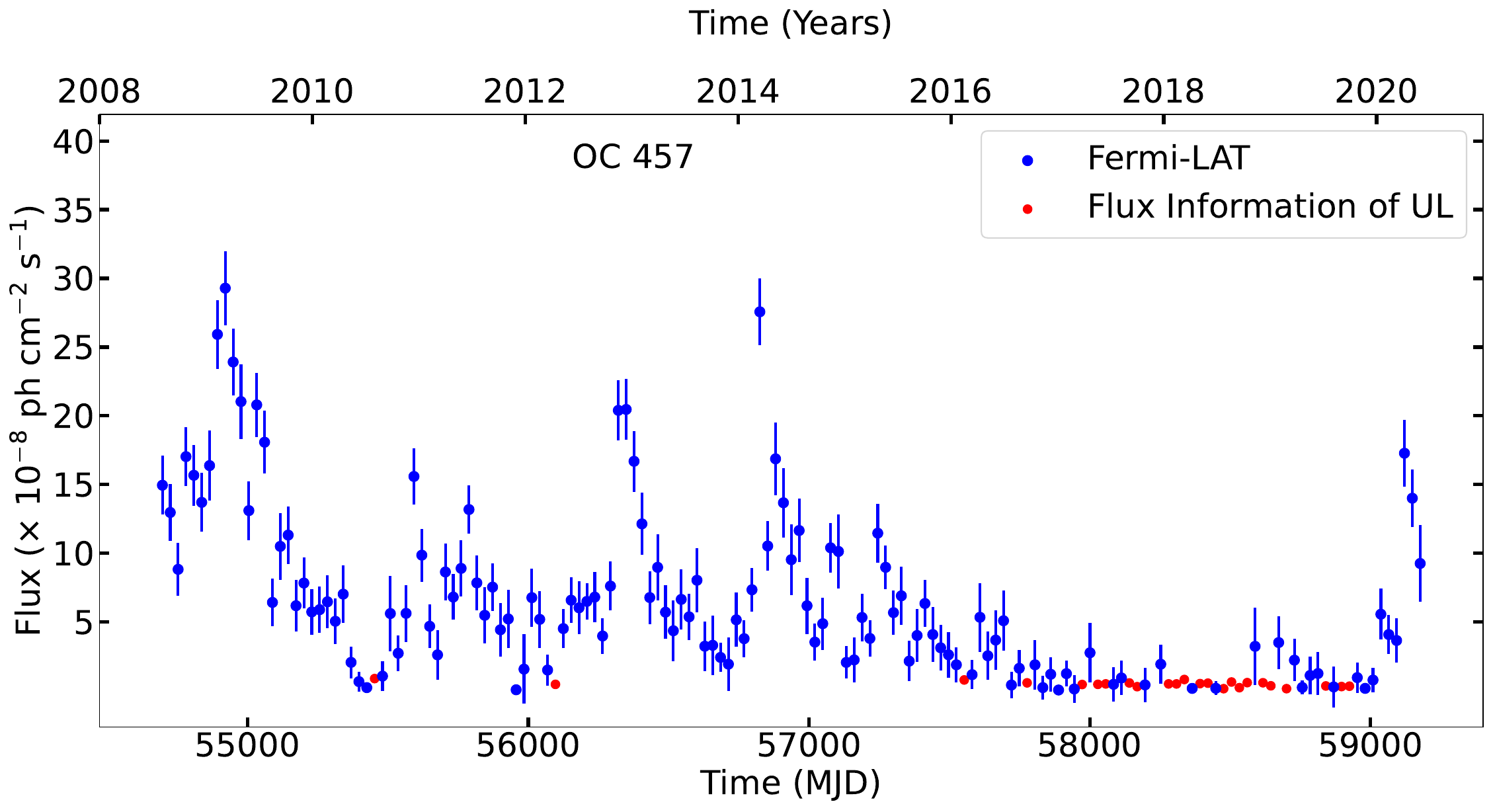}
	\caption{\textit{Left}: Light curve of the blazar OC 457. \textit{Right}: Light curve used for the trend-search analysis, where upper limits are replaced with flux values that maximize the likelihood function for each specific time bin (red points).}
	\label{fig:lc_ul}
\end{figure*}

Each LC is produced by splitting the data into 28-day bins and performing a full likelihood fit in each time bin. The best-fit ROI model obtained from the full-time interval analysis is employed to get the likelihood fit of each time bin. Initially, we try a fit the normalization of the target and of all sources in the inner $3^\circ$ of the ROI, along with the diffuse components. For a non-converging fit, the number of free parameters is progressively and iteratively restricted until the fit converges. This iterative process starts by fixing sources in the ROI that are weakly detected (i.e., with TS$<$4). After that, we fix sources with TS$<$9. Then, we fix sources up to $1^\circ$ from the ROI center and those with TS$<$25. Finally, all parameters except the target source’s normalization are fixed. We provide fluxes in each time bin when TS$>$1, yet when TS$<$1, we use the likelihood profile of the flux distribution for extracting the 95\% upper limit (see Figure \ref{fig:lc_ul}), as done by \citet{penil_2022_periodicity}.

Incorporating upper-limit data could introduce bias, leading to underestimation of real flux. An alternative is to exclude upper limits, treating them as missing data, but this creates unevenly distributed LCs. Irregular sampling might distort potential long-term trends, introduce artifacts, and affect the detection of dominant timescales in the LC analysis \citep{penil_2020}. By incorporating upper limits, we capture a broader range of variability in the blazar’s behavior, allowing for a more realistic understanding of the source's long-term variability behavior.

\begin{figure}[!ht]
	\centering
	\includegraphics[scale=0.223]{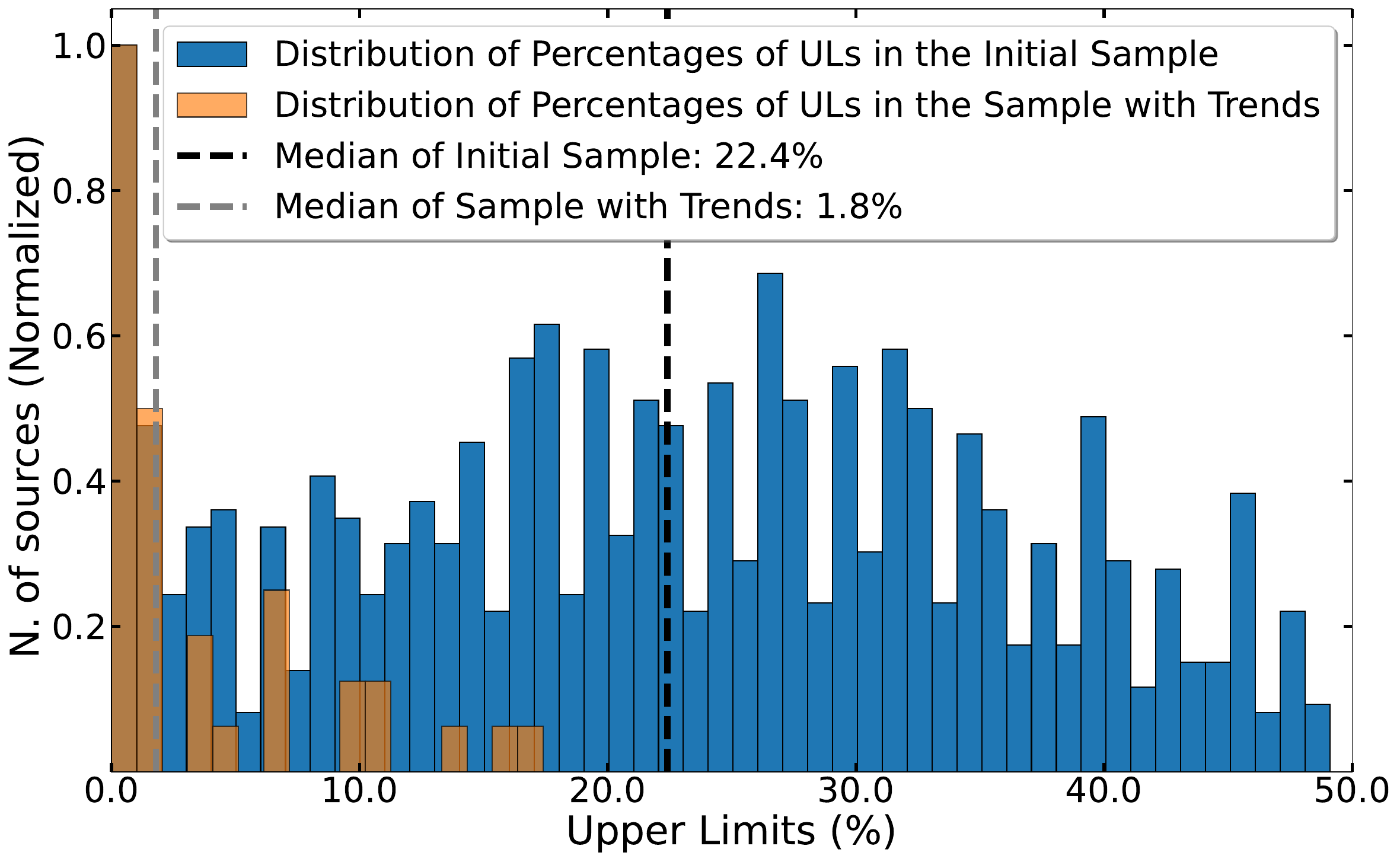}
        \caption{The distribution of upper limits across the analyzed LCs. The blue bars represent the normalized distribution of upper limits in the initial sample, with normalization based on the bin containing the highest number of sources. This distribution shows a peak concentration at 0\% (i.e., detections in all time bins) and a median of 22.4\%, indicating that most sources have moderate upper limits. In contrast, the orange bars illustrate the subset of sources exhibiting trends, also normalized based on the bin with the highest number of sources. This subset shows a lower overall percentage of upper limits, with a peak at 0\% (detections in all time bins) and a median of 1.8\%. It is evident that trends are detected more efficiently when there are no gaps in the LCs.}\label{fig:ul_distribution}
\end{figure}

\begin{figure*}[!ht]
	\centering
	\includegraphics[scale=0.62]{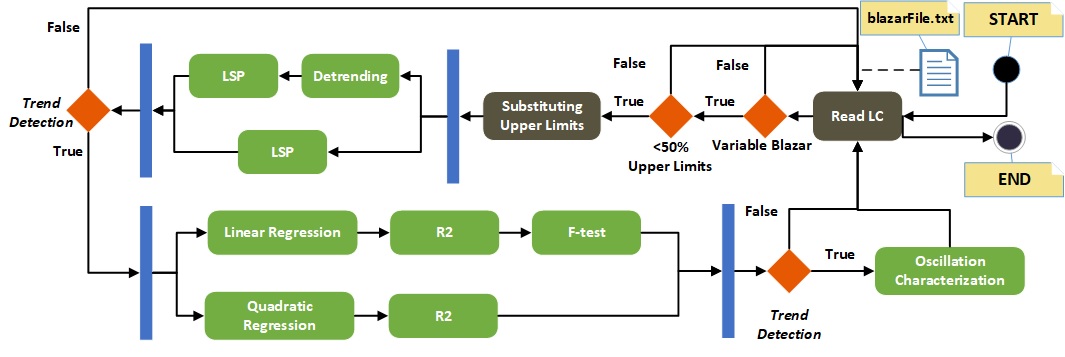}
	\caption{Trend-search pipeline depicted in an activity diagram of Unified Modeling Language.}
	\label{fig:study_flow}
\end{figure*}

\begin{figure*}[!ht]
	\centering
        \includegraphics[scale=0.223]{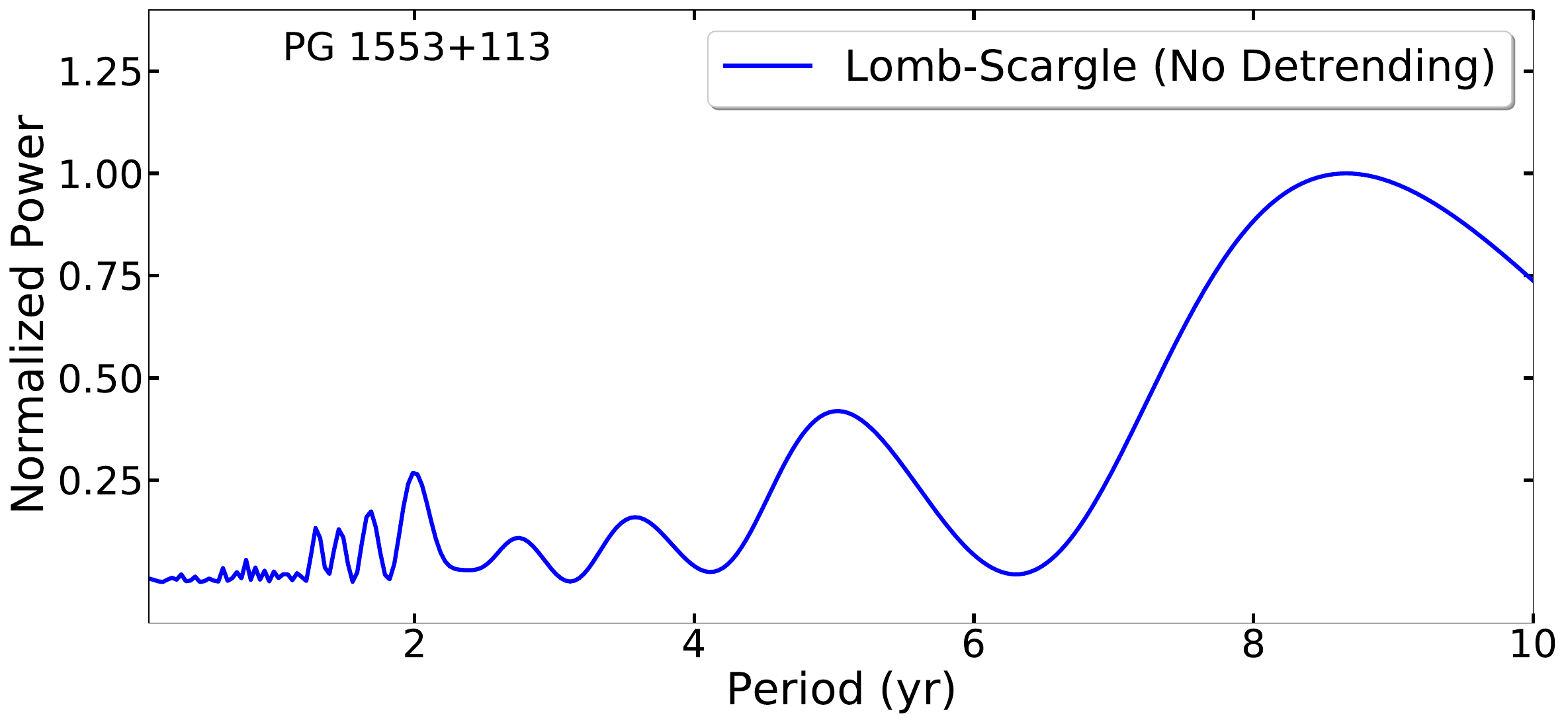}	
        \includegraphics[scale=0.223]{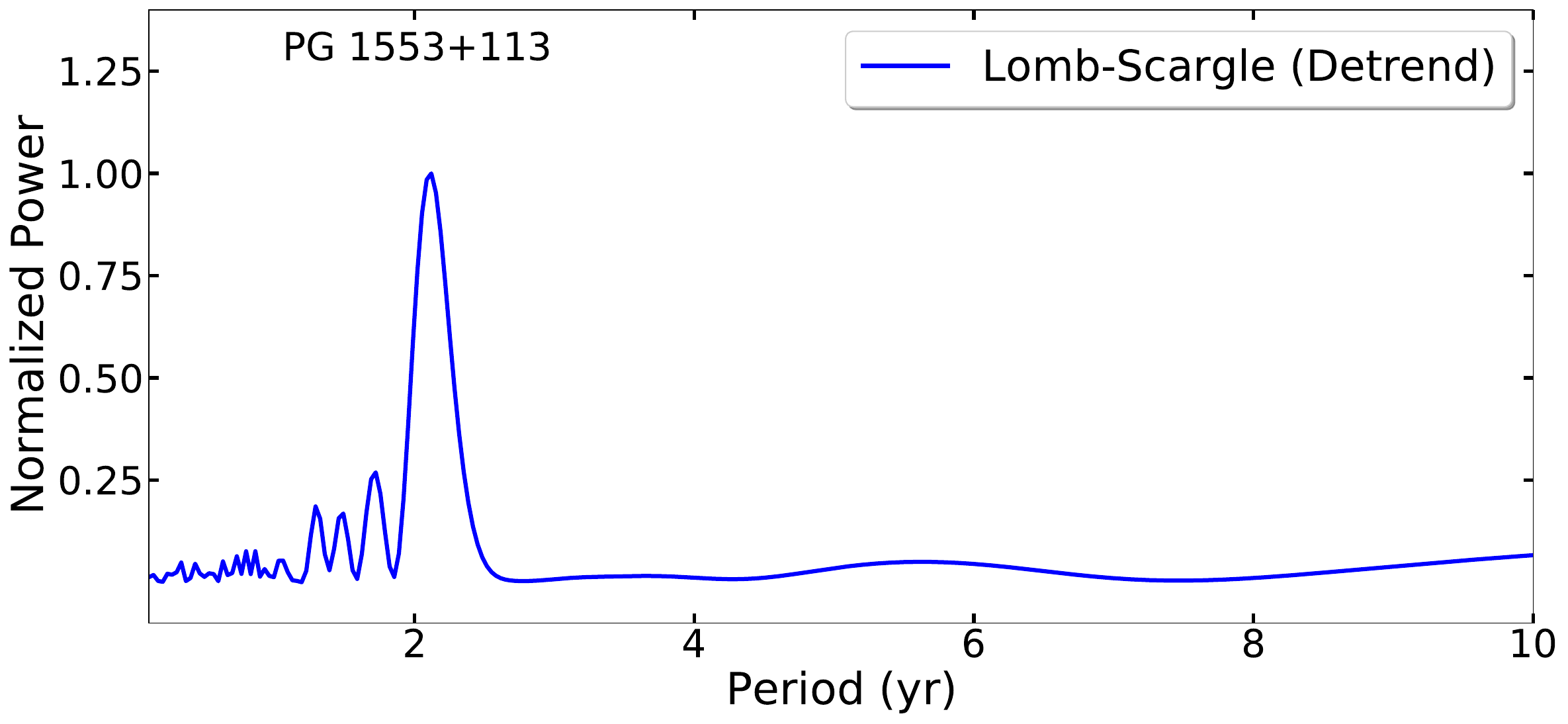}
	\caption{LSP analysis of PG~1553+113: ({\it Left}): Without detrending the LC. ({\it Right}): Detrending the LC. Note that the periodicity search is more efficient after detrending.}
	\label{fig:dft_analysis}
\end{figure*}

\section{Methodology} \label{sec:methodology}

\subsection {Pre-analysis Selection} 
We have developed a systematic methodology for the purpose of identifying long-term trends spanning the entire extension of the LC, which typically exceeds a decade in duration. An overview of our pipeline is illustrated in Figure \ref{fig:study_flow}. As a preliminary step, we filter the initial jetted AGN sample by selecting those that exhibit variability. This selection is based on the variability index, which is a numerical measure used to assess and quantify the level of variability exhibited by blazars according to changes in the blazar's brightness over time \citep[][]{4fgl_catalogue}. A jetted AGN is considered variable if the variability index is $\geq$18.48, as specified in the 4FGL-DR2 catalog \citep[][]{4fgl_catalogue}. Out of the initial sample, 1620 jetted AGN, constituting 48.9\% of the total, meet this criterion. Subsequently, we further filter this variable jetted AGN subset by retaining only those with $\leq$50\% of upper limits in their LCs, following the criteria established by \citet{penil_2020}. This filter results in a new subsample comprising 1492 jetted AGN, accounting for 45.1\% of the initial sample. In these sources, upper limits in their LCs are replaced with the flux value that maximizes the likelihood function for that specific time bin (see $\S$\ref{sec:fermianalysis}). This process is illustrated in Figure \ref{fig:study_flow} as the ``Substituting Upper Limits'' stage. 

The distribution of upper limits in our sample is shown in Figure \ref{fig:ul_distribution}, indicating that the majority of sources have upper limits ranging from 0\% to 30\% (70\% of the sample), with the largest concentration observed at 0\%. The distribution median is 22.4\%, suggesting that most datasets contain relatively low proportions of upper limits.

\begin{figure*}[!ht]
	\centering
        \includegraphics[scale=0.223]{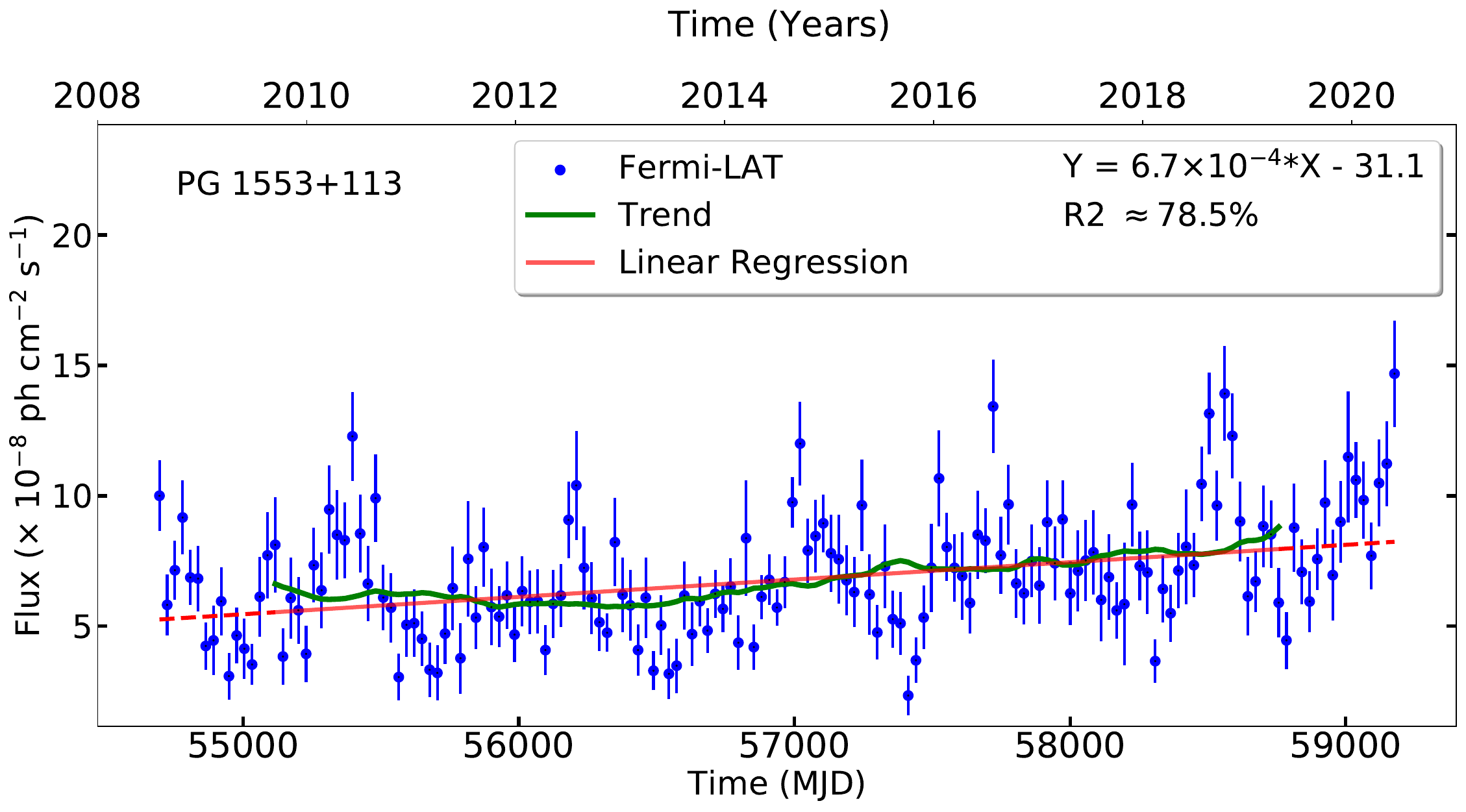}
	\includegraphics[scale=0.223]{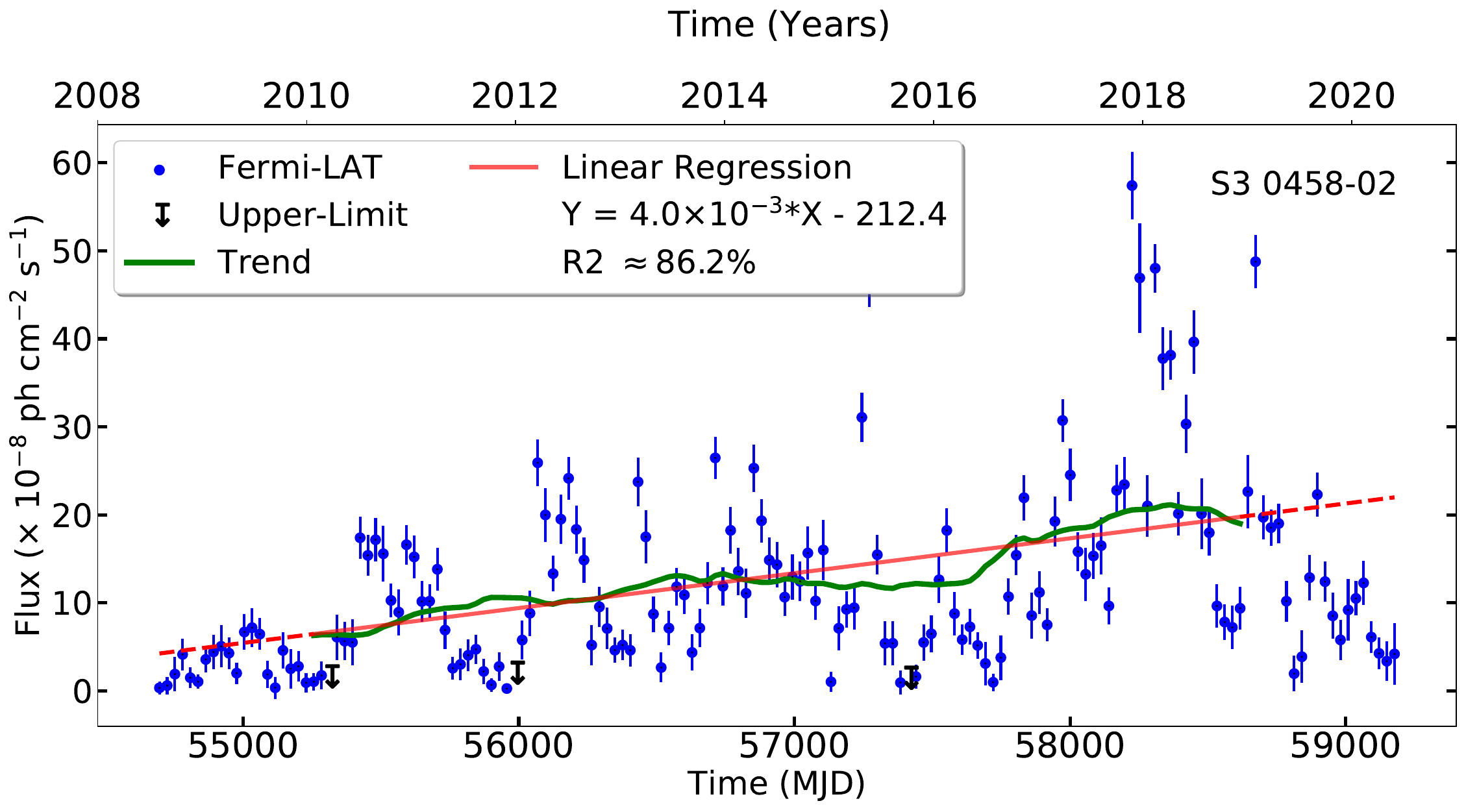}
        \includegraphics[scale=0.223]{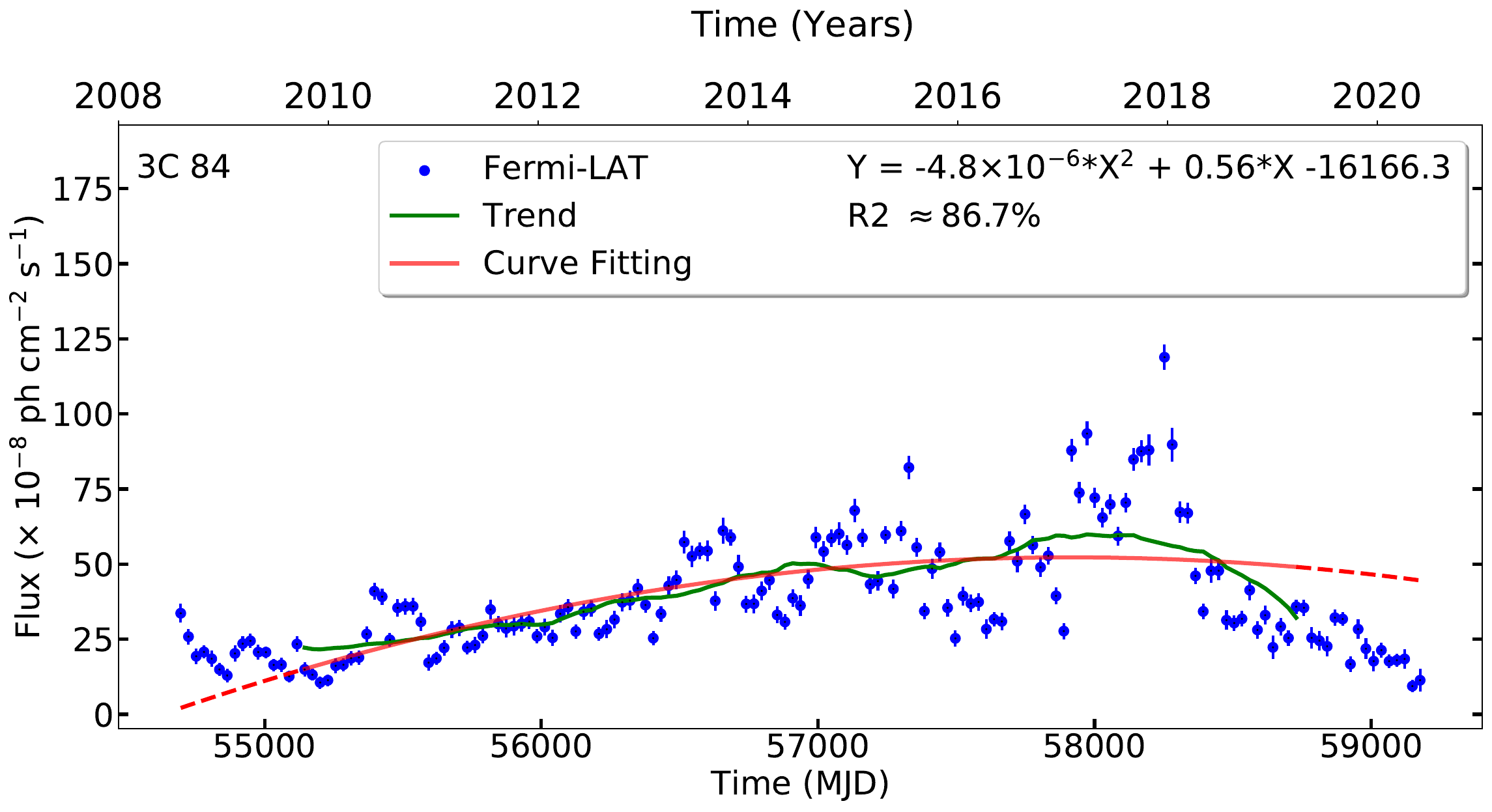}
	\includegraphics[scale=0.223]{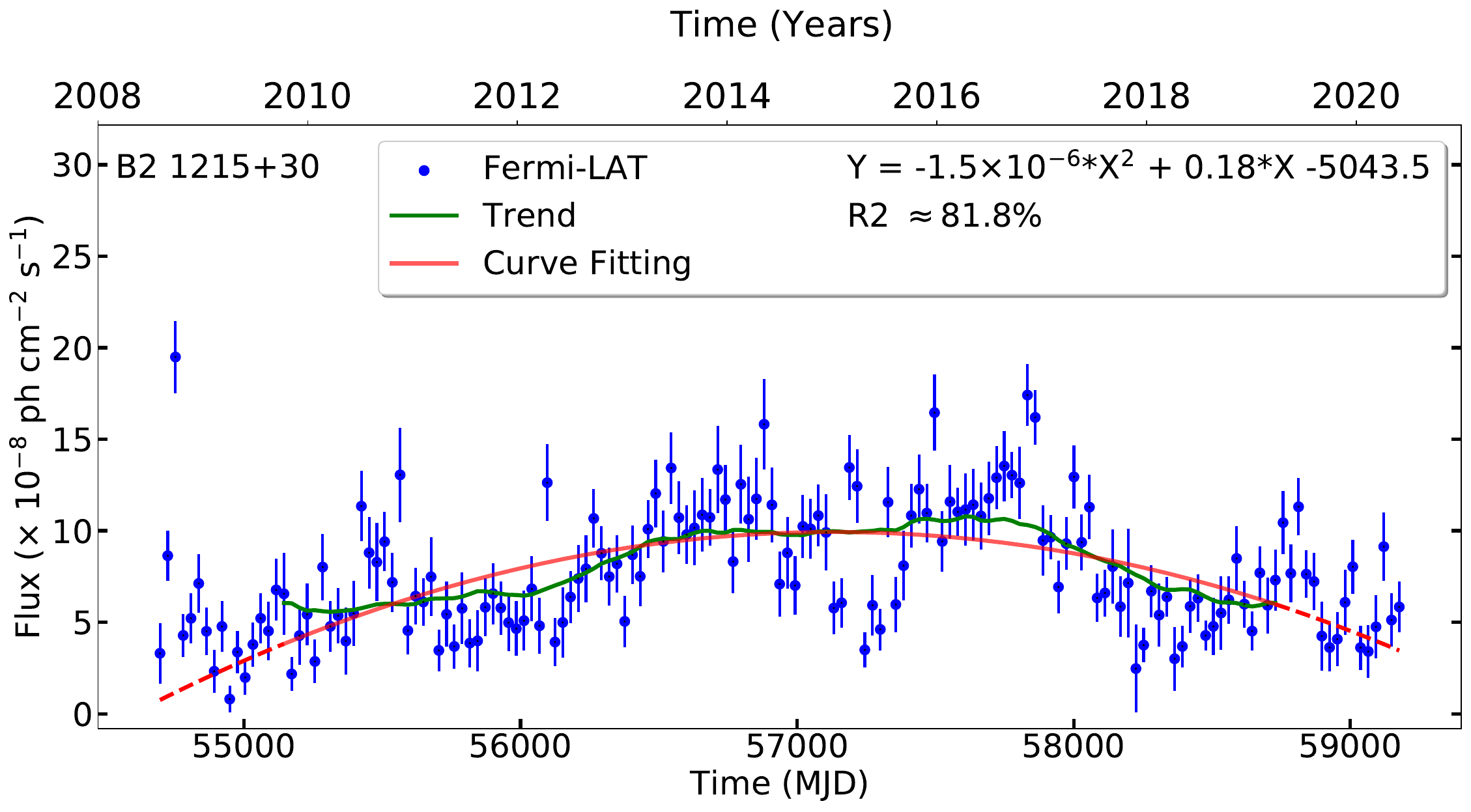}
	\caption{\textit{Top}: Two examples of LCs with long-term linear trends. \textit{Left}: PG~1553+113 with the additive oscillations (see $\S$\ref{sec:type_oscillations}). \textit{Right}: S3 0458$-$02 with multiplicative oscillations (see $\S$\ref{sec:type_oscillations}). \textit{Bottom}: Two examples of LCs with long-term quadratic trends. \textit{Left}: 3C 84 with multiplicative oscillations. \textit{Right}: B2 1215+30 with multiplicative oscillations. The upper limits are denoted by down arrows. The green line represents the underlying trend extracted by the function \textit{seasonal\_decompose}. Note that the green line covers a shorter time span than the full LC duration, a result of the trend decomposition applied by this function. The red line indicates the fit of the green line, with the dashed red line extending the fitted line across the entire LC.}.
	\label{fig:trend_types}
\end{figure*}

\subsection{Detection of Long-term trends}
In the initial analysis stage, we employ the Lomb-Scargle periodogram as our primary tool \citep[LSP,][]{lomb_1976, scargle_1982}. The LSP provides a dominant peak in the periodogram (associated with the dominant frequency inferred from the LC), regardless of whether this peak is associated with a QPO or not \citep[which have to be determined by specific statistical studies, e.g.,][]{ackermann_pg1553, penil_2022_periodicity}. In this methodology, we are not focused on inferring any presence of a potential QPO; we simply use the impact of the long-term trends in the LSP. The advantage of using the LSP lies in its rapid computation of the periodogram, facilitating the swift initial selection of objects with the potential to exhibit a trend. Specifically, we conduct the LSP in two distinct scenarios, as depicted in Figure \ref{fig:study_flow}: one with detrending of the LC and the other without detrending. This dual approach provides a fast means to identify changes in the frequency domain potentially produced by a trend, which is an essential feature when dealing with datasets comprising hundreds of sources. This initial selection process optimizes the handling of large samples, making it a practical necessity for efficiently managing the volume of data involved in our study.

As depicted in Figure \ref{fig:dft_analysis}, the application of the LSP is focused within the range of 1 to 10 years. In the context of periodicity-search analysis, it is advisable to detrend the signal as a preprocessing step \citep[e.g.,][]{detrend_welsh}. This step is crucial because a trend within the data has the potential to contaminate the low-frequency components of the Power Spectral Density (PSD), which measures the variability power concerning frequency. Trends can mislead and falsely indicate periodic patterns within the time series, potentially leading to the erroneous detection of false periodicity \citep[][]{mcquillan_trend_fake_detection}. Furthermore, LCs with finite lengths are susceptible to the influence of red noise leakage, especially at low frequencies. This red-noise effect can manifest itself as a rising or falling trend spanning the entire duration of the LC, and it can have an observable impact on the shape of the PSD \citep[e.g.,][]{vaughan_criticism}. 

Nevertheless, the process of detrending data can introduce certain correlations, particularly when transformations like differencing are applied. Differencing, for instance, might lead to the emergence of autocorrelations if the differenced data still retains some underlying patterns. In our analysis, we use linear detrending\footnote{We use \texttt{detrend} function of the Python package \texttt{Scipy.signal}}. Linear detrending involves subtracting a straight line from the original data to accentuate intrinsic behavior. The application of this function does not account for breakpoints and employs linear detrending. 

Linear detrending can create the appearance of increased noise correlation in the detrended data, even though it effectively reduces autocorrelation in the time series. This elevated noise correlation can occur when the linear component is not entirely removed or when the original data's noise contains non-random systematic patterns or fluctuations. In our specific scenario, we can confidently eliminate the latter possibility: blazar LCs, as demonstrated in prior research \citep[e.g.,][]{vaughan_fractional_variability}, are known for displaying red noise characteristics, which arise from intrinsic source randomness, in addition to observational errors.  

After applying our trend-detection procedure, we systematically excluded jetted AGN that showed no significant change in their PSD. We define a change as significant when the difference between the frequencies of the highest peaks of the LSPs obtained with and without detrending exceeds 20\%. We aim to set this threshold constraining enough to avoid identifying random fluctuations as significant. In datasets with high variability, as in our case, smaller thresholds might lead to too many false positives. This is particularly relevant in data with intense background noise, as in the case of $\gamma$-ray LCs. Additionally, in studies of periodicity, as \citet{penil_2022_periodicity}, the uncertainty of the periods is typically around 10\% of the period itself. We also choose the proposed threshold to mitigate the effects of changes caused by variations in the highest peak of the periodogram, which are associated with the inherent uncertainty in the frequencies. Conversely, it is necessary that the threshold is not so restrictive that it overlooks potentially meaningful trends. Consequently, this is a reasonable threshold for detecting any significant change within a period range [1-10] years in periodograms affected by the red noise and to have the contamination of spurious trend detection $<$0.5\% (see $\S$\ref{sec:fpd}).

This filtering process resulted in a refined sample size, reducing it to 337 jetted AGN, which constitutes 10.1\% of the initial sample. As an illustrative example of this long-term trend detection methodology, we consider the case of PG~1553+113, which is displayed in Figure \ref{fig:dft_analysis}. Our analysis revealed a discernible long-term trend in the LC of PG~1553+113 since the periodograms with/without detrending are notably different.

\subsection{Types of Long-term trend}
In the subsequent stage of our pipeline, depicted in Figure \ref{fig:study_flow}, we embark on a search for two distinct types of trends within the data: linear trends and quadratic trends, as illustrated in Figure \ref{fig:trend_types}. For linear trend analysis, we employ a linear regression\footnote{We use the \texttt{LinearRegression} function of the Python package \texttt{Scikit-learn}, which is optimized specifically for linear fitting}, characterizing the parameters of the linear trend\footnote{We use the \texttt{seasonal\_decompose} function of the Python package \texttt{Statsmodels}. The parameters for the function are set as``Multiplicative'' for the ``Model'' parameter and ``40'' for the ``Period'' parameter.\label{fn:method}} as Figure \ref{fig:trend_types} shows. As for the quadratic trend analysis, the underlying trend is obtained\footref{fn:method}, and the model $aX^2+bX+c$ is employed for the quadratic fitting\footnote{We use the Python package \texttt{Scipy} and the function \texttt{curve fit}}, as illustrated in Figure \ref{fig:trend_types}. The units of ``a'' are 10$^{-8}$ ph cm$^{-2}$ s$^{-1}$ days$^{-2}$), for ``b'' are 10$^{-8}$ ph cm$^{-2}$ s$^{-1}$ days$^{-1}$), and for ``c'' are 10$^{-8}$ ph cm$^{-2}$ s$^{-1}$. This approach allows us to thoroughly explore both linear and quadratic trends within the jetted AGN LCs, ensuring analysis of potential long-term variations.

In our trend analysis, we rely on the R-squared (R$^{2}$) criterion to assess the quality of trend fits. R$^{2}$ is a statistical measure that assesses the goodness of fit of a regression model. Specifically, R$^{2}$ is defined as the proportion of the variance in the dependent variable that is predictable from the independent variable(s) in a regression model. The value of R$^{2}$ ranges from 0 to 1, where the higher values denote a better fit and suggest that the model can better explain the variations in the data. Determining what constitutes an acceptable level of R$^{2}$ for a good fit can be context-dependent, and there is no universally defined threshold since it hinges on the specifics of the application. The choice of the R$^{2}$ value depends on the particular application scenario. In \citet{hair_r2_2011}, three different levels of R$^{2}$—25\%, 50\%, and 75\%—are outlined and categorized as \textit{weak}, \textit{moderate}, and \textit{substantial}, respectively. As a guiding principle, we opt for a more conservative approach, considering a regression model with R$^{2}$ $\geqslant$~75\% to ensure greater stringency in our analysis. This higher threshold is chosen to enhance the robustness of our results.

In our methodology, we initially perform linear and quadratic trend fits to the LC and select the fit that yields the higher R$^{2}$ value.

This approach ensures that we choose the most appropriate trend representation for the data. Subsequently, we evaluate whether the goodness of fit for the selected trend is $\geqslant$~75\%. If it meets this criterion, the jetted AGN is considered for further analysis and potential inclusion in our study.

When a linear trend is identified, we conduct an additional assessment to verify if the slope of the linear fit significantly differs from zero. To accomplish this, we employ the F-test\footnote{The F-test is equivalent to the likelihood ratio test \citep[][]{lu_ftest_LRT}}. The F-test is a statistical method used to compare the fits of different models. Typically, it entails comparing the fit of a more complex model to that of a simpler model in order to assess whether the increased complexity significantly enhances the fit. The null hypothesis in this test posits that the trend can be adequately represented by a constant. If the p-value associated with the F-test is $\rm{\leq0.01}$, we reject the null hypothesis. The selected p-value emphasizes the caution to prevent the misinterpretation of a trend as merely the median of the flux, which can be represented by a horizontal line. In that way, the inferred linear trends are significantly different from a no-trend scenario. Therefore, this p-value indicates that the linear regression model provides a better fit to the data than a constant, further justifying the inclusion of the jetted AGN in our analysis. This approach ensures that jetted AGN with meaningful and statistically significant trends are selected for further investigation.

\subsection{Oscillations Evolution with the Long-term Trend}\label{sec:type_oscillations}
In some cases, the LCs of certain sources exhibit oscillations, which can be either quasi-periodic or aperiodic in nature, as illustrated in Figure \ref{fig:trend_types}. These oscillations exhibit distinct behaviors based on their relationship with the underlying trend and amplitude, leading to the identification of two primary types: additive and multiplicative: 
\begin{enumerate}[label={[\arabic*]}]
\item Additive Oscillation: In this case, the amplitude of the flux remains independent of any changes in the long-term trend, as exemplified in Figure \ref{fig:trend_types}. 
\item Multiplicative Oscillation: Conversely, multiplicative oscillations are characterized by the amplitude of the flux changing proportionally to variations in the trend, as depicted in Figure \ref{fig:trend_types}. 
\end{enumerate}

Indeed, the variability in the amplitude of the oscillations due to the trend can mask the presence of QPOs when using conventional techniques. Therefore, it is crucial to recognize and account for this specific property when analyzing blazars exhibiting multiplicative oscillations. Specialized methodologies may be required to effectively identify and characterize QPOs in these cases, taking into consideration the complex interplay between the long-term trend and oscillatory behavior.

In our trend-search pipeline, as illustrated in Figure \ref{fig:study_flow}, we have incorporated the ``Oscillation Characterization'' stage dedicated to identifying and distinguishing between additive and multiplicative oscillations within the LCs. Initially, we apply a detrending procedure to the LC, removing the long-term trend component. This step is essential to isolate the oscillatory behavior from the overall LC variation. Within the detrended LC, we identify the high-flux states\footnote{We use the \texttt{find\_peaks} function of the Python package \texttt{Scipy.signal}. This function identifies local maxima in an array based on specific conditions. In our case, we set the threshold to 40\%-60\% of the maximum flux of the LC, ensuring at least three peaks are included within this range.}. Then, we perform a linear regression analysis of the flux associated with one of these peaks. The key criterion for classifying the type of oscillation is the slope ($s$) obtained from the previous linear regression. Specifically, if the absolute value of the slope is $\mid$$s$$\mid\approx 0$, we categorize the oscillation as additive. Conversely, if $\mid$$s$$\mid$ is significantly different from zero, we classify the oscillation as multiplicative. 

The distinction is established through the utilization of the F-test. In this test, the null hypothesis assumes that the peaks can be sufficiently described by a constant, implying that the oscillation is additive. In our analysis, we employ a significance level with a p-value of $\rm{\leq0.05}$ to reject the null hypothesis. The chosen p-value warrants consideration of an important factor: even in situations with potential additive oscillations, the presence of short-term local variability could introduce the possibility of incorrect inferences regarding the type of these oscillations.

\section{Results} \label{sec:results}

\subsection{Emerging Trends: Search Findings}
Following our comprehensive analysis, the sample of jetted AGN exhibiting long-term trends in $\gamma$ rays comprises a total of 40 sources, which are detailed in Table \ref{tab:candidates_list}\footnote{Information obtained from the 4FGL-DR2 catalog \url{https://fermi.gsfc.nasa.gov/ssc/data/access/lat/10yr_catalog/}}, along with relevant trend properties provided in Table \ref{tab:trend_properties}. 
This refined sample represents just 1.3\% of the initial jetted AGN sample. Upon closer examination of these sources, it becomes evident that a significant majority of them are characterized by linear trends (e.g., Figure \ref{fig:trend_types} and Figure \ref{fig:lcs_blazars}). Specifically, 32 out of the 40 sources display linear trends, underscoring the prevalence of this type of behavior. Therefore, The remaining 8 sources within this subset exhibit quadratic trends (e.g., Figure \ref{fig:trend_types} and Figure \ref{fig:lcs_blazars}).

Analyzing the oscillatory behavior within this sample, about 80\% of these sources manifest multiplicative oscillations, indicating a complex interplay between the long-term trend and amplitude variations in the oscillations.

Finally, the distribution of upper limits in our sample with trends is shown in Figure \ref{fig:ul_distribution}, indicating that the sources have upper limits ranging from 0.0\% to 17.4\%, with the largest concentration observed at 0\%. The distribution median is 1.8\%, suggesting that most datasets contain relatively low proportions of upper limits. This limited upper limits ratio suggests a reduced likelihood that upper limits significantly distort the trend characterization of our data.

\subsection{Evaluation of methods against noise} \label{sec:methods_against_noise}
We also assess the methodology's performance when dealing with time series data that exhibit long-term trends contaminated by noise. Through statistical analysis, we aim to evaluate how our methodology performs across various noise conditions, particularly when these long-term trends are present. We consider four different noises: ``white noise'', ``pink noise'' (generating random power-law indices in the range [$0.8 - 1.2$])\footnote{This range is defined based on the findings of \citet{penil_2022_periodicity}\label{fn:ranges}}, ``red noise'' (generating random with power-law indices in the range [$1.8 - 2.2$]), and ``bending power-law'' according to the model \citet{chakraborty_bending_power_law} defined by a bending frequency and a spectral index. We randomly generate values for the spectral index in the range [$1.4 - 1.8$]\footref{fn:ranges} and for the bending frequency in the range [$0.1-0.8$]\footref{fn:ranges}. Our analysis examines the two categories of long-term trends. For linear trends, we randomly select slope values from the range of $1\mathrm{x}10^{-3}$ to $9\mathrm{x}10^{-3}$, as most of the detected linear trends fall within this range (Table \ref{tab:trend_properties}). Meanwhile, for quadratic trends, we set the values of the $a$ parameter within the range of $-1\mathrm{x}10^{-6}$ to $-9\mathrm{x}10^{-6}$, and the $b$ parameter within the range of 0.1 to 1 (as detailed in Table \ref{tab:trend_properties}). To conduct a comprehensive assessment, we simulate a total of 100,000 artificial LCs for each type of trend and noise condition. First, we apply the detrending-non-detrending procedure in the LSP to denote a change in the periodogram due to the presence of a slope in such noisy LCs. As a result, the LSP infers a change in the periodogram in 98.9\% of the cases for ``white noise'', 93.3\% for ``pink noise'', 96.5\% for ``red noise'', and 95.7\% for the ``bending power-law'' model. These values represent the extent to which the LSP successfully detects a modification in the frequency domain when there is a trend in noisy LCs.

Then, for the LCs where the LSP indicates a change in the periodogram, we estimate the slope, the uncertainty and the R$^{2}$. We use the same criteria previously defined: R$^{2}$$\geq$75\% to assess the quality of trend fit. As a result, we obtain a liner-trend detection of 100\% for all the noise cases.

We determine the slope of the noisy LC and compare it with the original trend shown in the simulated LCs with R$^{2}$$\geq$75\%. To ensure accurate selection, the gap between the slopes of the noisy LC and the original trend should remain within a $\pm1.0\times10^{-4}$ margin of error as most of the uncertainties of detected linear trends are within this range (Table \ref{tab:trend_properties}). This approach allows us to assess how our methodology performs in detecting trends in the presence of noise while maintaining a reasonable level of accuracy and correlation. Consequently, in the case of ``white noise'', we obtained a compatible slope of 91.2\%, for ``pink noise'', it was 52.7\%, for ``red noise'', 88.5\%, and for the ``bending power-law'' model, it stood at 86.8\%.

The total efficiency rate is determined by the proportion of correctly inferring the same trend as the one generating the LC. Hence, the detection rate for ``white noise'' stands at 90.2\%, for ``pink noise'' at 49.2\%, for ``red noise'' at 85.4\%, and for the ``bending power-law'' at 83.1\%. These values serve as indicators of how the methodology is robust against distortion created by the noise.

Regarding quadratic trends, the LSP identifies a variation in the periodogram due to the presence of a trend in 100\% of the cases for ``white noise'', 98.6\% for ``pink noise'', 98.3\% for ``red noise'', and 97.6\% for ``bending power-law''. Among these LCs where the LSP indicates the presence of a trend, the R$^{2}$ criterion selects 94.8\% for ``white noise'', 82.7\% for ``pink noise'', 94.9\% for ``red noise'', and 93.9\% for the ``bending power-law''. We utilize a margin of error of $\pm1.0\times10^{-7}$ and $\pm1.0\times10^{-4}$ for inferring the parameters $a$ and $b$ of the quadratic trend, respectively, based on the majority of uncertainty values outlined in Table \ref{tab:trend_properties}. We apply this detection criterion to the LCs determined by the R$^{2}$ criterion. As a result, in 100\% of cases for each noise type, the estimated parameters $a$ and $b$ correspond to those of the simulated LC with the quadratic trend. Lastly, to evaluate the overall effectiveness of the methodology in maintaining consistency in the inferred parameters despite the noise, we calculate the total detection rates. The total detection rate for ``white noise'' is 94.8\%, for ``pink noise'' is 81.5\%, for ``red noise'' is 93.3\%, and for the ``bending power-law'' is 91.6\%.

Quadratic trends exhibit a higher detection rate than their linear counterparts. This distinction can be attributed to the nature of noise --particularly pink and red noise-- which predominantly affects lower frequencies and is more likely to obscure slow-changing linear trends, particularly in datasets exhibiting significant variability. Quadratic trends, by contrast, require a specific pattern of change that is less likely to be randomly replicated by noise, making them more robust from the background noise and facilitating their identification.

Interestingly, the specific type of noise pattern does not appear to have a critical impact on the results, as the outcomes for red and bending power-law noises are fairly similar for both linear and quadratic trends. However, when considering "pink noise," its impact is more substantial compared to other types of noise. Upon comparing the detection rates for each type of noise, we observe that the influence of "pink noise" is $\approx$45\% higher than the rest for linear trends and $\approx$12\% for quadratic trends. This could be because ``pink noise'' resembles many natural astrophysical processes, including variations in jetted AGN emissions that can occur due to accretion disk dynamics or jet instabilities \citep[e.g.,][]{abdo_variability, rieger_2019, bhatta_variability}. Therefore, the similarity between the ``pink noise'' and potential signals makes it difficult to differentiate between intrinsic trends and noise-induced variations, leading to a lower detection rate.

All the results of the previous tests are summarized in Table \ref{tab:experiment_1}.

\subsection{Evaluation of methods against flares}\label{sec:flares}
We conduct an experiment by randomly injecting flares into 100,000 LCs with linear trends and another 100,000 LCs with quadratic trends. Each LC was then contaminated with white noise, chosen due to its high detection rates (see $\S$\ref{sec:methods_against_noise}). This type of noise serves as a baseline to isolate the impact of the flares, minimizing interference from other stochastic phenomena. In each LC, a flare is randomly introduced, characterized by two key parameters: amplitude and duration.

Regarding amplitude, we draw from the \textit{Fermi} All-sky Variability Analysis (FAVA) catalog\footnote{\url{https://fermi.gsfc.nasa.gov/ssc/data/access/lat/fava_catalog/}} \citep{fava_catalog}, considering two amplitude levels: 
\begin{enumerate}
\item A1: the median-flux of flares in the FAVA catalog, which is 26.5$\times$ 10$^{-8}$ ph cm$^{-2}$ s$^{-1}$. 
\item A2: the median of the brightest flares in the FAVA catalog, which includes flares with a flux greater than the previous median, which is 70.1$\times$ 10$^{-8}$ ph cm$^{-2}$ s$^{-1}$.
\end{enumerate} 

For the flare duration, we considered two cases: 2 months and 12 months. These durations allow us to examine the effect of flares across a range of timescales, from short-term events that might resemble transient bursts to year-long events that could mask or mimic the trend.

The results are summarized in Table \ref{tab:experiment_2}. For the linear trend scenario, the trend detection rate drops to 15\% when a flare with amplitude A1 and a duration of 2 months is injected. For all other flare scenarios, the detection rate is effectively 0\%, indicating a substantial reduction in the trend detection capability under these conditions. In the quadratic trend scenario, detection capacity is similarly reduced to nearly zero, with the exception of one case, amplitude A1 and 2-month duration, where 57.7\% of the simulated trends are successfully recovered.

These results highlight the significant impact of bright flares on trend detection in LCs. Specifically, the presence of flares with substantial amplitude and moderate-to-long durations can overwhelm the underlying trend signals, reducing the likelihood of accurately detecting linear or quadratic trends. Consequently, the results suggest that some genuine trends may go undetected due to the contaminating effects of bright flares.

\subsection{False-Positive Detection Rate}\label{sec:fpd}
To assess the potential for false-positive detections in our results, we calculate the False-Positive Detection Rate (FPDR). This rate quantifies the level of contamination in our findings that can be attributed to stochastic effects. To calculate this rate, we simulate artificial LCs. These artificial LCs are generated using 100 randomly selected blazars from the subsample that do not exhibit any evidence of trends (filtered in the first stage of the pipeline depicted in Figure \ref{fig:study_flow}) and with the same PSD and probability density function as real blazar LC \citep{emma_lc}. We estimate the PSD by considering the model $\propto$$A*f^{-\beta}$, where $A$ is the normalization, $\rm f$ is the frequency, and $\beta$ is the power-law index \citep[e.g.,~][]{gao_power_law}. The specific model of each PSD is estimated using the maximum likelihood and Markov Chain Monte Carlo simulations\footnote{We use the Python package emcee} \citep[][]{penil_2022_periodicity}. As a result, we observe a range of power spectral indices falling within the interval of [0.59-1.56]\footref{fn:ranges}. In total, we simulate a total of 500,000 artificial LCs, generating 5,000 artificial LCs for each of the originally selected LCs. We then apply the trend-search methodology to these simulated datasets.

The FPDR is calculated as the ratio of the number of trends detected to the number of LCs simulated. In our case, this calculation yields an FPDR of 0.47\%. This indicates that, based on our simulations, we can expect approximately 0.47\% of trend detections to be spurious or false positives. Applying this FPDR to our sample of 1492 jetted AGN, it suggests that around 7 trend detections within our dataset may be spurious due to stochastic effects. This helps us account for the potential impact of random noise and ensure the reliability of our trend detection results.

We also evaluate the FPDR due to the presence of flares in the LC, using the flare characterization values described in $\S$\ref{sec:flares}. For flares with amplitude A1 and a duration of 2 months, the FPDR is 0.57\%. When the duration is extended to 12 months, the FPDR is $\approx$0.00\%. Similarly, for flares with amplitude A2 and durations of 2 and 12 months, the FPDR is $\approx$0.00\%. These results show that the presence of flares in the LC is unlikely to induce the detection of a spurious trend, highlighting the robustness of our methodology against flare-induced variability. 

\subsection{Physical Interpretation}
The long-term trends identified in this study do not have a straightforward association with any specific physical phenomena. There are several possible sources of variability that could give rise to such trends, including noise and stochastic processes within the jetted AGN itself, as highlighted in prior research \citep[e.g.,][]{mcquillan_trend_fake_detection, kaaz_accretion_disk_trends}. However, despite the challenges posed by these noise and stochastic processes, researchers have proposed physical models to interpret and explain the observed long-term trends. According to the model put forth by \citet{giannios_trend_magnetic}, long-term trends may result from magnetic reconnections. These reconnections are triggered by instabilities that disrupt the collimated jet flow in AGNs. As a consequence, the magnetic field fragments into small plasmoids that interact and coalesce into larger plasmoids within the reconnection region. These large plasmoids then move along the reconnection region, generating flares. The duration of this reconnection event is associated with the duration of the observed trend. In \citet{sarkar_curved_jet}, they explain the long-term trend observed in 3C 454.3, as well as the detected QPOs, using a geometrical jet model. In this model, plasma inhomogeneities, or ``blobs'', originating from the accretion disk, move helically within a curved jet. The QPOs are attributed to the motion of these blobs within the jet. Meanwhile, the long-term trend is generated by changes in the Doppler factor, stemming from variations in the viewing angle caused by spatial curvature in the jet. \citet{fichet_trends_shocks} propose a model where long-term trends are reproduced through hydrodynamic simulations. These simulations take into account the interaction between moving and stationary shocks in over-pressured jets. The complex dynamics of shock interactions within the jet environment can give rise to the observed long-term trends.

The interpretation of long-term trends in the context of SMBH binary hypotheses adds another layer of complexity to our understanding of these phenomena. Specifically, the observed trends may be indicative of a section of a long-period oscillation generated by the interaction of an SMBH binary system. In the case of OJ 287, there has been evidence of a long-term periodicity of 60 years, in addition to a 12-year periodicity, both of which are believed to be associated with the SMBH binary. The 12-year periodicity corresponds to the orbital period of the binary, while the 60-year periodicity is linked to orbital precession \citep[][]{Sillanpaa_oj284, dey_oj_287_orbital_precession}. 

In the model presented by \citet{orazio_lump_definition}, the presence of an SMBH binary creates an asymmetrical central cavity within the surrounding accretion disk. The movement of this binary system leads to the formation of an overdense lump orbiting at the ridge of this cavity. The critical aspect of this model is that the overdense lump significantly influences the accretion process. It modifies the accretion flux into the central cavity, introducing additional material into the accretion flow. This altered accretion flow, in turn, affects the emission from the SMBHs, resulting in a periodic behavior. The relationship between the lump and orbital periods, as revealed through high-resolution hydrodynamics simulations of SMBH binaries \citep[][]{ryan_simulations}, indicates that for orbital periods on the order of years, such as those appropriate for PG 1553+113 \citep[2.2-year period,][]{ackermann_pg1553}, the lump period is estimated to be approximately 5-10 times the orbital period \citep[e.g.,][]{farris_trend_binary, sagar_pg1553}. The observation of long-term periods in the optical band, such as the 22-year period in PG 1553+113 \citep[][]{sagar_pg1553}, has been explained through this lump scenario. In \citet{quadratic_trend_binary_combi}, quadratic trends are obtained considering the lump scenario. 

In other sources listed in Table \ref{tab:candidates_list}, a QPO in their $\gamma$-ray emission has been previously claimed as, for instance, PKS 0250$-$225, PKS 0447$-$439, OJ 014 \citep[presented in][]{penil_2022_periodicity}, PKS 0405$-$385 \citep[][]{gong_pks_0405_385}, 3C 66A \citep[][]{ren_s5_1044+71}, or BL Lacertae \citep[e.g.,][]{jorge_2022}. Furthermore, it is plausible that future research and observations may unveil QPOs in additional jetted AGNs, some of which could be listed in Table \ref{tab:candidates_list}. These jetted AGNs should be analyzed considering the trends. Finally, it is crucial to recognize and address the potential challenges associated with the multiplicative nature of oscillations in the emission of these jetted AGNs, which introduce complexity in detecting QPOs.

\section{Summary} \label{sec:summary}
In this study, we have conducted a systematic search for long-term trends present in the $\gamma$-ray emissions of 1492 jetted AGNs from the {\it Fermi}-LAT jetted AGN from the 4FGL-DR2 catalog. Our research involved the development of a comprehensive pipeline designed to detect and characterize these long-term trends. Specifically, we sought out both linear and quadratic trends, categorizing them as either additive or multiplicative based on how oscillations evolve in relation to these trends. Our analysis revealed that among the 1492 jetted AGNs studied, 40 jetted AGNs exhibited long-term trends. Of these, 8 displayed quadratic trends, meanwhile 32 displayed linear trends. Furthermore, 9 jetted AGNs exhibited additive oscillations alongside their trends. The remaining 31 jetted AGNs showed multiplicative oscillations in conjunction with their trends.
It is worth noting that this represents the first known sample of $\gamma$-ray emitters with long-term trends of this nature. Understanding the mechanisms driving these long-term trends is essential for unraveling the complex behaviors of jetted AGNs and gaining insights into the underlying physical processes responsible for their emissions.

\software{
	astropy \citep{astropy_2013, astropy_2018}, 
	fermipy software package \citep{Wood:2017yyb},
	statsmodels \citep{seabold2010statsmodels} \url{https://www.statsmodels.org/stable/index.html},
        scikit-learn \url{https://scikit-learn.org/stable/},
	SciPy \citep {SciPy},
	Simulating light curves \citep{connolly_code},
}

\section{Acknowledgements}
P.P and M.A acknowledge funding under NASA contract 80NSSC20K1562. S.B. acknowledges financial support by the European Research Council for the ERC Starting grant MessMapp under contract no. 949555, and by the German Science Foundation DFG, research grant “Relativistic Jets in Active Galaxies” (FOR 5195, grant No. 443220636). A.D. is thankful for the support of the Ram{\'o}n y Cajal program from the Spanish MINECO, Proyecto PID2021-126536OA-I00 funded by MCIN / AEI / 10.13039/501100011033, and Proyecto PR44/21‐29915 funded by the Santander Bank and Universidad Complutense de Madrid.

The \textit{Fermi}-LAT Collaboration acknowledges generous ongoing support
from a number of agencies and institutes that have supported both the
development and the operation of the LAT as well as scientific data analysis.
These include the National Aeronautics and Space Administration and the
Department of Energy in the United States, the Commissariat \`a l'Energie Atomique and the Centre National de la Recherche Scientifique / Institut National de Physique Nucl\'eaire et de Physique des Particules in France, the Agenzia Spaziale Italiana and the Istituto Nazionale di Fisica Nucleare in Italy, the Ministry of Education, Culture, Sports, Science and Technology (MEXT), High Energy Accelerator Research Organization (KEK) and Japan Aerospace Exploration Agency (JAXA) in Japan, and the K.~A.~Wallenberg Foundation, the Swedish Research Council and the Swedish National Space Board in Sweden.

Additional support for science analysis during the operations phase is gratefully acknowledged from the Istituto Nazionale di Astrofisica in Italy and the Centre National d'\'Etudes Spatiales in France. This work was performed in part under DOE Contract DE-AC02-76SF00515.

\bibliography{literature.bib} 
\bibliographystyle{aasjournal}
\input{table_1.tex}

\input{table_2.tex}

\input{table_3.tex}
\input{table_4.tex}

\clearpage

\appendix \label{sec:appendix}
\renewcommand{\thefigure}{A\arabic{figure}}
\setcounter{figure}{1}
This section reports the LCs of the jetted AGN with a trend in their $\gamma$-ray emission. 

\begin{figure*}[ht!]
	\centering
        \ContinuedFloat
	\includegraphics[scale=0.2295]{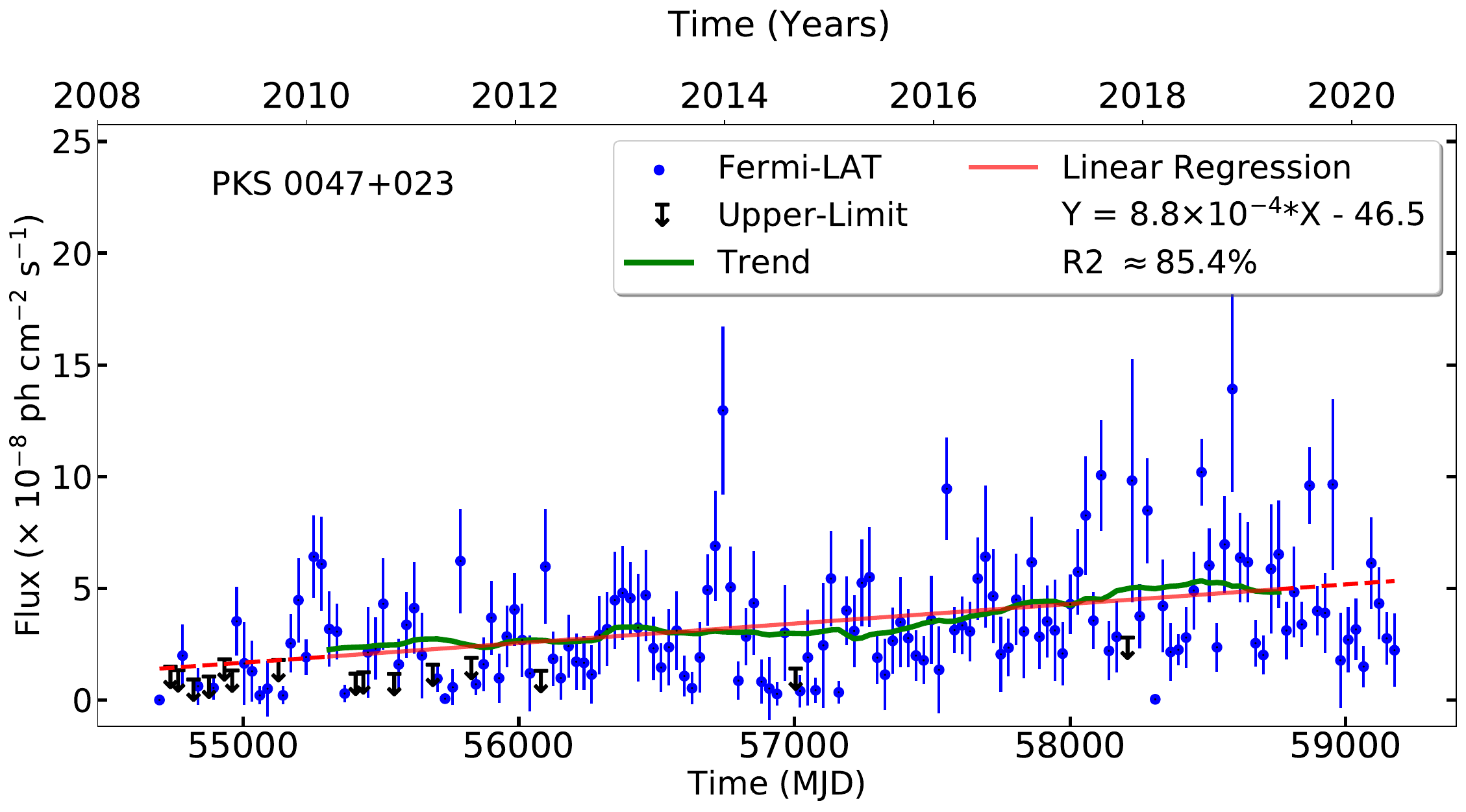}
	\includegraphics[scale=0.2295]{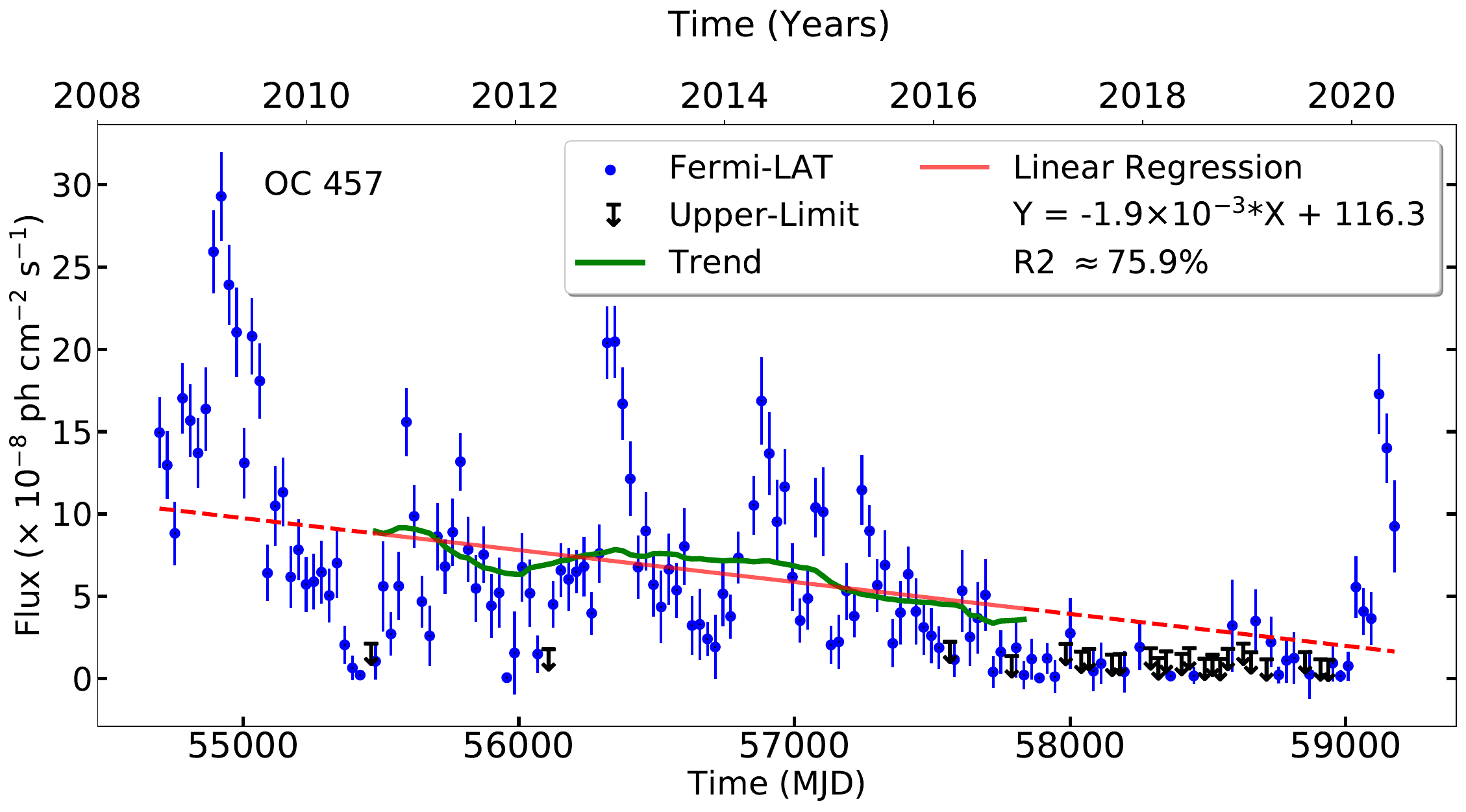}
	\includegraphics[scale=0.2295]{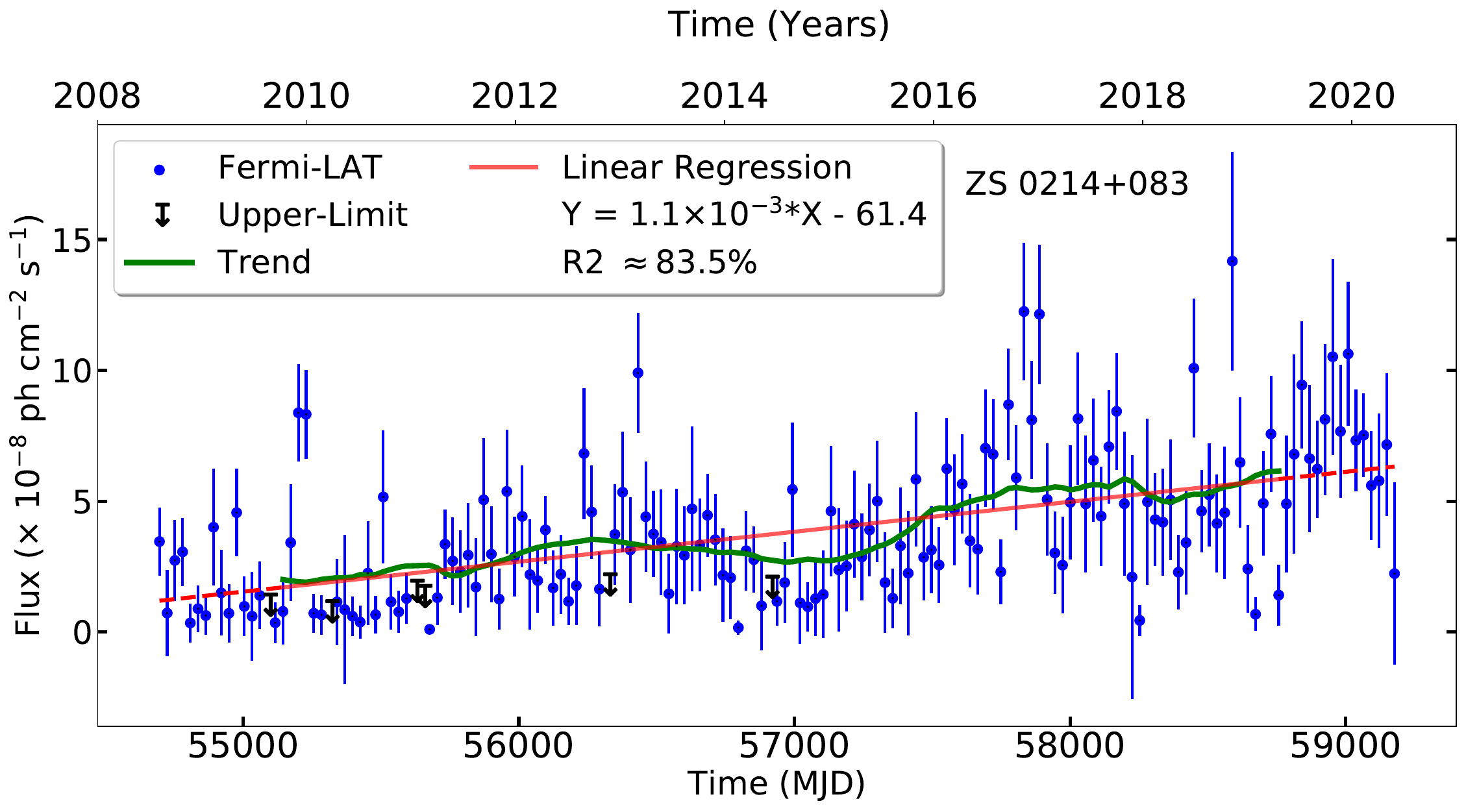}
	\includegraphics[scale=0.2295]{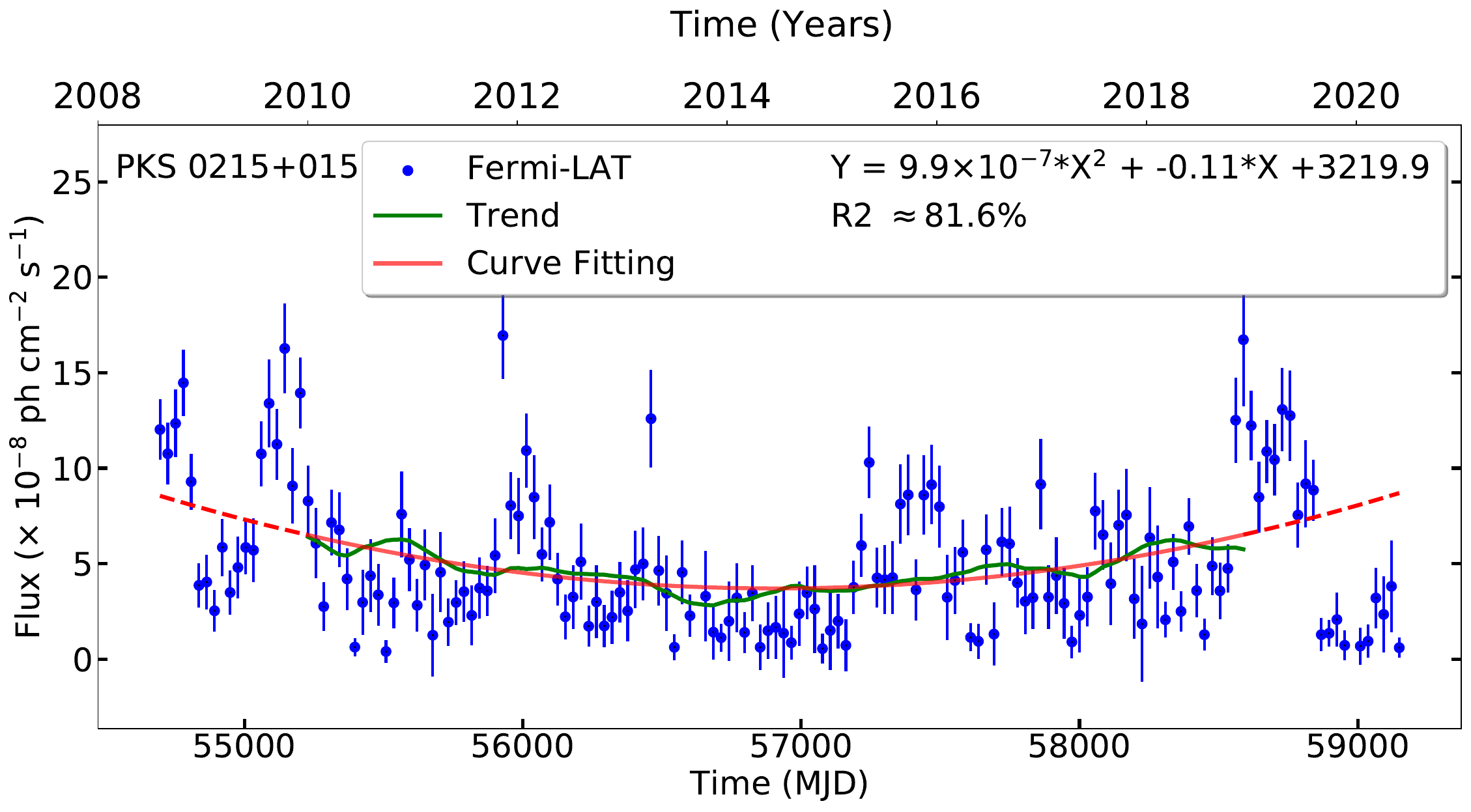}
	\includegraphics[scale=0.2295]{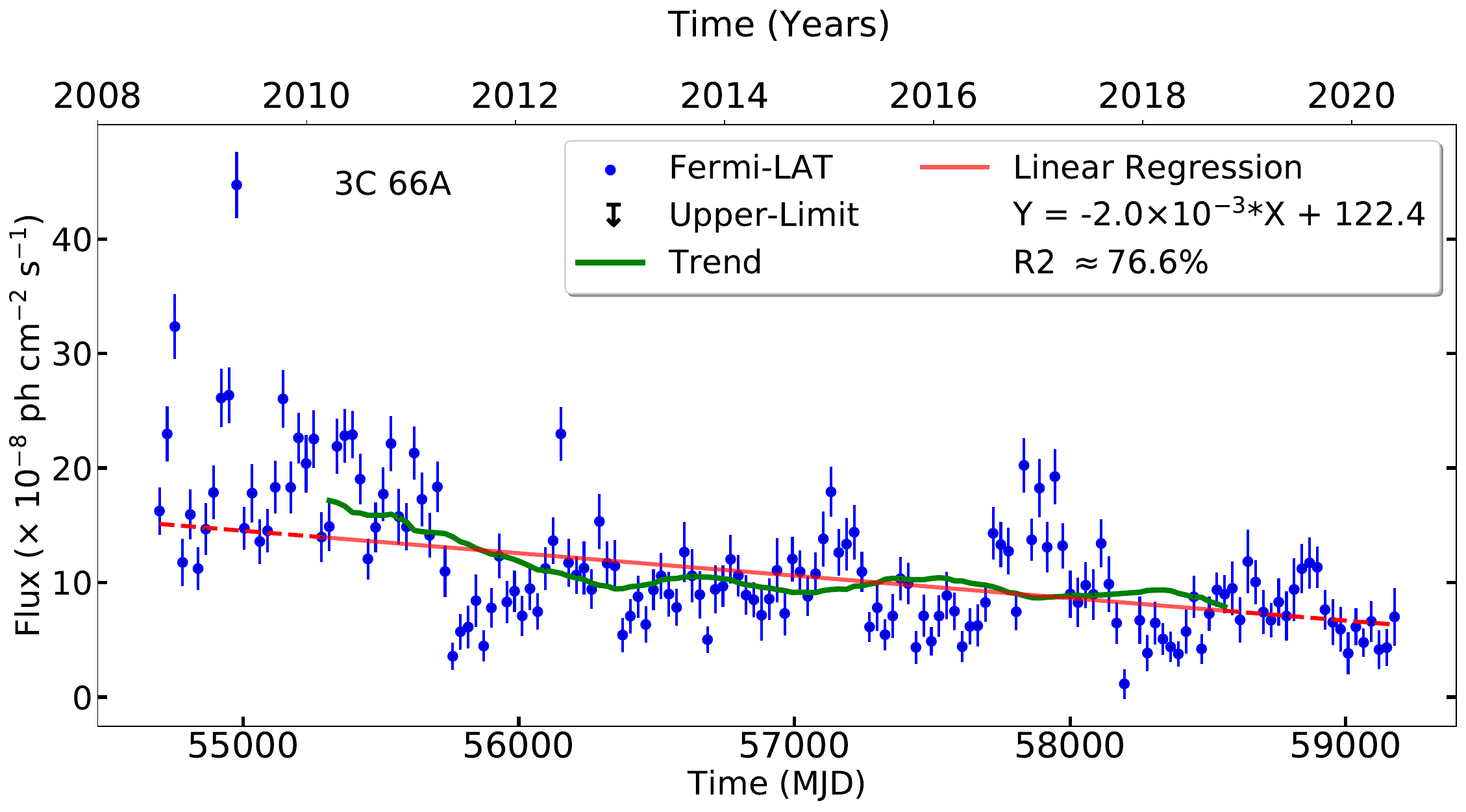}
	\includegraphics[scale=0.2295]{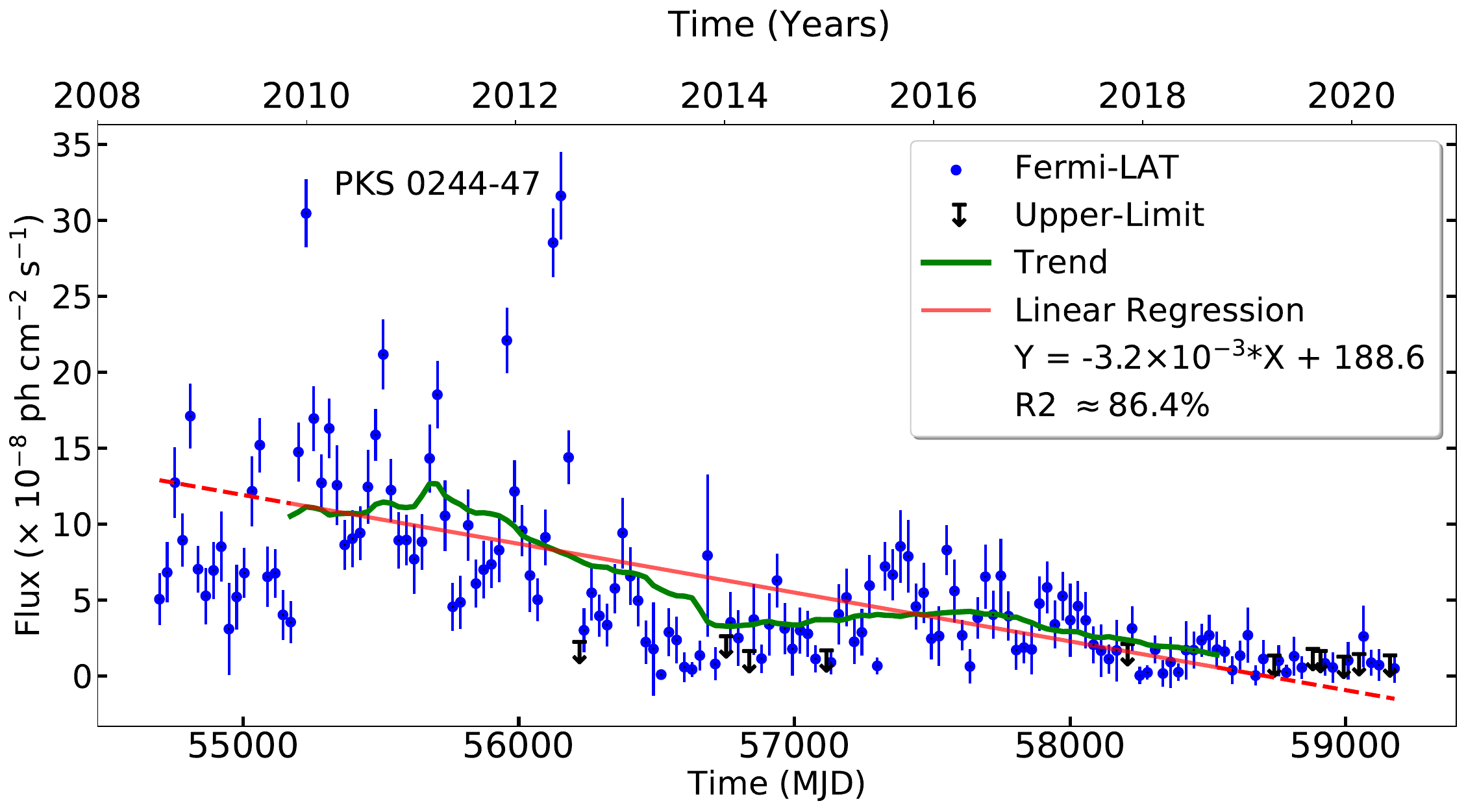}
	\includegraphics[scale=0.2295]{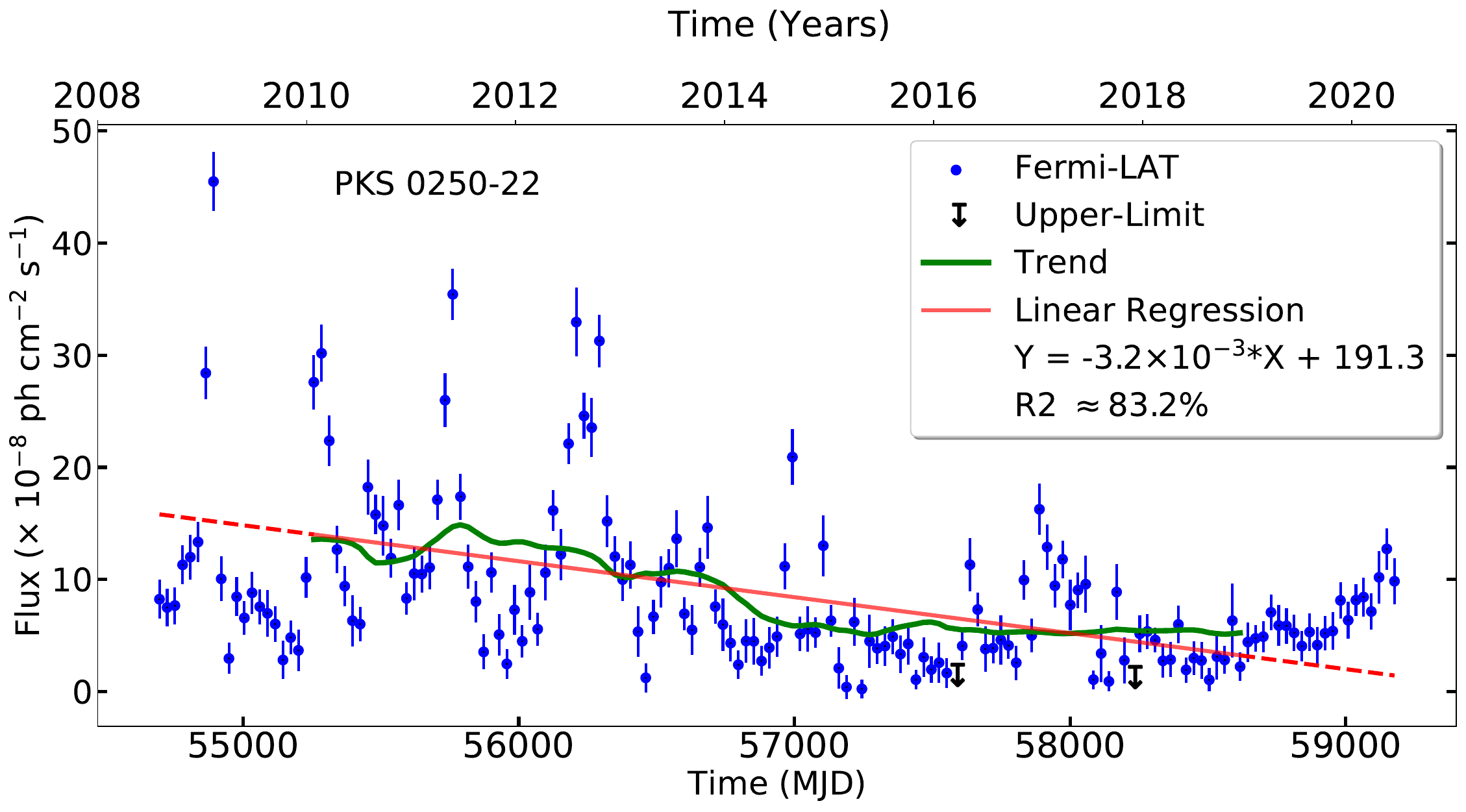}
 	\includegraphics[scale=0.2295]{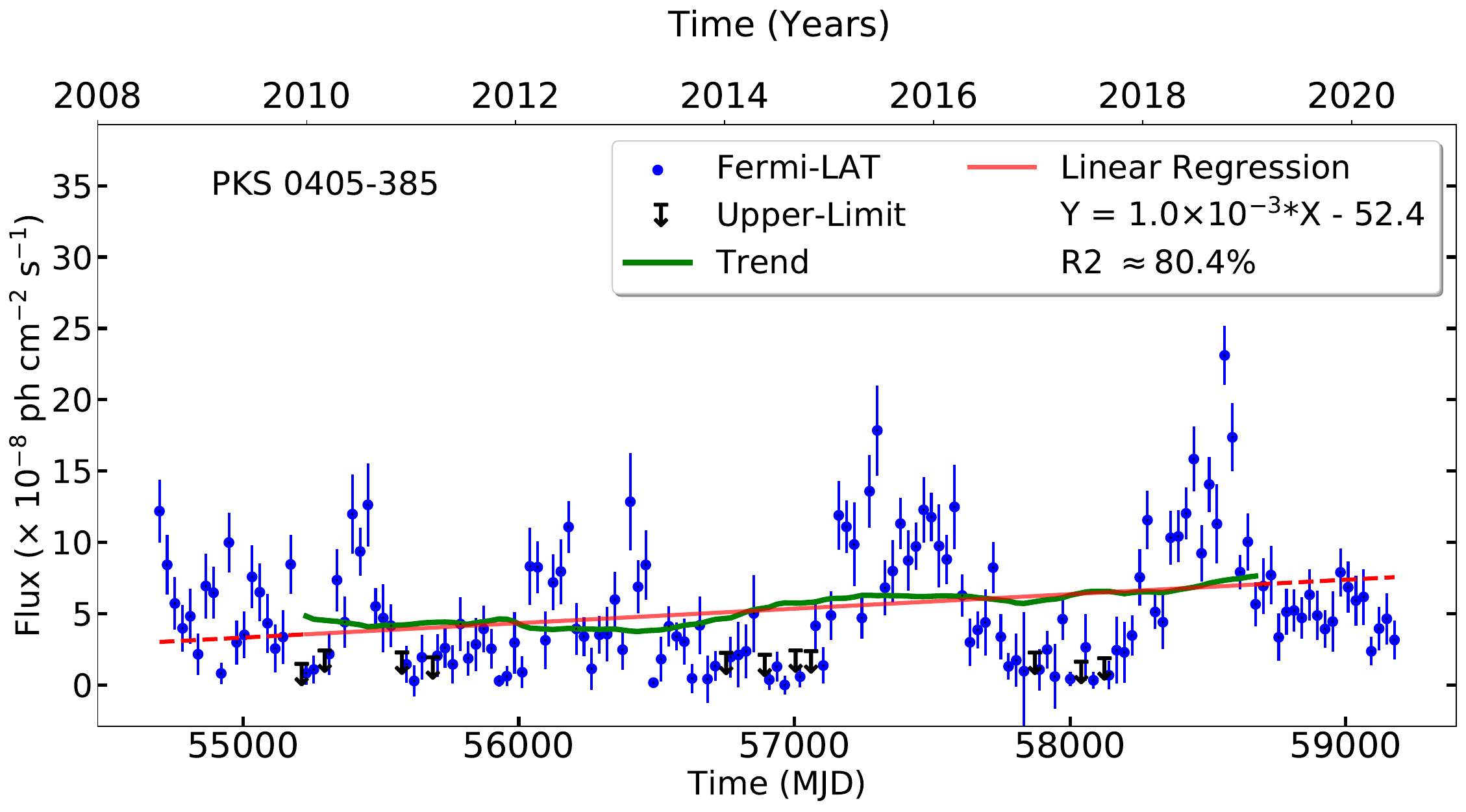}
	\caption{Light curves of the jetted AGN presented in Table \ref{tab:candidates_list}. The upper limits are denoted by down arrows. The green line represents the underlying trend extracted by the function \textit{seasonal\_decompose}. Note that the green line covers a shorter time span than the full LC duration, a result of the trend decomposition applied by this function. The red line indicates the fit of the green line, with the dashed red line extending the fitted line across the entire LC.}
	\label{fig:lcs_blazars}
\end{figure*}

\begin{figure*}[ht!]
	\centering
	\ContinuedFloat 	
	\includegraphics[scale=0.2295]{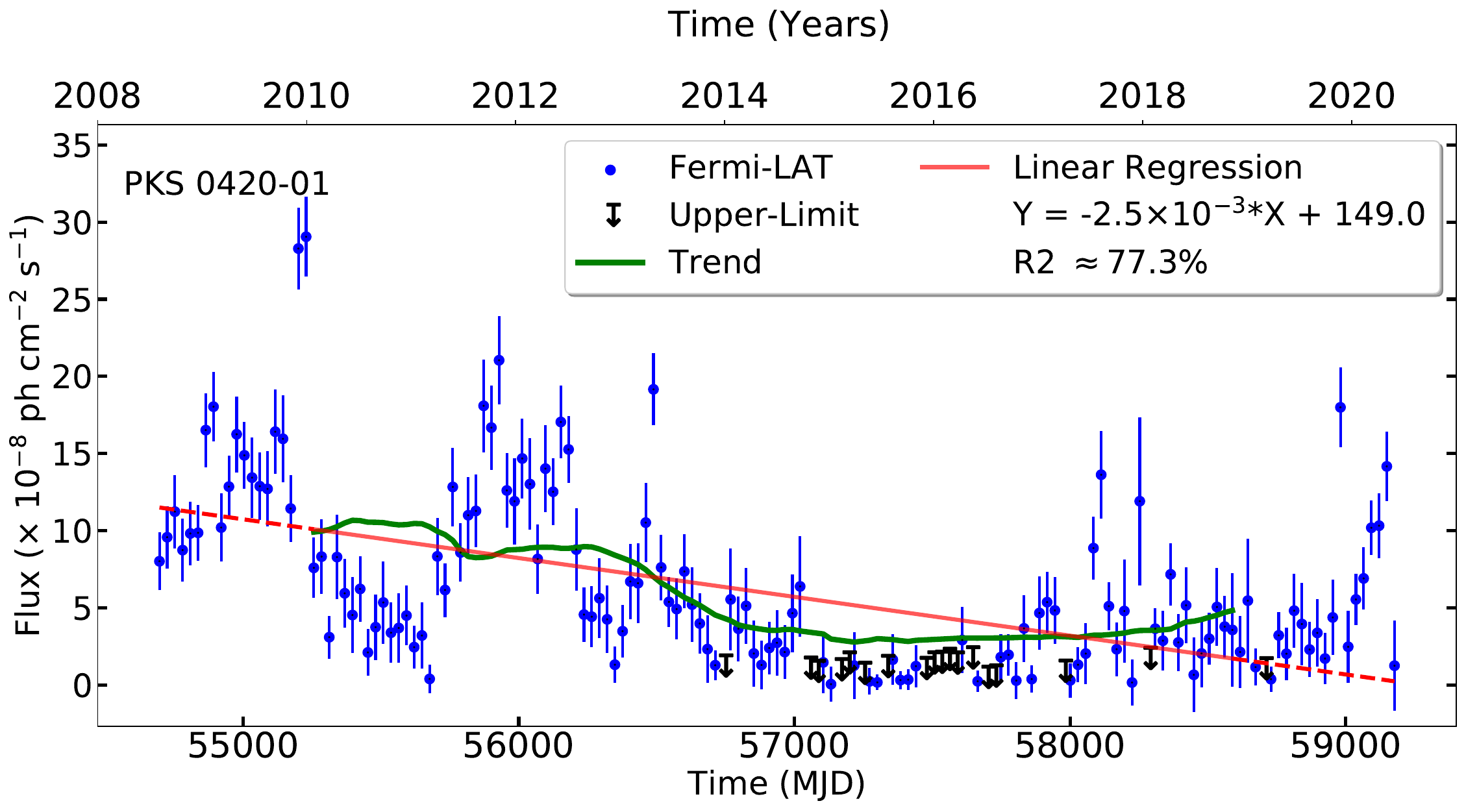}
	\includegraphics[scale=0.2295]{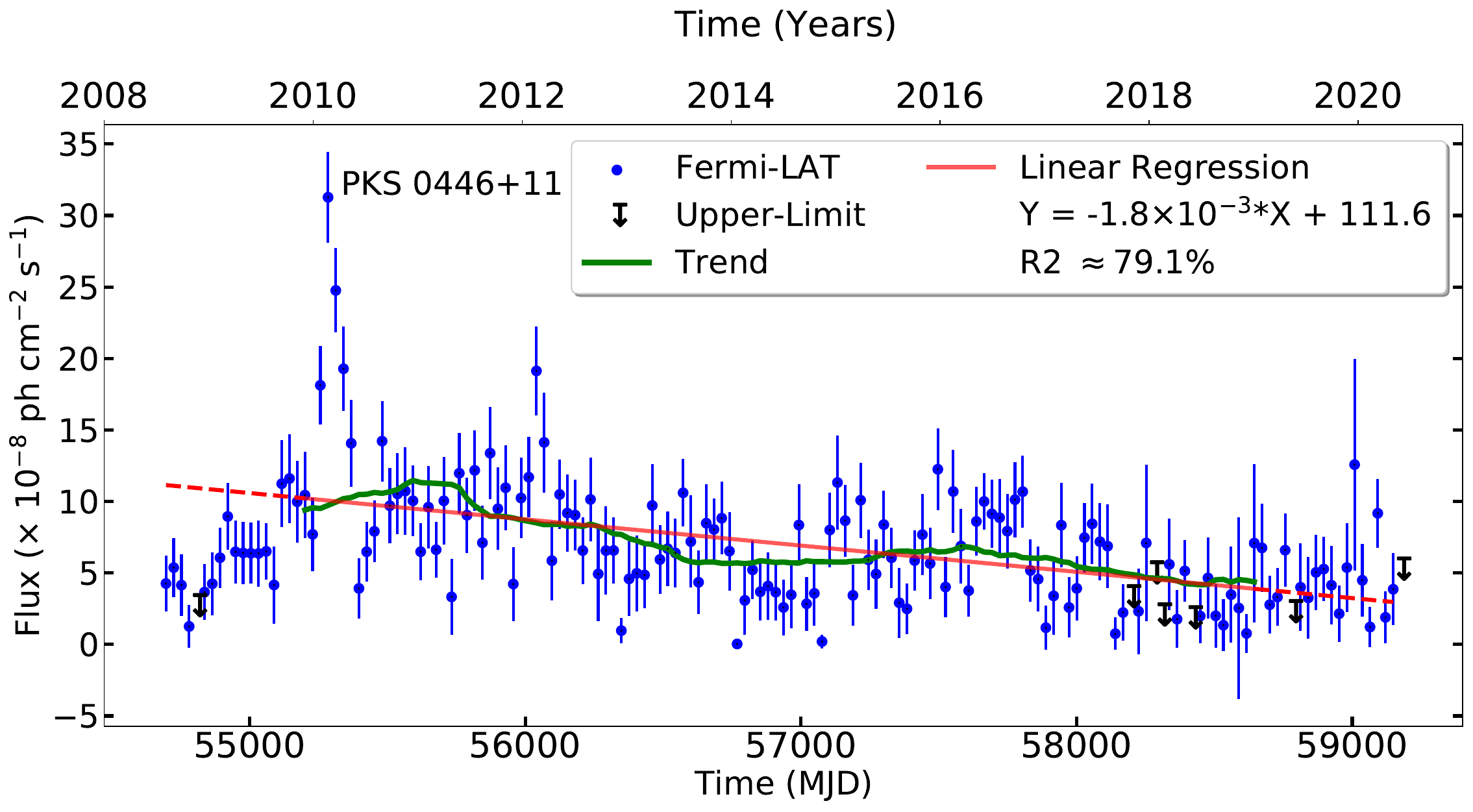}
	\includegraphics[scale=0.22]{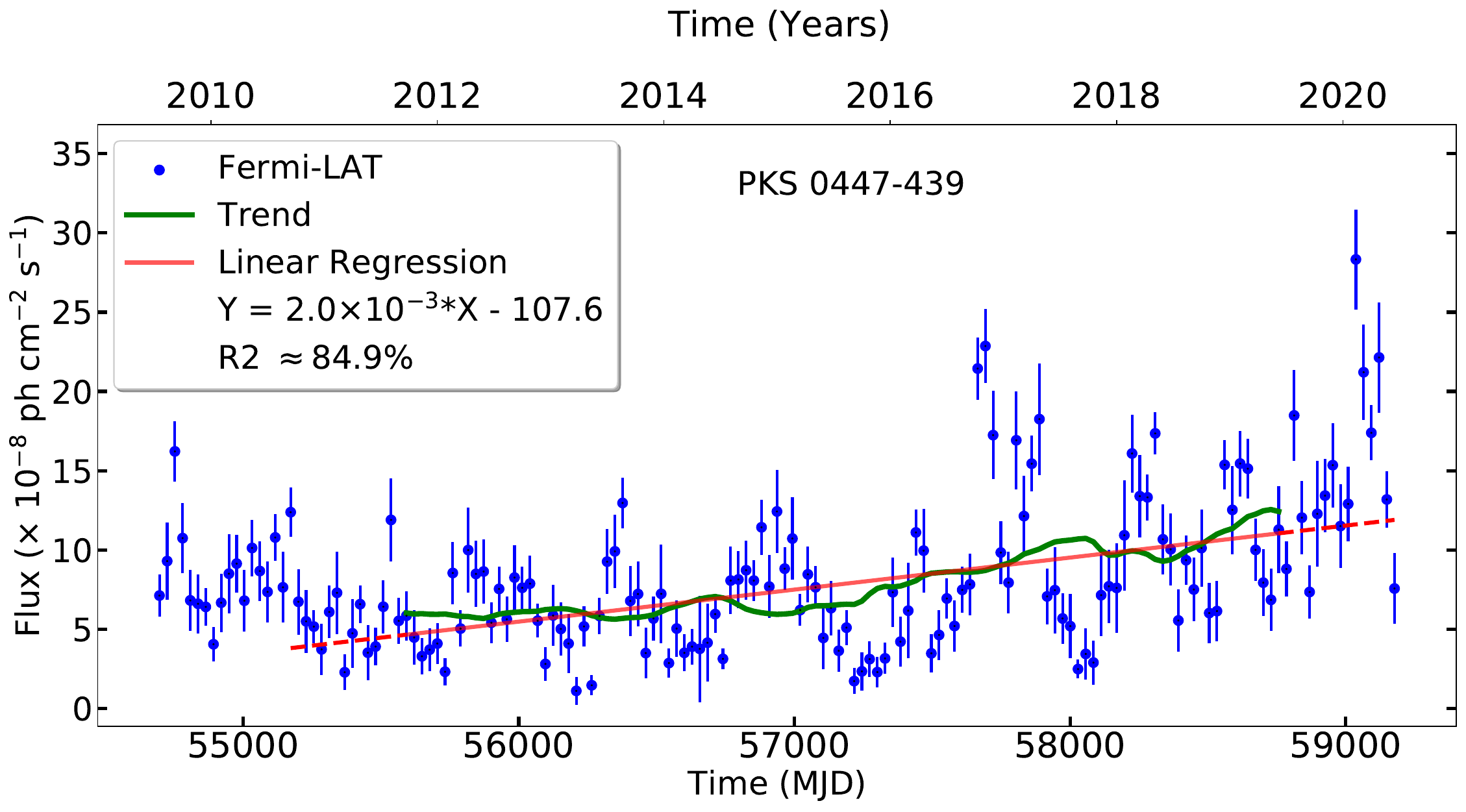}
	\includegraphics[scale=0.2295]{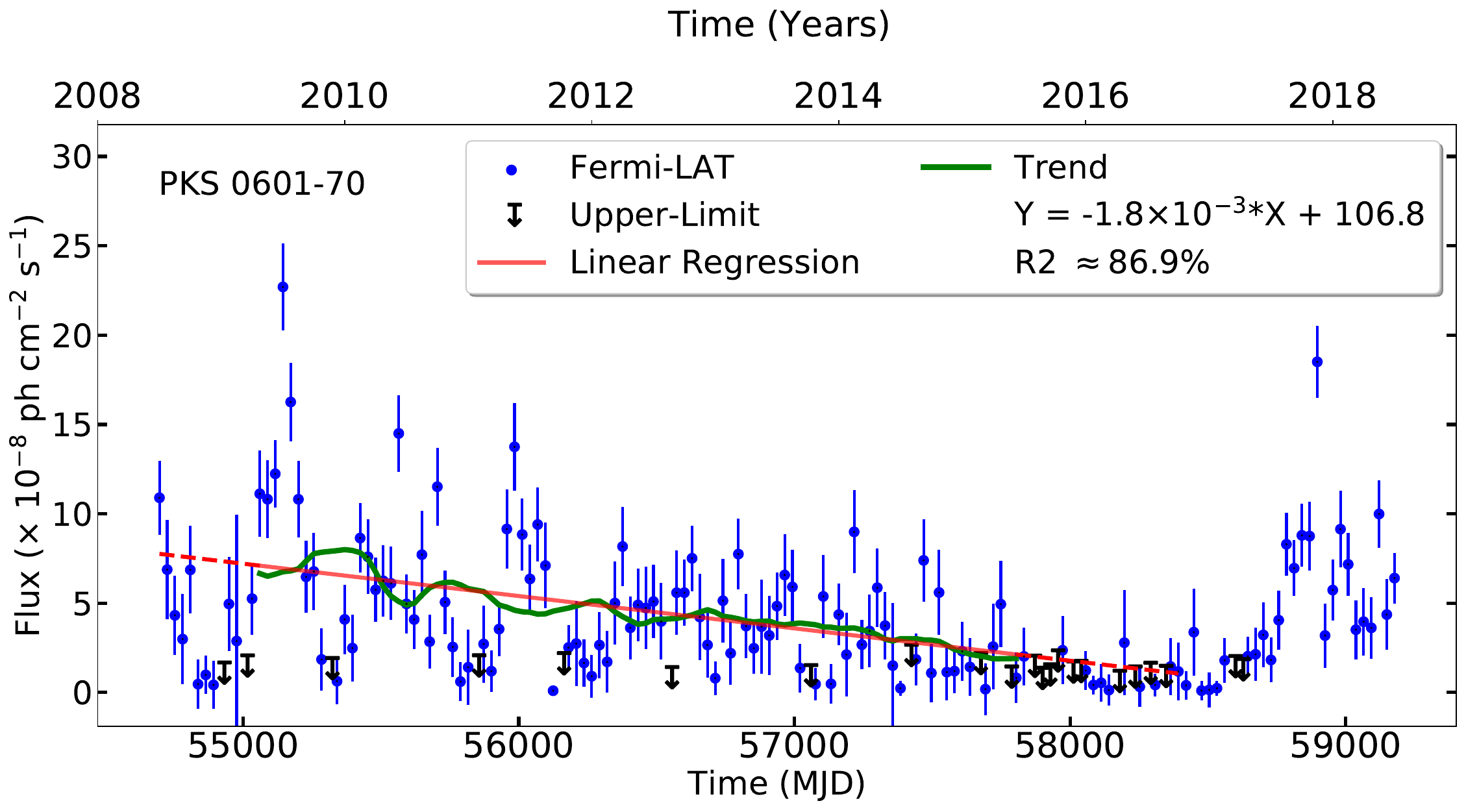}
	\includegraphics[scale=0.2295]{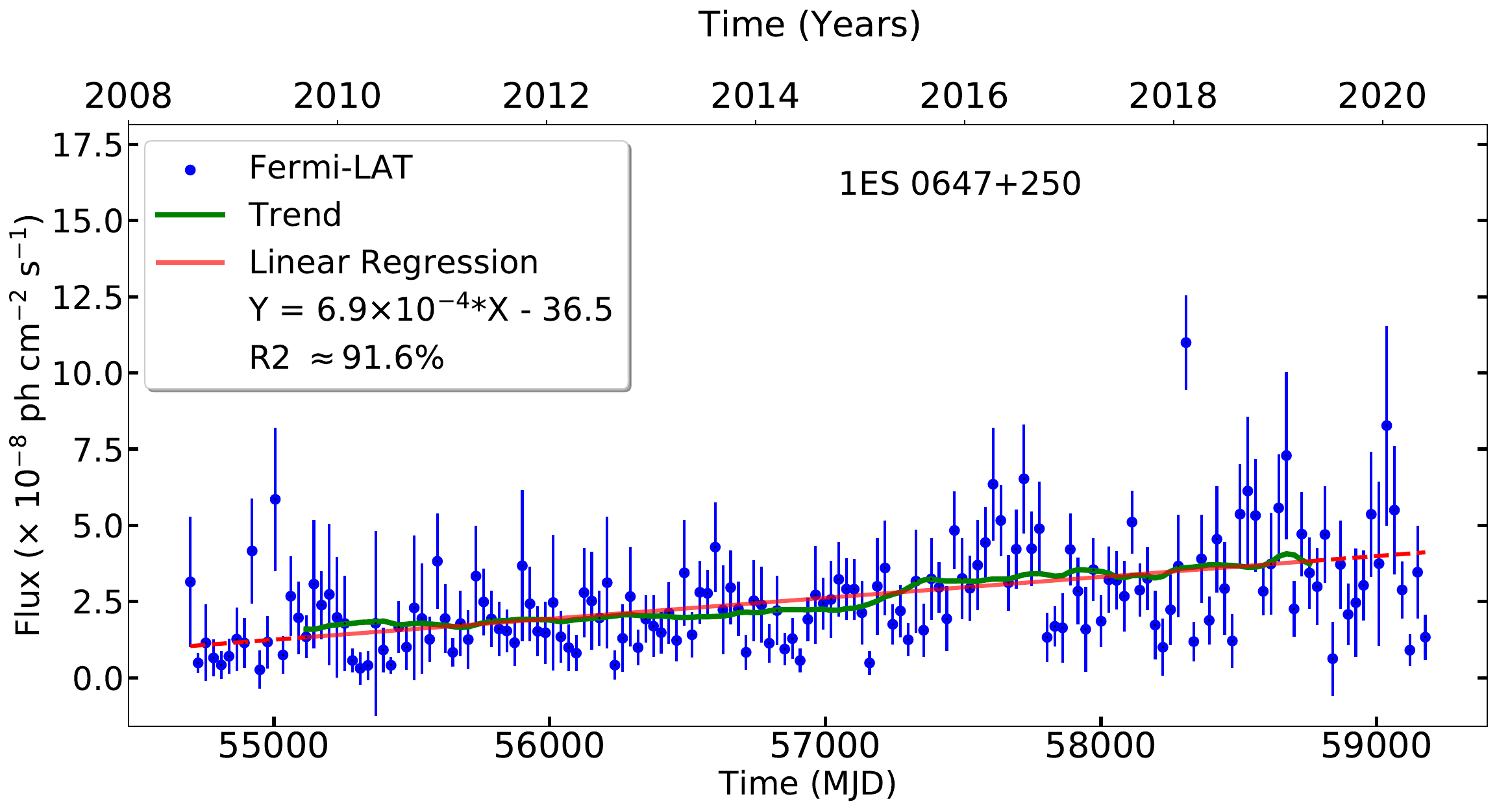}
	\includegraphics[scale=0.2295]{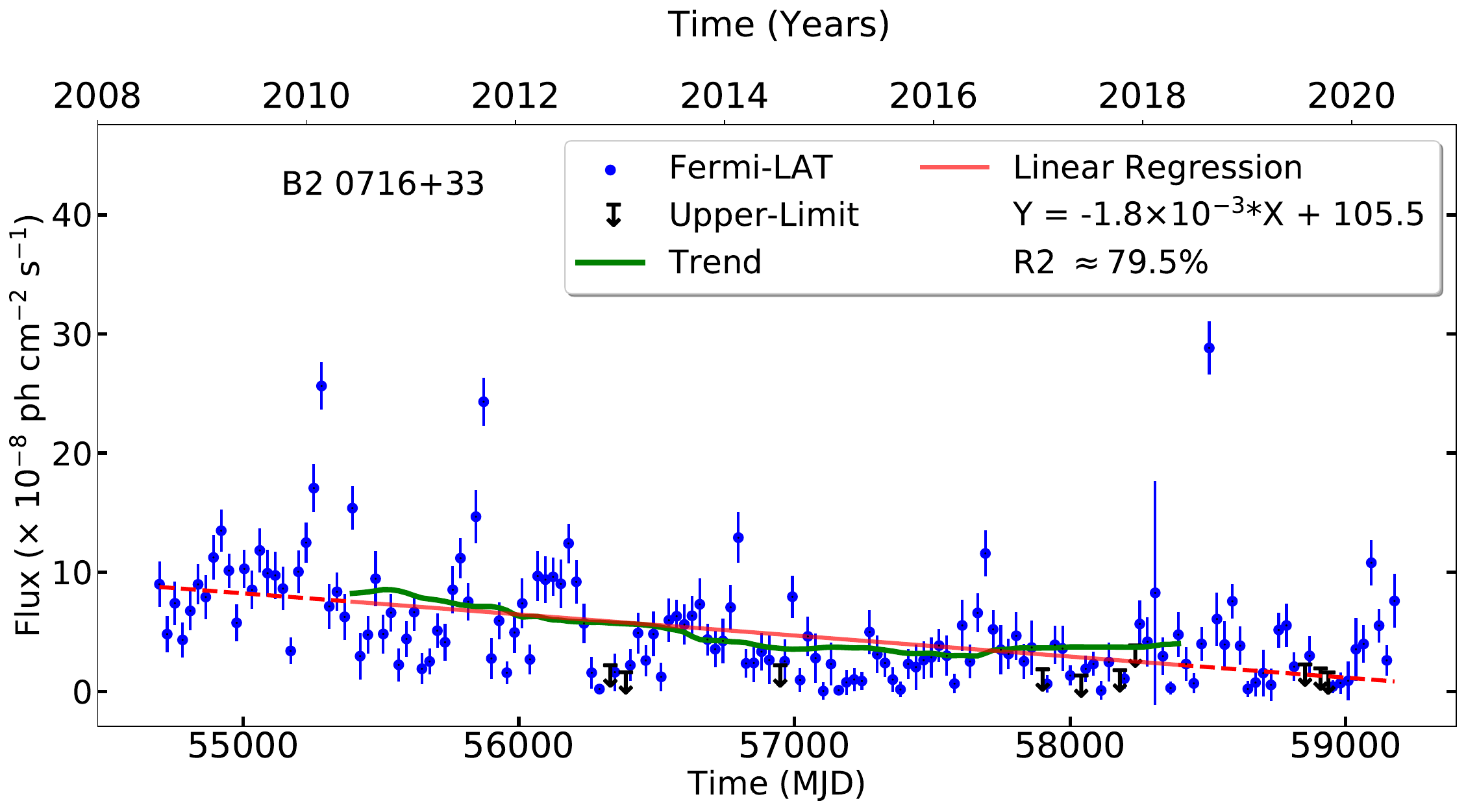}
  	\includegraphics[scale=0.2295]{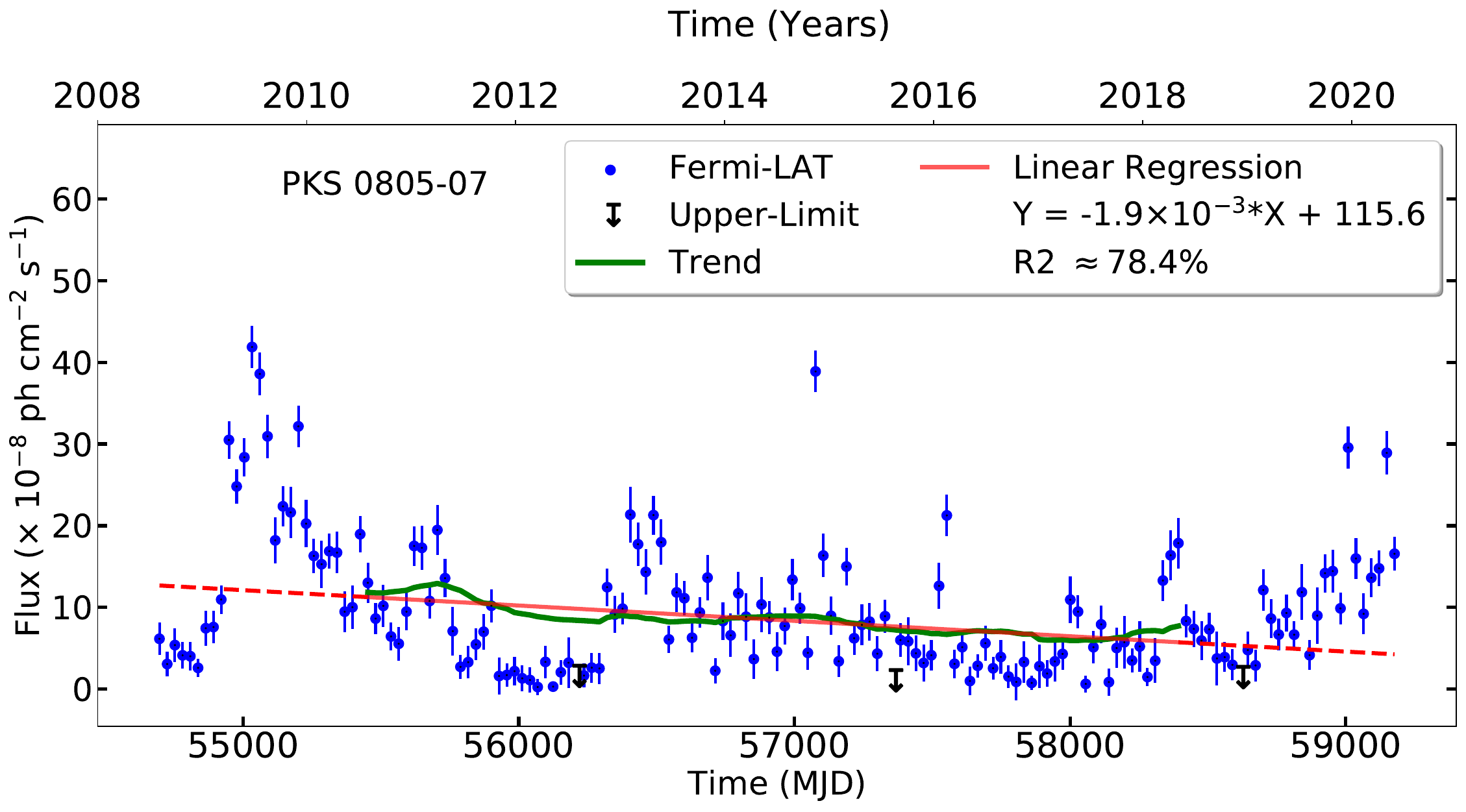}
  	\includegraphics[scale=0.2295]{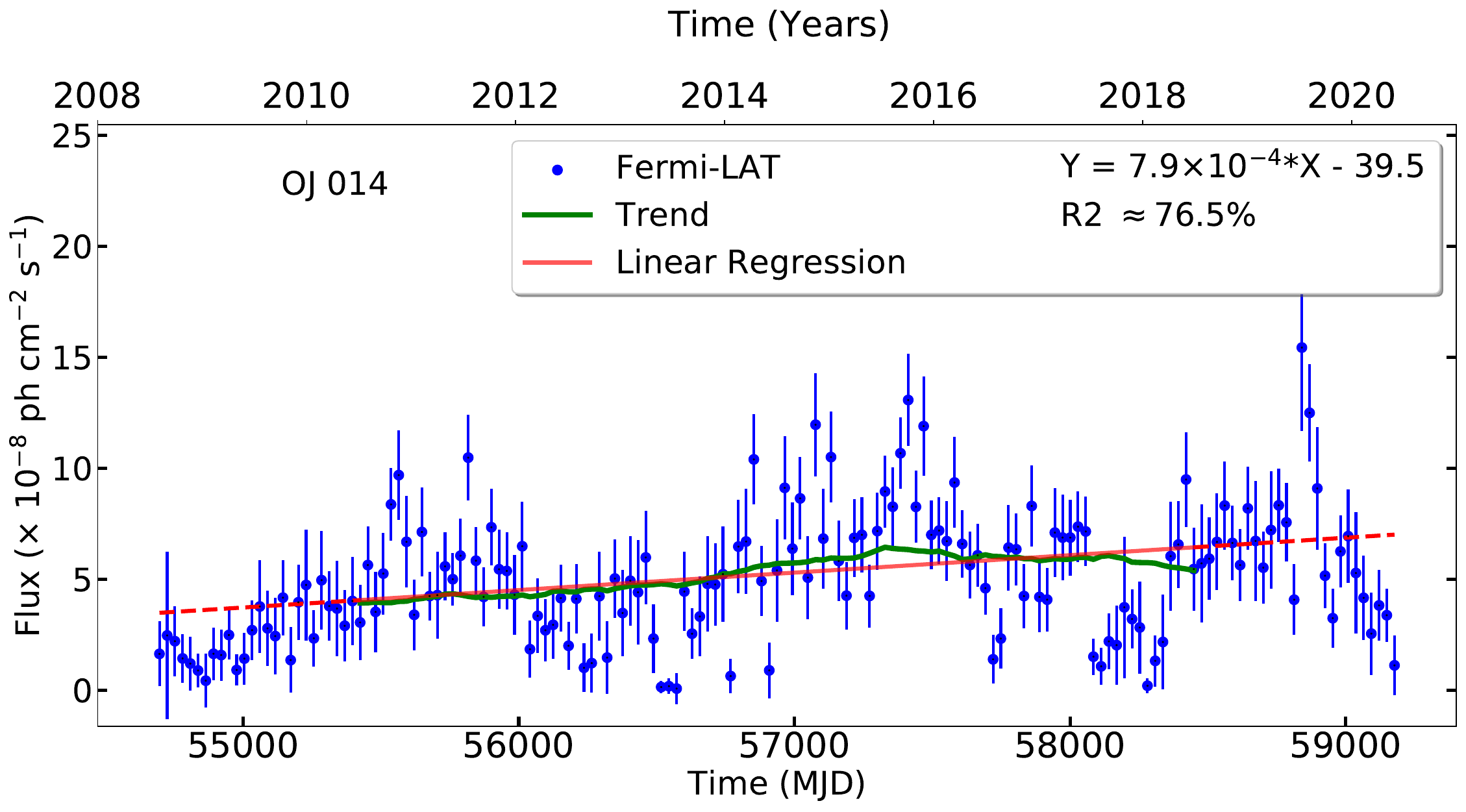}
	\caption{(Continued).}
\end{figure*}

\begin{figure*}[ht!]
	\centering
	\ContinuedFloat
	\includegraphics[scale=0.2295]{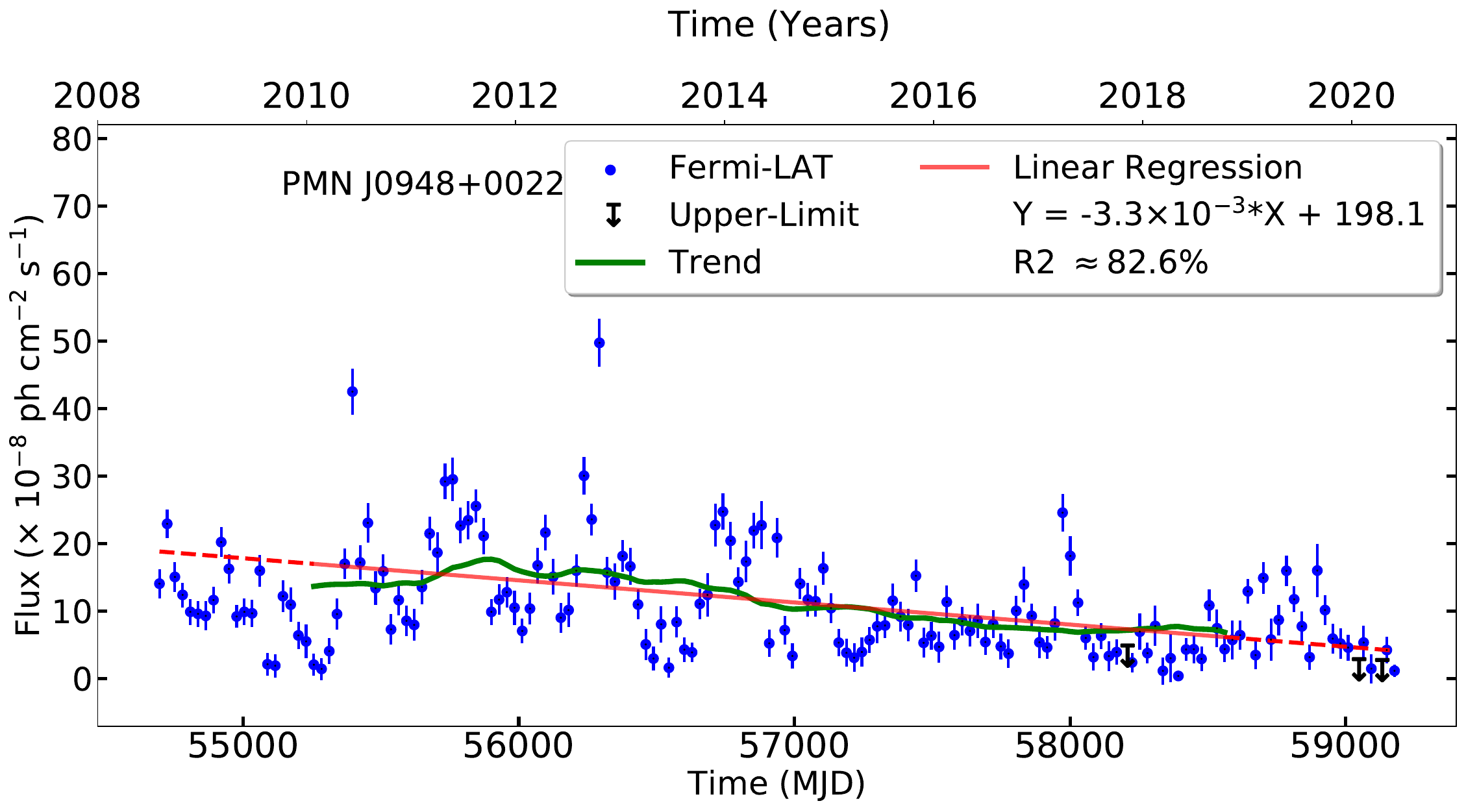}
	\includegraphics[scale=0.22]{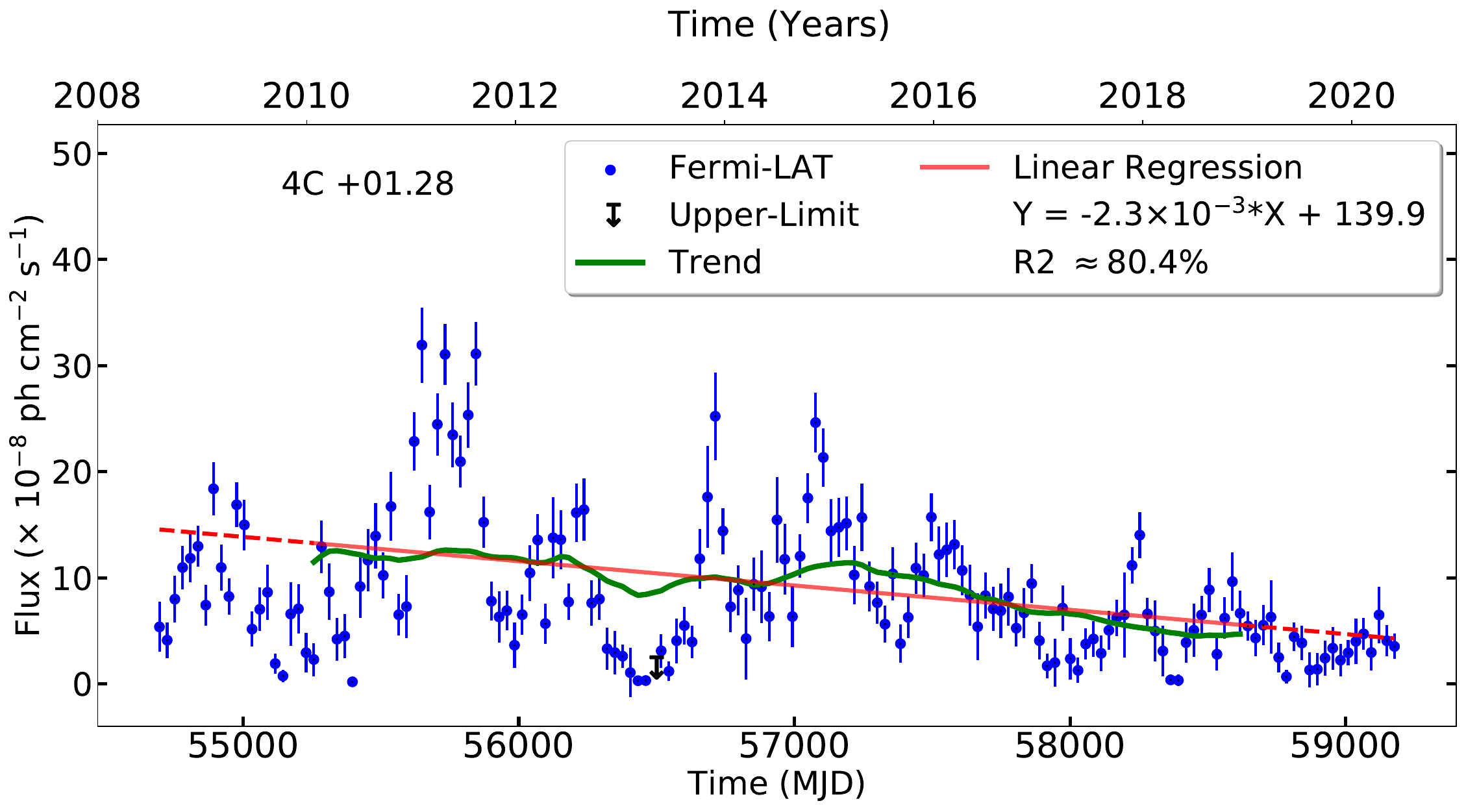}
        \includegraphics[scale=0.22]{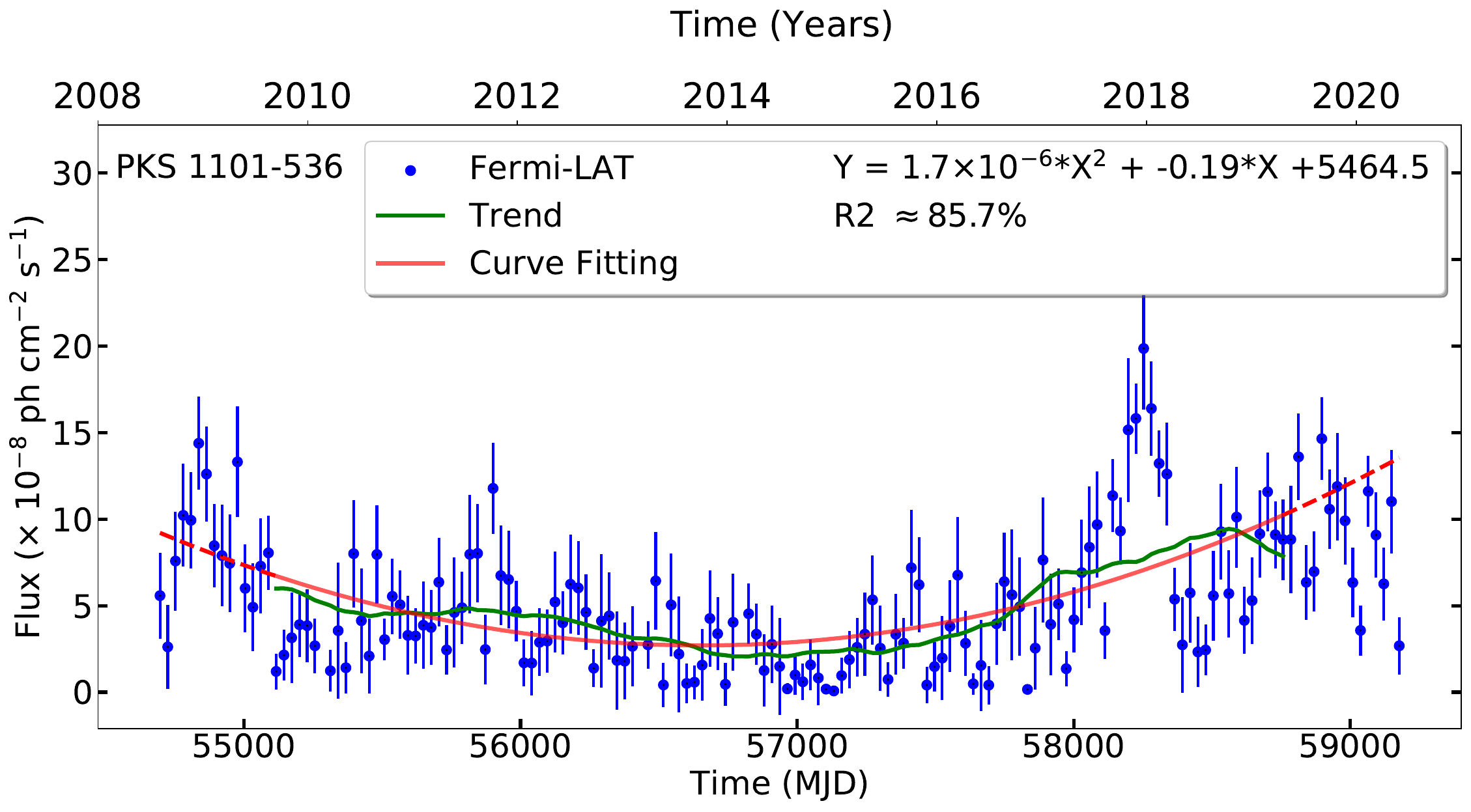}
	\includegraphics[scale=0.2295]{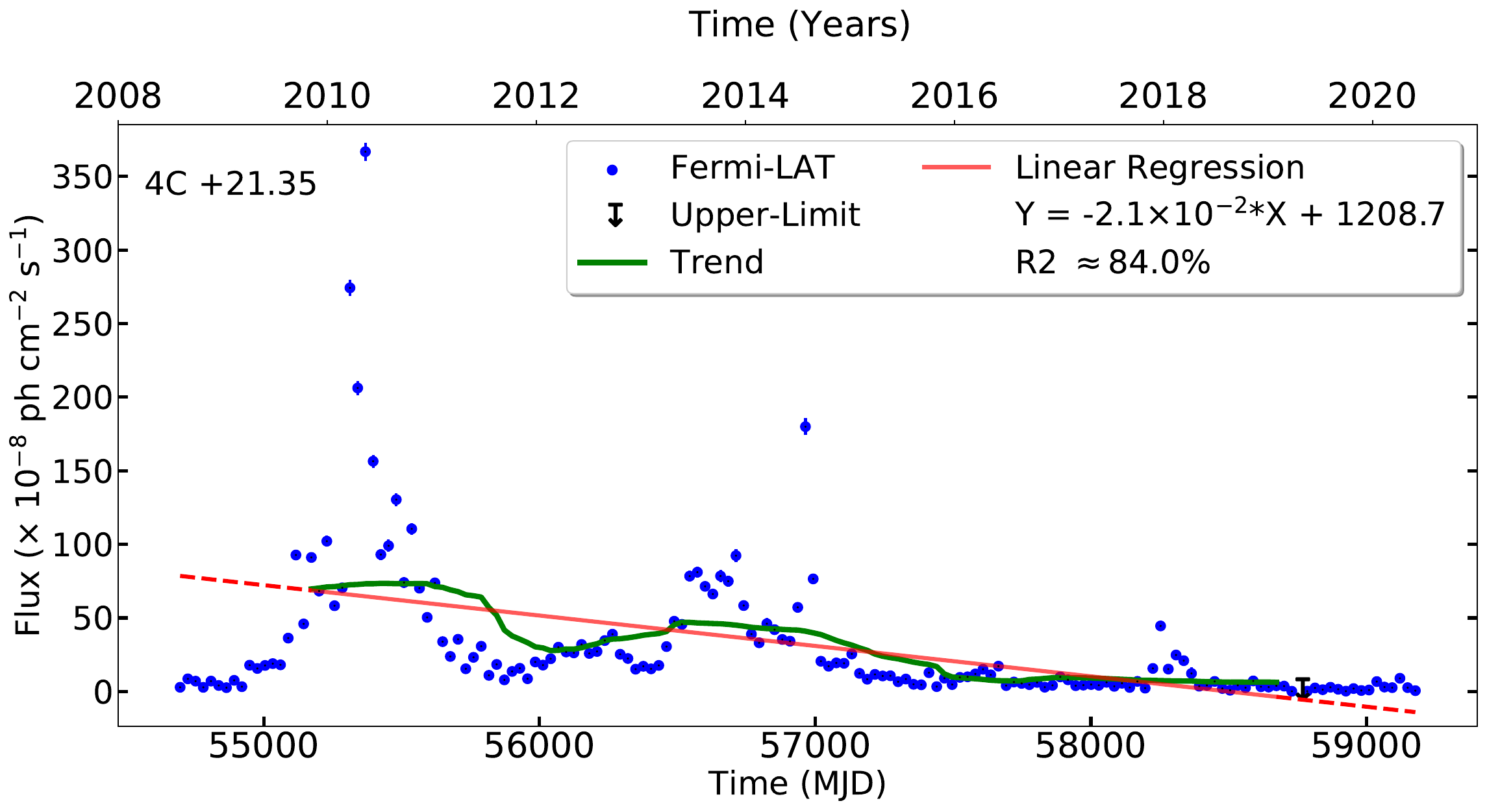}
        \includegraphics[scale=0.2295]{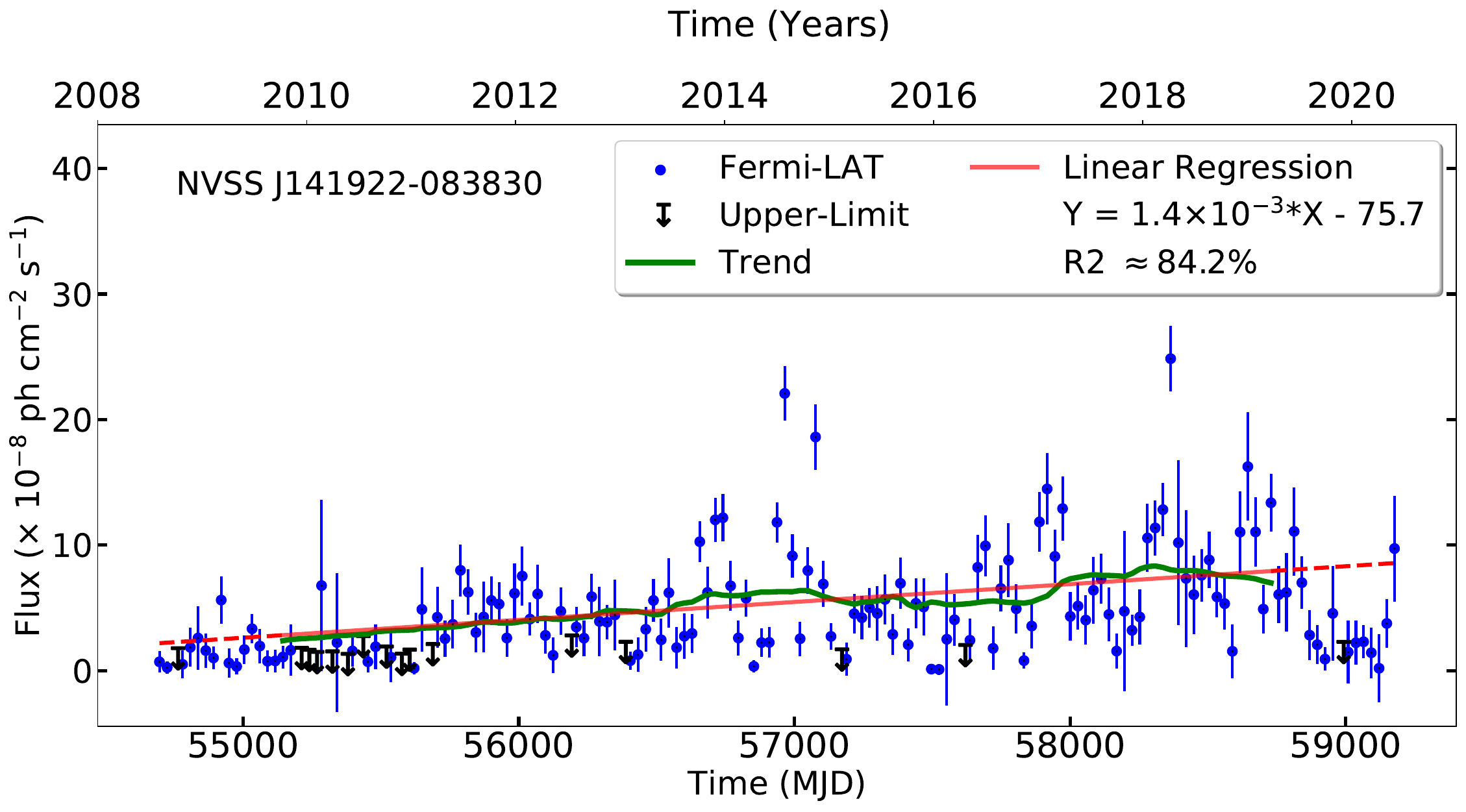}
	\includegraphics[scale=0.2295]{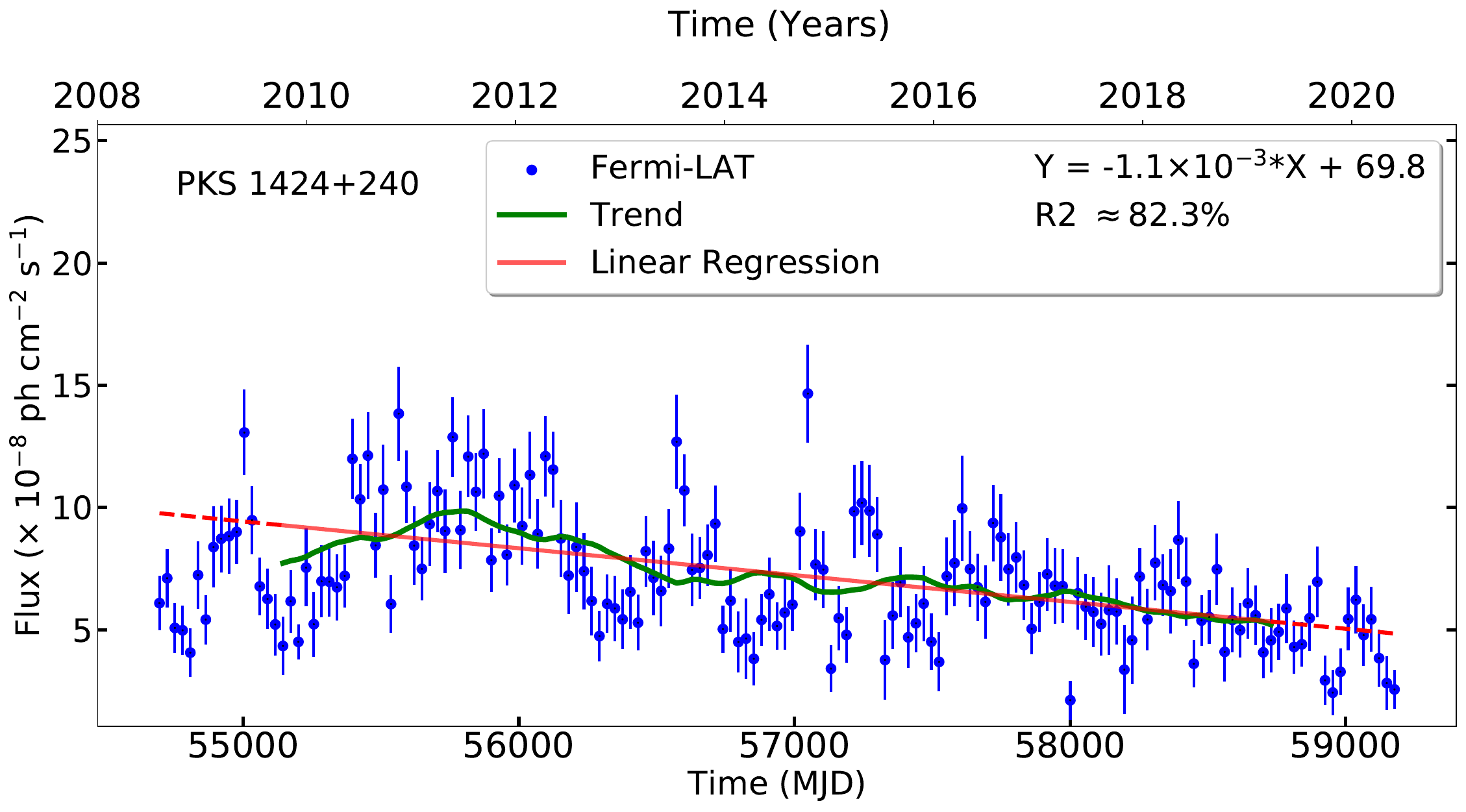}
  	\includegraphics[scale=0.2295]{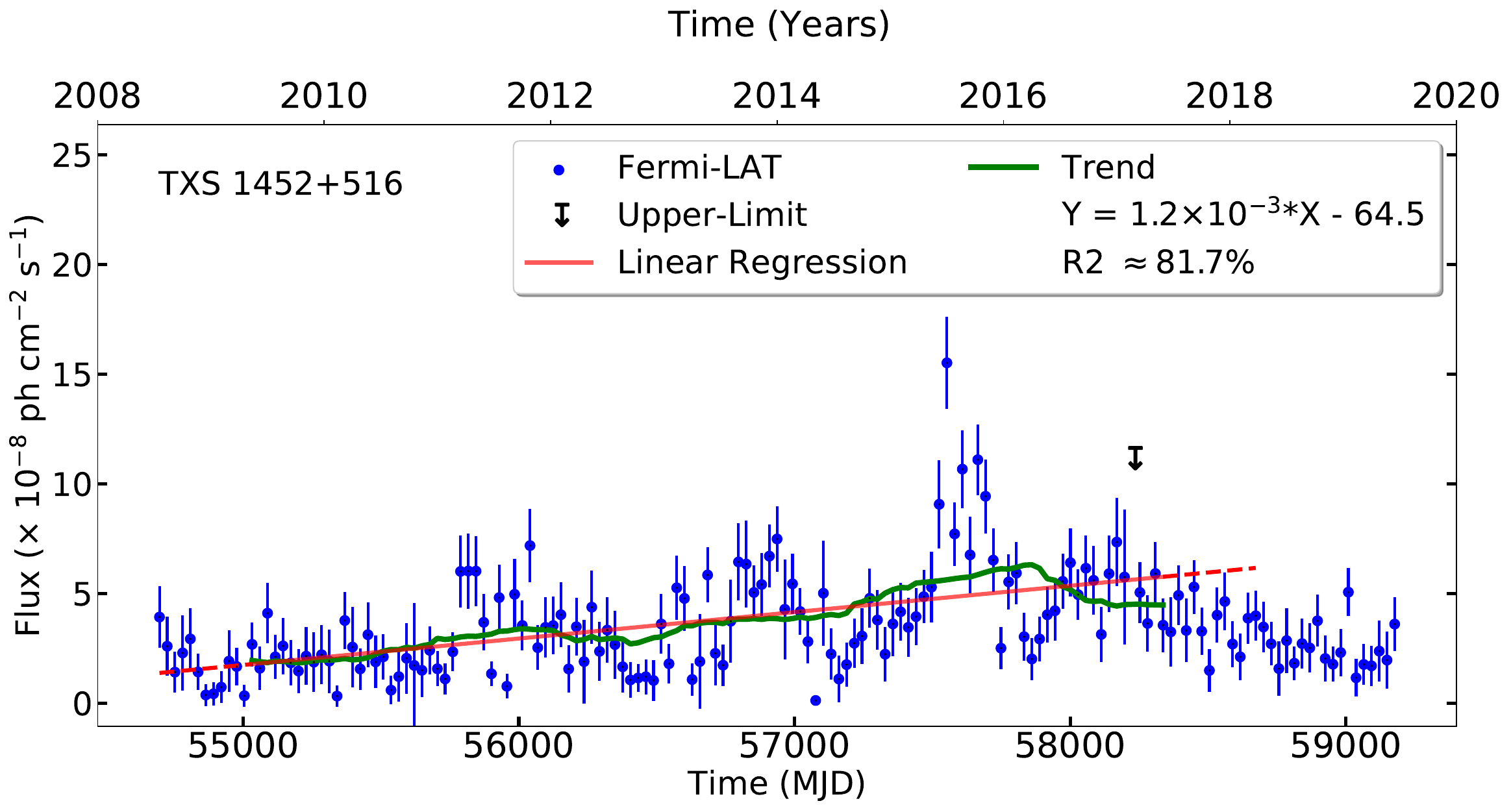}
   	\includegraphics[scale=0.2295]{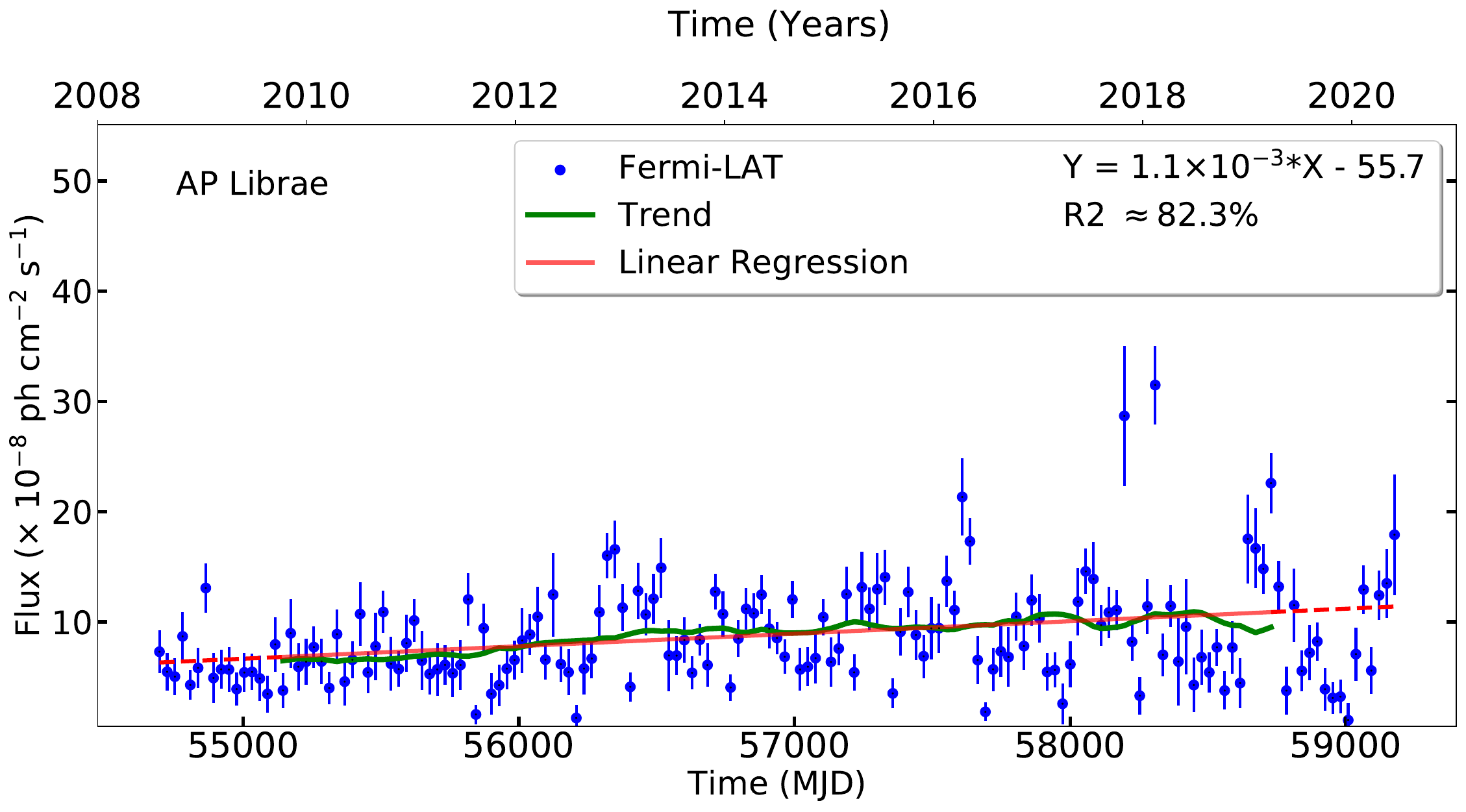}
	\caption{(Continued).}
\end{figure*}

\begin{figure*}[ht!]
	\centering
	\ContinuedFloat
        \includegraphics[scale=0.2265]{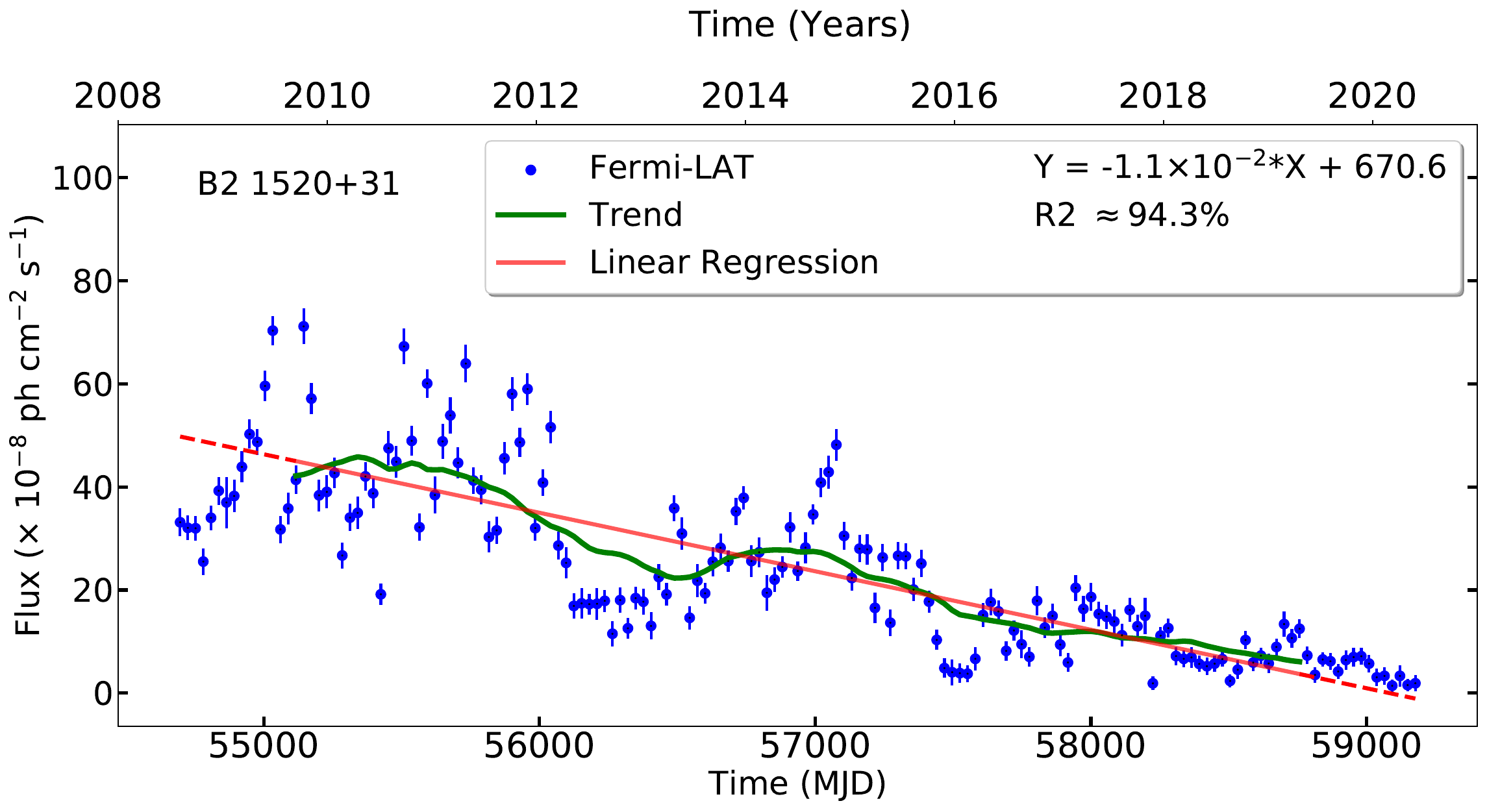} 
	\includegraphics[scale=0.2295]{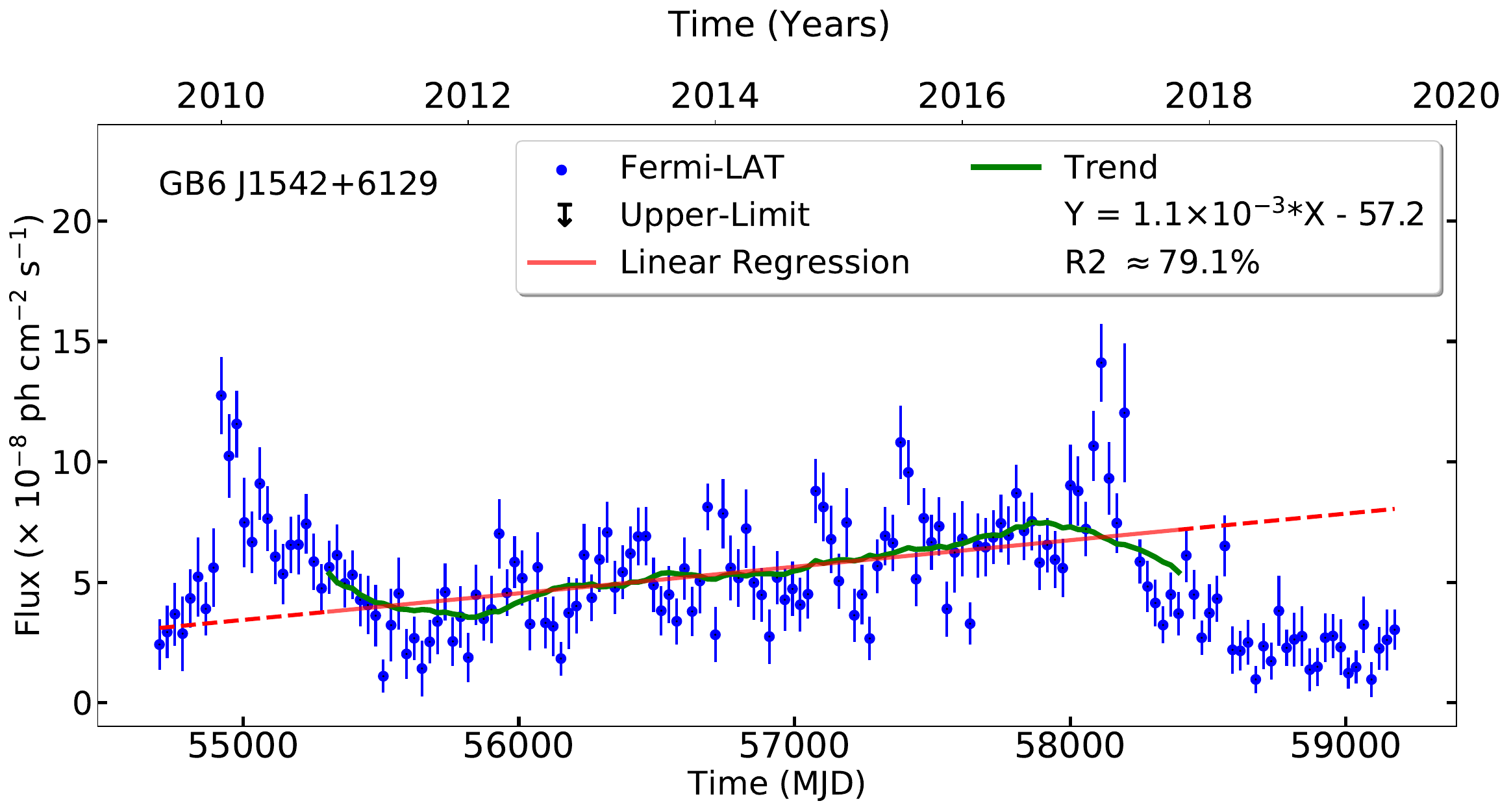}
        \includegraphics[scale=0.2295]{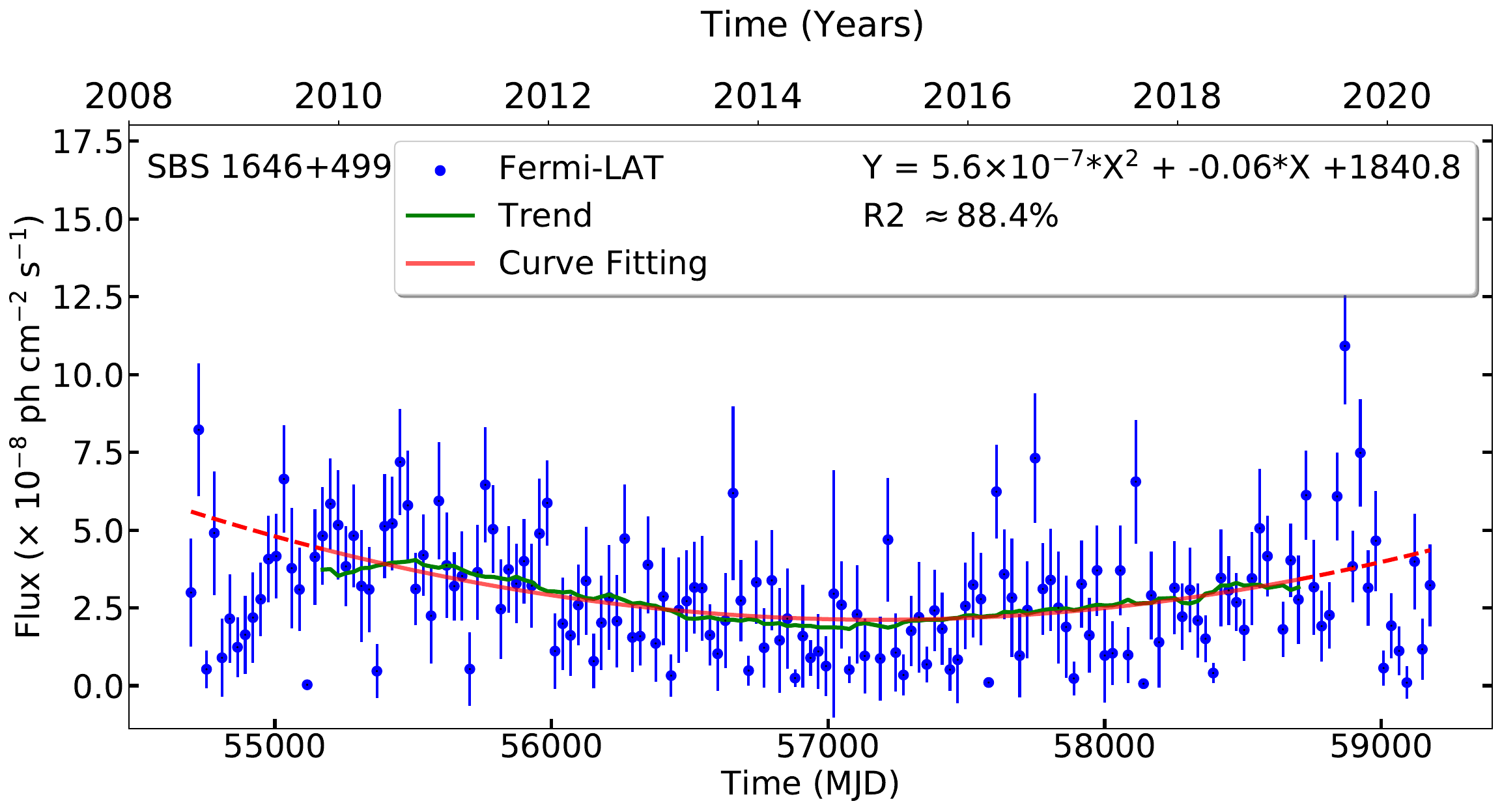}
        \includegraphics[scale=0.2295]{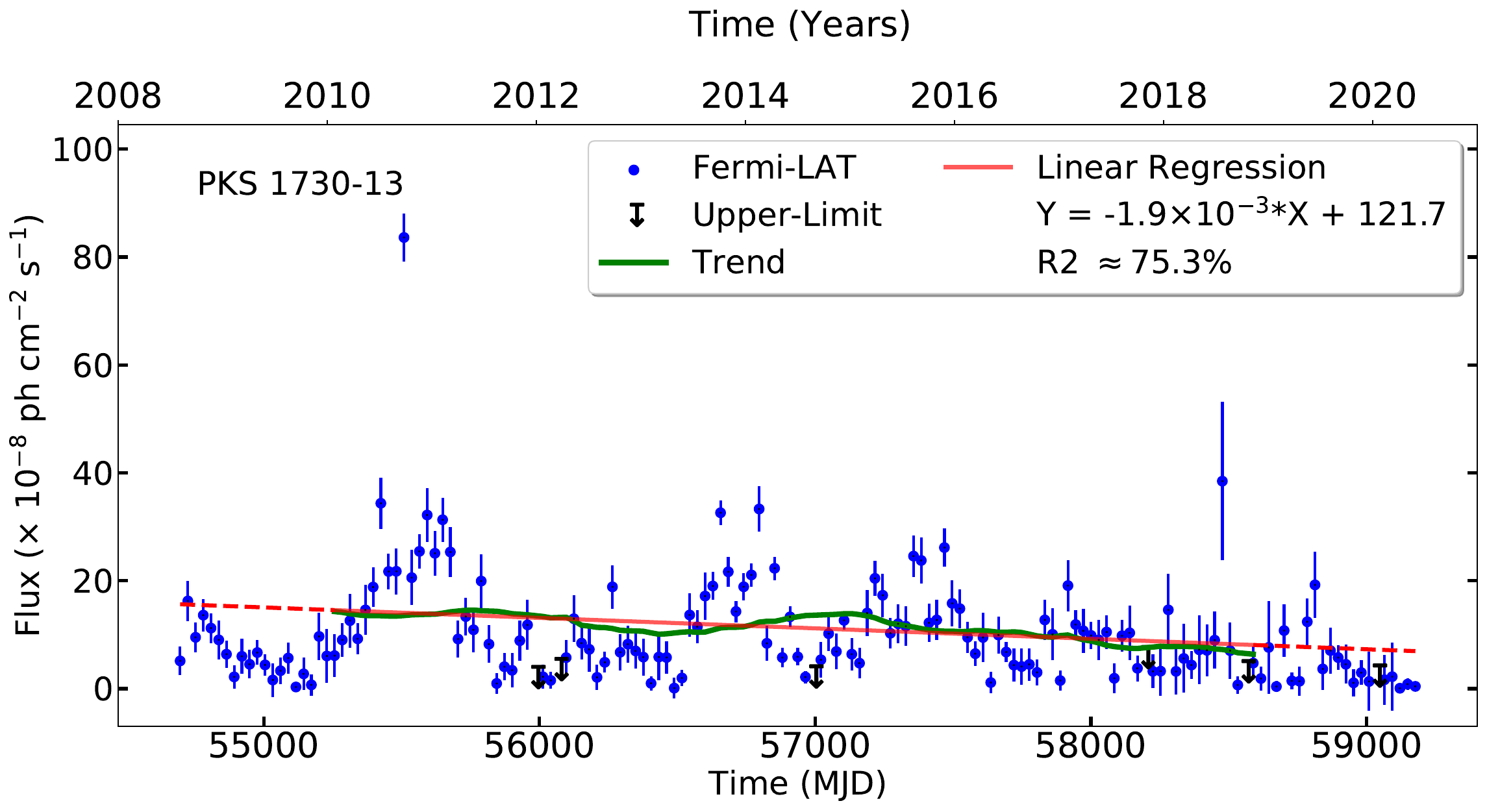}
	\includegraphics[scale=0.2295]{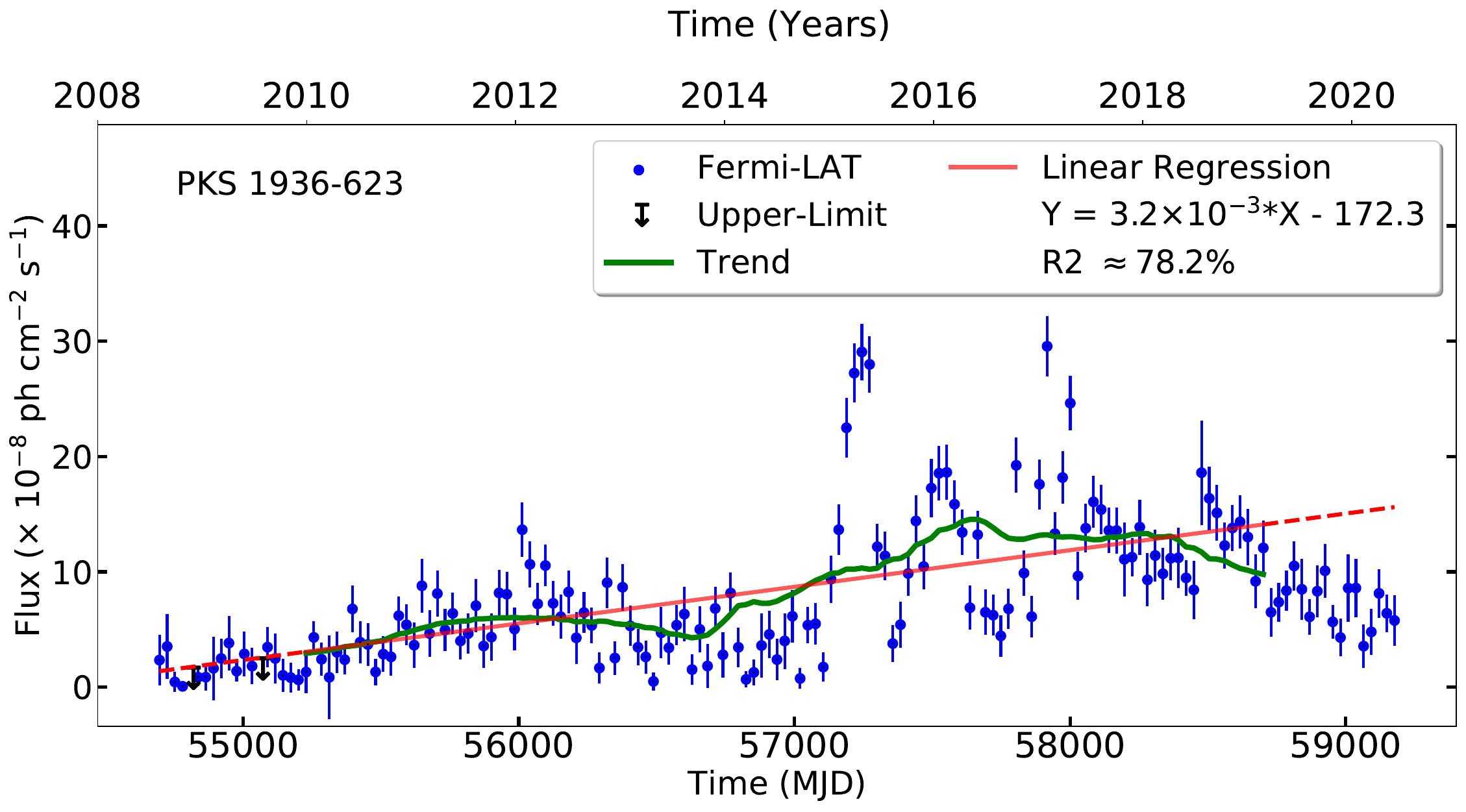}
	\includegraphics[scale=0.2295]{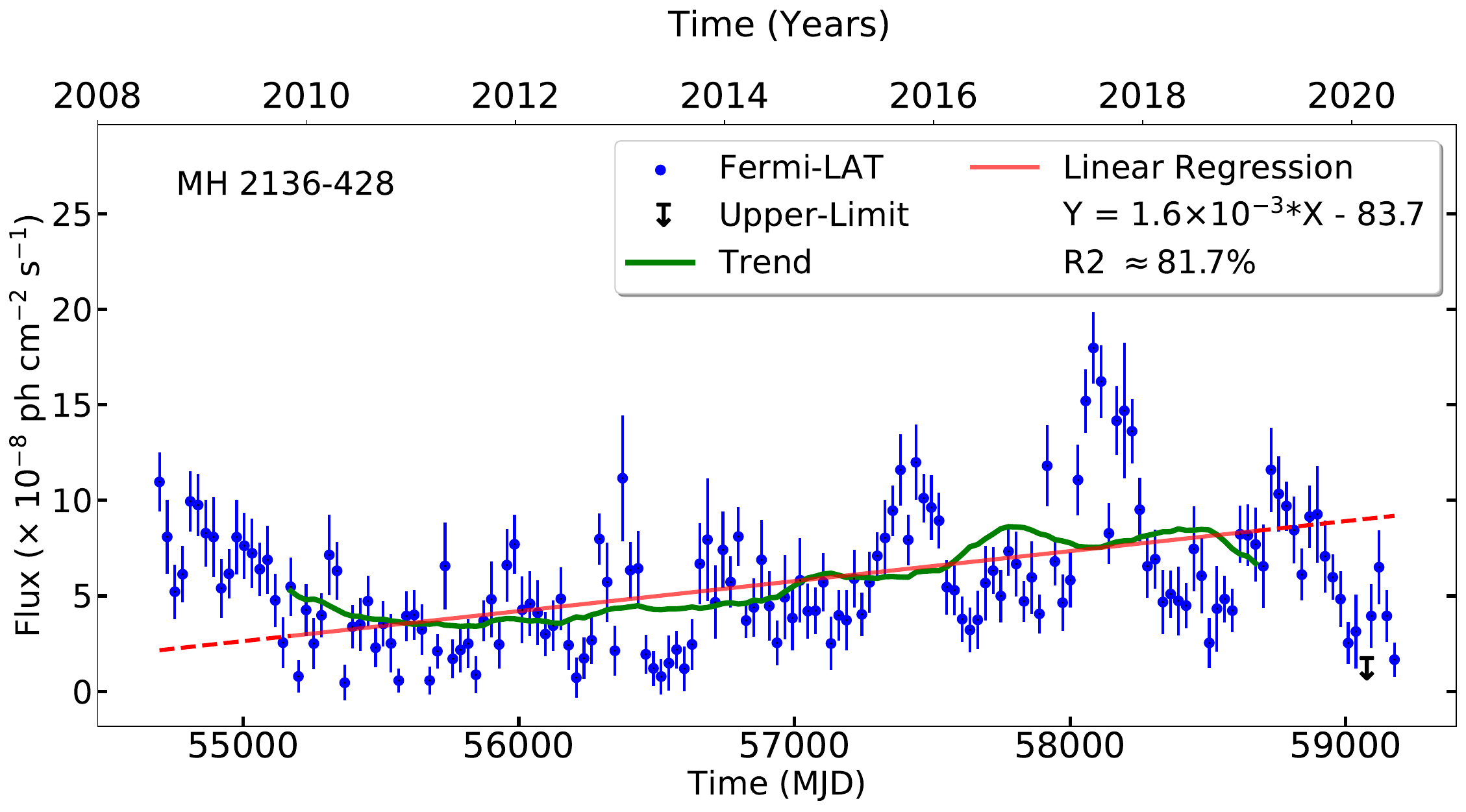}
         \includegraphics[scale=0.2295]{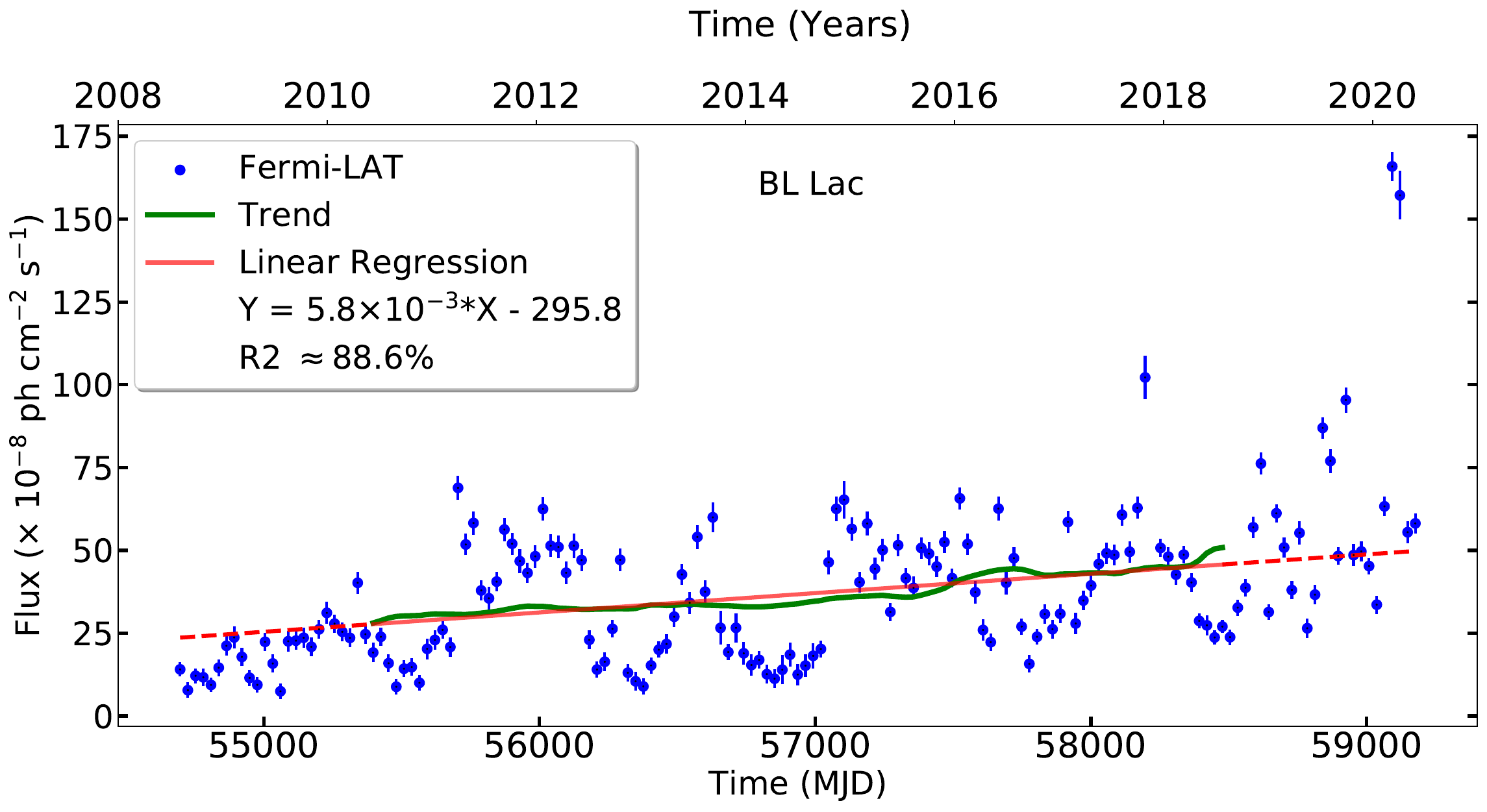}
   	\includegraphics[scale=0.2295]{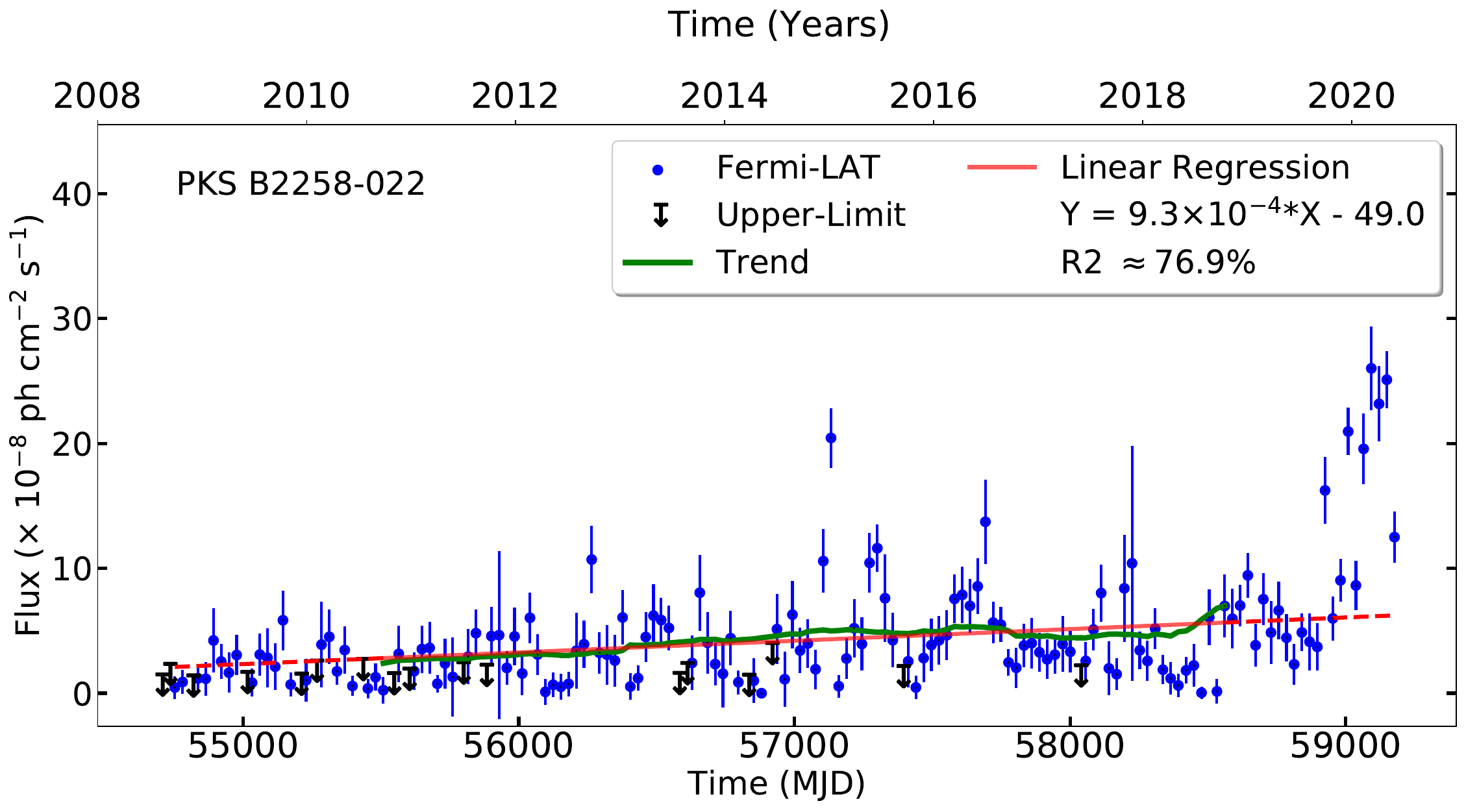}
	\caption{(Continued).}
\end{figure*}
\begin{figure*}[ht!]
	\centering
	\ContinuedFloat
        \includegraphics[scale=0.2295]{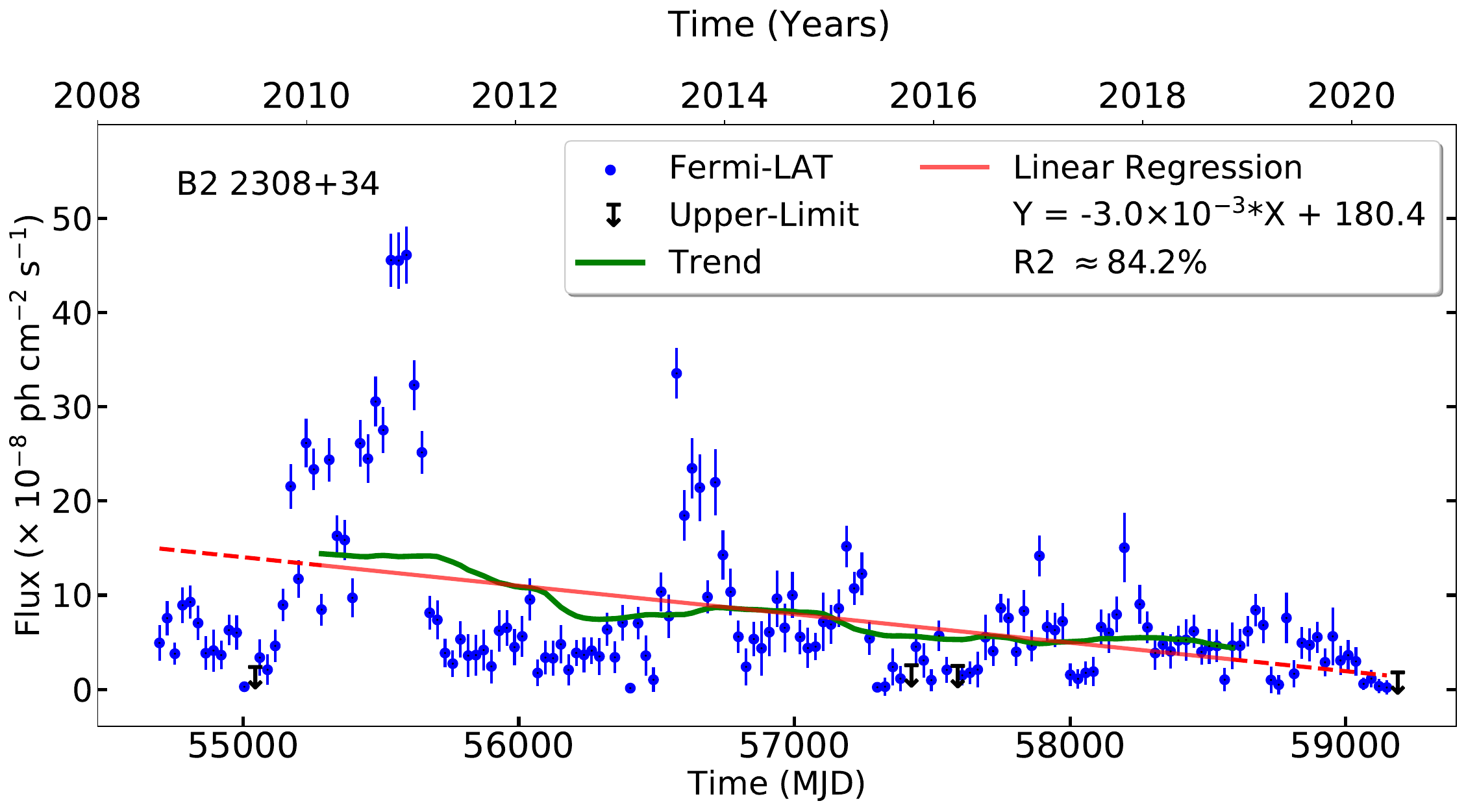}
        \includegraphics[scale=0.2295]{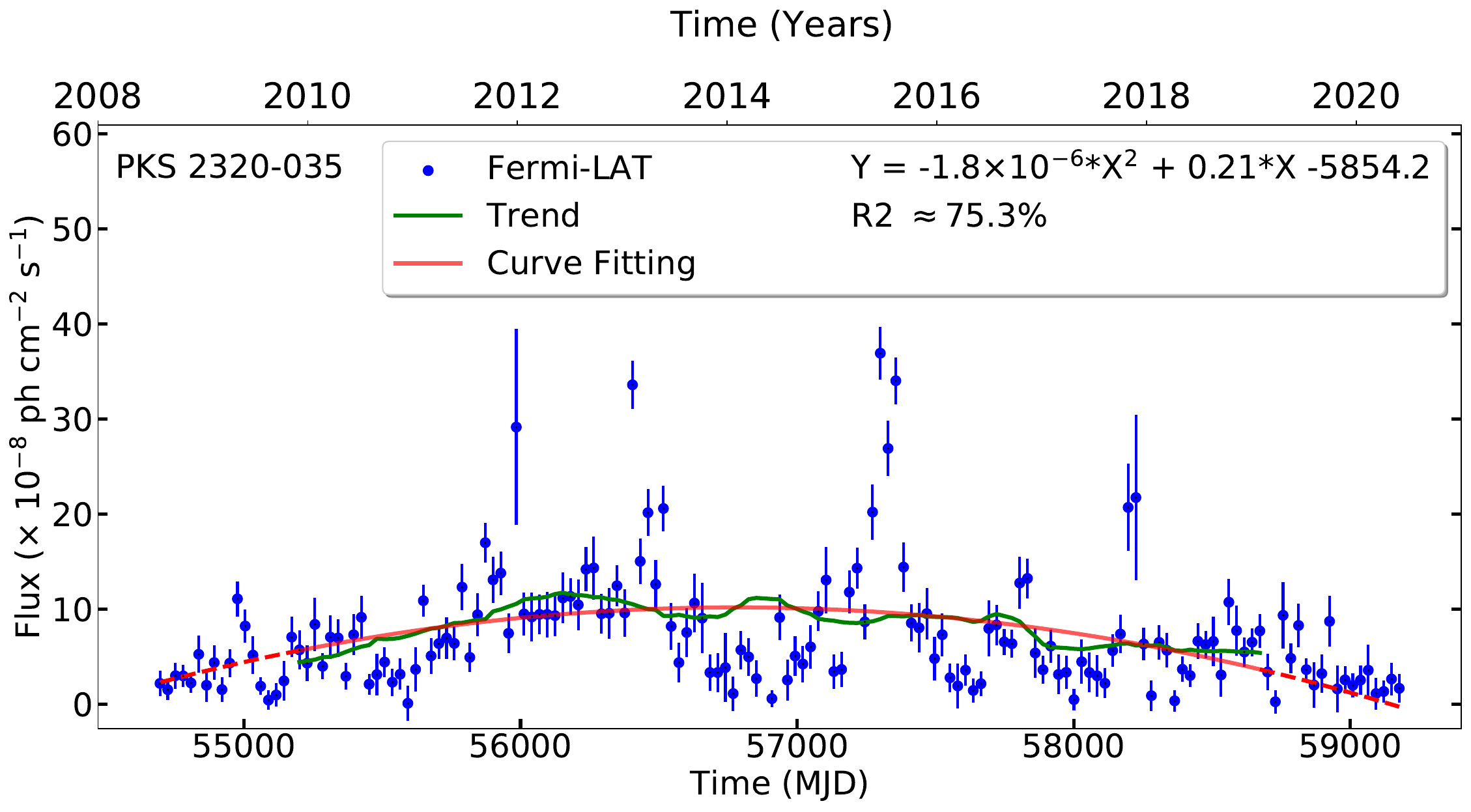}
        \includegraphics[scale=0.2295]{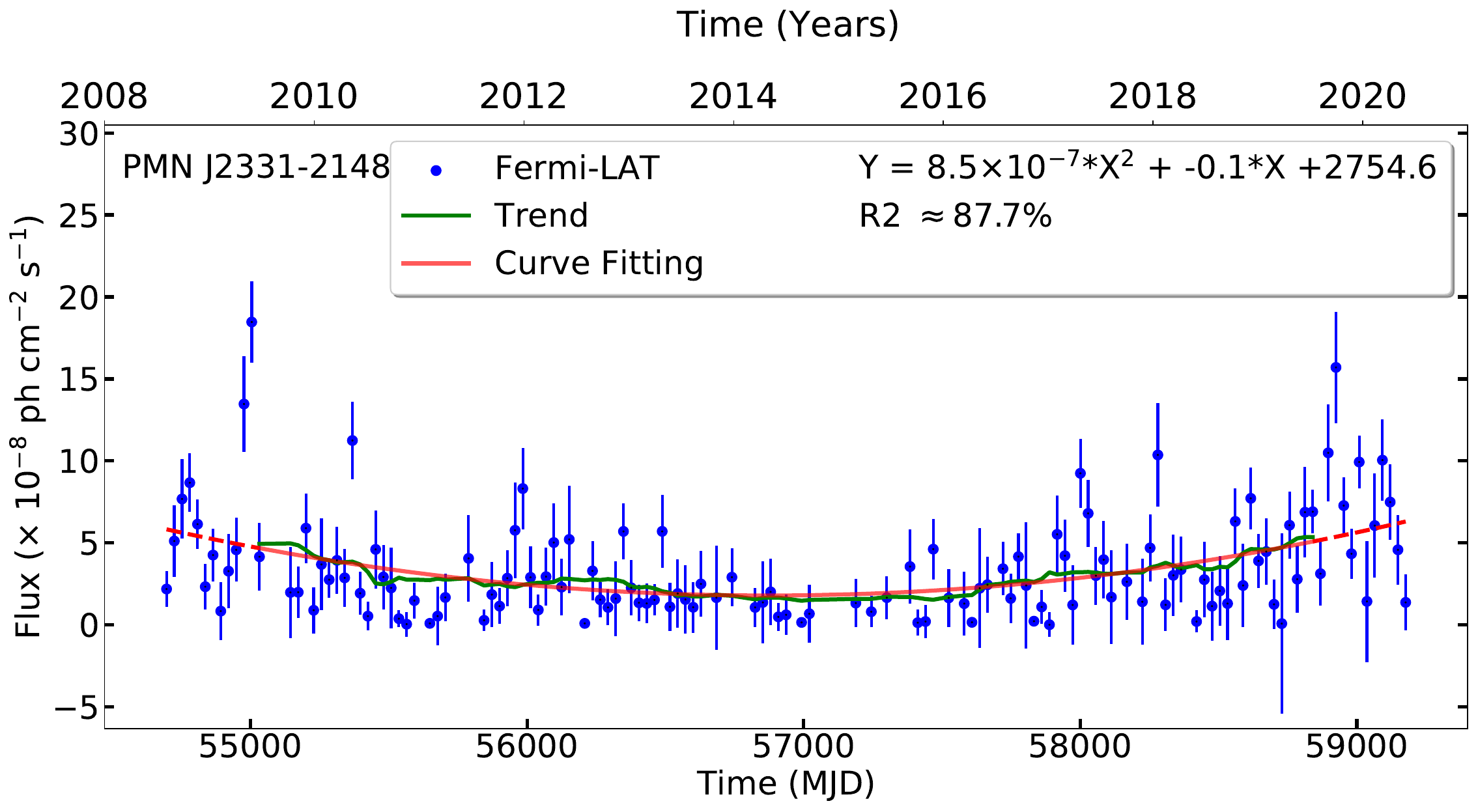}
        \includegraphics[scale=0.2295]{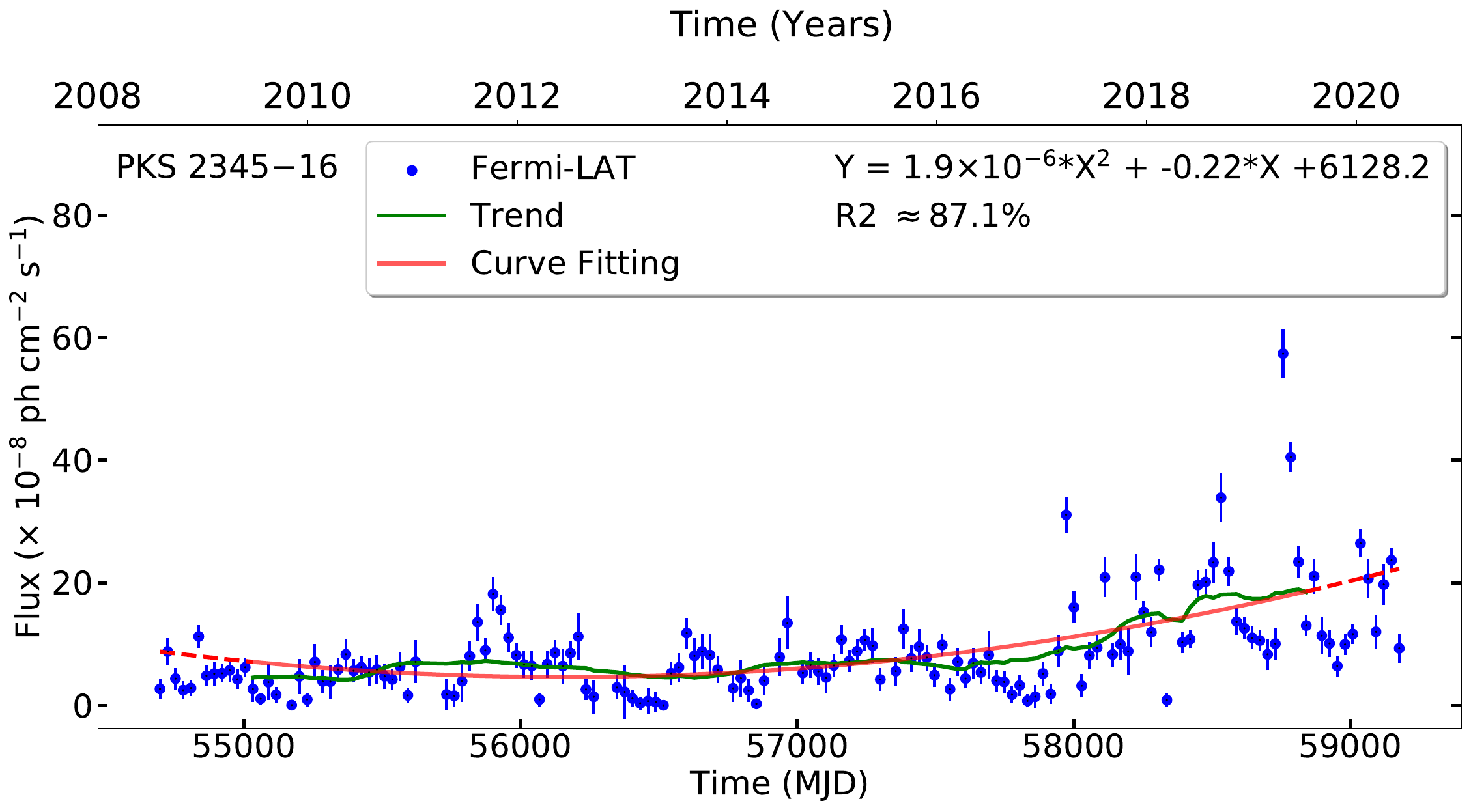}
	\caption{(Continued).}
\end{figure*}
\end{document}

%% file: table_1.tex
\begin{deluxetable*}{ccccccccccc}
	\tablecaption{List of AGN with trends in their $\gamma$-ray emission. The AGN are characterized by their \textit{Fermi}-LAT name, coordinates, AGN type, redshift, and the association name. We obtain this information from the 4FGL-DR2 catalog. The types of sources included in the 4FGL are BL Lacertae (bll), flat-spectrum radio quasar (fsrq), narrow-line Seyfert 1 (nlsy1), and radio galaxy (rdg). We include the percentage of upper limits (ULs) in the LC. 
	\label{tab:candidates_list}} 
	\tablewidth{0pt}
	\tablehead{
		\colhead{4FGL Source Name} &
		\colhead{RAJ2000} &
		\colhead{DEJ2000} &
		\colhead{Type} &
		\colhead{Redshift} & 
		\colhead{Association Name} & 
        \colhead{ULs (\%)} &
		\\
		\colhead{} &
		\colhead{} &
		\colhead{} &
		\colhead{} &
		\colhead{} &
		\colhead{} &
		\\
	}
	\startdata        
	J0049.7+0237 & 12.43 & 2.62 & bll & 1.474 & PKS 0047+023 & 9.3\% \\
	J0137.0+4751 & 24.26 & 47.86 & fsrq & 0.859 & OC 457 & 15.5\% \\ 
	J0217.2+0837 & 34.31 & 8.62 & bll & 0.085 & ZS 0214+083 & 3.7\% \\
	J0217.8+0144 & 34.46 & 1.73 & fsrq & 1.715 & PKS 0215+015 & 1.9 \% \\
	J0222.6+4302 & 35.66 & 43.03 & bll & 0.444 & 3C 66A & 0.6\% \\
	J0245.9$-$4650 & 41.49 & -46.84 & fsrq & 1.385 & PKS 0244$-$470 & 6.8\% \\
	J0252.8$-$2219 & 43.20 & -22.32 & fsrq & 1.419 & PKS 0250$-$225 & 1.2\% \\
	J0319.8+4130 & 49.95 & 41.51 & rdg & 0.018 & 3C 84 & 0.6\% \\
	J0407.0$-$3826 & 61.76 & -38.43 & fsrq & 1.285 & PKS 0405$-$385 & 6.8 \%\\
	J0423.3$-$0120 & 65.82 & -1.33 & fsrq & 0.916 & PKS 0420$-$01 & 11.2\% \\
	J0449.1+1121 & 72.28 & 11.35 & fsrq & 2.153 & PKS 0446+11 & 4.3\% \\
	J0449.4$-$4350 & 72.35 & -43.83 & bll & 0.205 & PKS 0447$-$439 & 0.0\% \\
	J0501.2$-$0158 & 75.30 & -1.97 & fsrq & 2.291 & S3 0458$-$02 & 1.8\% \\
	J0601.1$-$7035 & 90.29 & -70.58 & fsrq & 2.409 & PKS 0601$-$70 & 13.6\% \\
	J0650.7+2503 & 102.69 & 25.05 & bll & 0.203 & 1ES 0647+250 & 0.0\%\\
	J0719.3+3307 & 109.84 & 33.12 & fsrq & 0.779 & B2 0716+33 & 6.2\%\\
	J0808.2$-$0751 & 122.06 & -7.85 & fsrq & 1.837 & PKS 0805$-$07 & 1.8\%\\
	J0811.4+0146 & 122.86 & 1.77 & bll & 1.148 & OJ 014 & 0.0\%\\
	J0948.9+0022 & 147.24 & 0.37 & nlsy1 & 0.585 & PMN J0948+0022 & 1.8\%\\
	J1058.4+0133 & 164.62 & 1.56 & bll & 0.89 & 4C +01.28 & 0.6\%\\
        J1103.9$-$5357 & 165.97 & -53.96 & bll & -- & PKS 1101$-$536 & 1.8\% \\
	J1217.9+3007 & 184.47 & 30.11 & bll & 0.13 & B2 1215+30 & 0.0\% \\
	J1224.9+2122 & 186.22 & 21.38 & fsrq & 0.434 & 4C +21.35 & 0.6\% \\
        J1419.4$-$0838 & 214.86 & -8.64 & fsrq & 0.903 & NVSS J141922$-$083830 & 9.9\% \\
	J1427.0+2348 & 216.75 & 23.80 & bll & 0.6035 & PKS 1424+240 & 0.0\% \\
	J1454.4+5124 & 223.62 & 51.40 & bll & -- & TXS 1452+516 & 0.6\% \\
	J1517.7$-$2422 & 229.42 & -24.37 & bll & 0.048 & AP Librae & 0.0\% \\
        J1522.1+3144 & 230.54 & 31.73 & fsrq & 1.489 & B2 1520+31 & 0.0\% \\
	J1543.0+6130 & 235.75 & 61.50 & bll & 0.117 & GB6 J1542+6129 & 0.6\% \\
	J1555.7+1111 & 238.93 & 11.18 & bll & 0.433 & PG 1553+113 & 0.0\% \\
        J1647.5+4950 & 251.89 & 49.83 & bll & 0.049 & SBS 1646+499 & 1.8\% \\
	J1733.0$-$1305 & 263.26 & -13.08 & fsrq & 0.902 & PKS 1730$-$13 & 3.7\% \\
        J1941.3$-$6210 & 295.34 & -62.17 & bll & -- & PKS 1936$-$623 & 1.2\% \\
	J2139.4$-$4235 & 324.85 & -42.58 & bll & -- & MH 2136$-$428 & 0.6\% \\
	J2202.7+4216 & 330.69 & 42.28 & bll & 0.069 & BL Lacertae & 0.0\% \\
        J2301.0$-$0158 & 345.26 & -1.97 & fsrq & 0.778 & PKS B2258$-$022 & 10.5\% \\
        J2311.0+3425 & 347.76 & 34.42 & fsrq & 1.817 & B2 2308+34 & 3.1\% \\
        J2323.5$-$0317 & 350.88 & -3.29 & fsrq & 1.393 & PKS 2320$-$035 & 0.6\% \\
        J2331.0$-$2147 & 352.76 & -21.79 & fsrq & 0.563 & PMN J2331$-$2148 & 17.4\% \\
        J2348.0$-$1630 & 357.01 & -16.51 & fsrq & 0.576 &  PKS 2345$-$16 & 6.8\% \\
	\\
	\enddata
\end{deluxetable*}

%% file: table_2.tex
\begin{deluxetable*}{cccrccccccccc}
	\tablecaption{Results of the trend characterization of the AGN presented in table \ref{tab:candidates_list}. Each AGN is characterized by the type of trend (linear or Quadratic) and type of oscillations (additive or multiplicative). ``a'' (10$^{-8}$ ph cm$^{-2}$ s$^{-1}$ days$^{-2}$), ``b'' (10$^{-8}$ ph cm$^{-2}$ s$^{-1}$ days$^{-1}$), and ``c'' (10$^{-8}$ ph cm$^{-2}$ s$^{-1}$) are the fitting parameters. Finally, the R$^{2}$ criterion is included.  
	\label{tab:trend_properties}} 
	\tablewidth{0pt}
	\tablehead{
		\colhead{Association Name} & 
		\colhead{Trend} &
		\colhead{Topology} &
		\colhead{\textit{a}} &
        \colhead{\textit{b}} &
        \colhead{\textit{c}} &
		\colhead{R$^{2}$} &
		\\
	}
	\startdata        
	PKS 0047+023 & Linear & Additive & -- & $\rm{88x10^{-5}}\pm\rm{3x10^{-5}}$ & -46.5$\pm$1.9 & 85.4\% \\
	OC 457 & Linear & Multiplicative & -- & -$\rm{19x10^{-4}}\pm\rm{1x10^{-4}}$ & 116.3$\pm$7.6 & 75.9\% \\ 
	ZS 0214+083 & Linear & Multiplicative & -- & $\rm{11x10^{-4}}\pm\rm{4x10^{-4}}$ & -61.4$\pm$3.3 & 83.5\% \\
	PKS 0215+015 & Quadratic & Multiplicative & $\rm{99x10^{-8}}\pm\rm{4x10^{-8}}$ & $\rm{-11x10^{-2}}\pm\rm{5x10^{-2}}$ & 3219.2$\pm$148.3 & 81.6\% \\
	3C 66A & Linear & Additive & -- & -$\rm{20x10^{-4}}\pm\rm{1x10^{-4}}$ & 122.4$\pm$8.2 & 76.6\% \\
	PKS 0244$-$470 & Linear & Multiplicative & -- & -$\rm{32x10^{-4}}\pm\rm{1x10^{-4}}$ & 188.6$\pm6.7$ & 86.4\% \\
	PKS 0250$-$225 & Linear & Multiplicative & -- & -$\rm{32x10^{-4}}\pm\rm{1x10^{-4}}$ & 191.3$\pm7.9$ & 83.2\% \\
	3C 84 & Quadratic & Multiplicative & -$\rm{48x10^{-7}}\pm\rm{3x10^{-7}}$ & $\rm{56x10^{-2}}\pm\rm{4x10^{-2}}$ & -16166.3$\pm$115.3 & 86.7\% \\
	PKS 0405$-$385 & Linear & Additive & -- & $\rm{10x10^{-4}}\pm\rm{1x10^{-4}}$ & -52.4$\pm2.8$ & 80.4\% \\
	PKS 0420$-$01 & Linear & Multiplicative & -- & -$\rm{25x10^{-4}}\pm\rm{1x10^{-4}}$ & 149.3$\pm$8.1 & 77.3\% \\
	PKS 0446+11 & Linear & Multiplicative & -- & -$\rm{188x10^{-5}}\pm\rm{9x10^{-5}}$ & 116.6$\pm$4.8 & 79.1\% \\
	PKS 0447$-$439 & Linear & Additive & -- & $\rm{207x10^{-5}}\pm\rm{8x10^{-5}}$ & -107.6$\pm$4.6 & 84.9\% \\
	S3 0458$-$02 & Linear & Multiplicative & -- & $\rm{40x10^{-4}}\pm\rm{1x10^{-4}}$ & -212.4$\pm$8.8 & 86.2\% \\
	PKS 0601$-$70 & Linear & Multiplicative & -- & -$\rm{182x10^{-5}}\pm\rm{8x10^{-5}}$ & 106.8$\pm$3.6 & 86.9\% \\
	1ES 0647+250 & Linear & Multiplicative & -- & $\rm{69x10^{-5}}\pm\rm{2.0x10^{-5}}$ & -36.5$\pm$1.2 & 91.6\% \\
	B2 0716+33 & Linear & Multiplicative & -- & -$\rm{17x10^{-4}}\pm\rm{1x10^{-4}}$ & 105.5$\pm$5.4 & 79.5\% \\
	PKS 0805$-$07 & Linear & Multiplicative & -- & -$\rm{19x10^{-4}}\pm\rm{2x10^{-4}}$ & 115.6$\pm$8.5 & 78.4\% \\
	OJ 014 & Linear & Additive & -- & $\rm{79x10^{-5}}\pm\rm{4x10^{-5}}$ & -39.5$\pm$2.5 & 76.5\% \\
	PMN J0948+0022 & Linear & Multiplicative & -- & -$\rm{33x10^{-4}}\pm\rm{1x10^{-4}}$ & 198.1$\pm$8.9 & 82.6\% \\
	4C +01.28 & Linear & Multiplicative & -- & -$\rm{23x10^{-4}}\pm\rm{2x10^{-4}}$ & 139.9$\pm$7.2 & 80.4\% \\
        PKS 1101$-$536 & Quadratic & Multiplicative & $\rm{174x10^{-8}}\pm\rm{6x10^{-8}}$ & $\rm{-193x10^{-3}}\pm\rm{8x10^{-3}}$ & 5464.5$\pm$204.8 & 85.7\% \\
	B2 1215+30 & Quadratic & Additive & -$\rm{158x10^{-8}}\pm\rm{4x10^{-8}}$ & $\rm{180x10^{-3}}\pm\rm{6x10^{-3}}$ & -5043.5$\pm$236.7 & 81.8\% \\
	4C +21.35 & Linear & Multiplicative & -- & -$\rm{21x10^{-3}}\pm\rm{2x10^{-3}}$ & 1208.7$\pm$46.1 & 84.0\% \\
        NVSS J141922$-$083830 & Linear & Multiplicative & -- & $\rm{141x10^{-5}}\pm\rm{5x10^{-5}}$ & -75.7$\pm$3.1 & 84.2\% \\
	PKS 1424+240 & Linear & Multiplicative & -- & -$\rm{118x10^{-5}}\pm\rm{4x10^{-5}}$ & 68.9$\pm$2.3 & 82.3\% \\
	TXS 1452+516 & Linear & Multiplicative & -- & $\rm{12x10^{-4}}\pm\rm{2x10^{-4}}$ & -64.5$\pm$3.7 & 81.7\% \\
	AP Librae & Linear & Multiplicative & -- & $\rm{112x10^{-5}}\pm\rm{4x10^{5}}$ & -55.7$\pm$2.5 & 82.3\% \\
        B2 1520+31 & Linear & Multiplicative & -- & -$\rm{116x10^{-4}}\pm\rm{2x10^{4}}$ & -670.6$\pm$11.6 & 94.3\% \\
	GB6 J1542+6129 & Linear & Additive & -- & $\rm{15x10^{-4}}\pm\rm{1x10^{4}}$ & -57.2$\pm$4.2 & 79.1\% \\
	PG 1553+113 & Linear & Additive & -- & $\rm{67x10^{-5}}\pm\rm{3x10^{-5}}$ & -31.4$\pm$1.8 & 78.5\% \\
        SBS 1646+499 & Quadratic & Multiplicative & $\rm{56x10^{-8}}\pm\rm{2x10^{-8}}$ & $\rm{-62x10^{-3}}\pm\rm{2x10^{-3}}$ & 1840.4$\pm69.1$ & 88.4\% \\
        PKS 1730$-$13 & Linear & Multiplicative	& -- & -$\rm{19x10^{-4}}\pm\rm{1x10^{-4}}$ & 121.7$\pm$8.1 & 75.3\% \\
        PKS 1936$-$623 & Linear & Multiplicative & -- & $\rm{33x10^{-4}}\pm\rm{1x10^{-4}}$ & -179.9$\pm$8.6 & 78.2\% \\
	MH 2136$-$428 & Linear & Multiplicative & -- & $\rm{163x10^{-5}}\pm\rm{7x10^{-5}}$ & -83.7$\pm$3.8 & 81.7\% \\
	BL Lacertae & Linear & Additive & -- & $\rm{580x10^{-5}}\pm\rm{3x10^{-5}}$ & -295.8$\pm$19.3 & 88.6\% \\
        PKS B2258-022 & Linear & Multiplicative & -- & $\rm{93x10^{-5}}\pm\rm{6x10^{-5}}$ & -49.0$\pm$3.8 & 76.9\% \\
        B2 2308+34 & Linear & Multiplicative & -- & -$\rm{30x10^{-4}}\pm\rm{1x10^{-4}}$ & 180.4$\pm$9.1 & 84.2\% \\
        PKS 2320$-$035 & Quadratic & Multiplicative & -$\rm{18x10^{-7}}\pm\rm{110^{-7}}$ & $\rm{21x10^{-2}}\pm\rm{1x10^{-2}}$ & -5854.2$\pm$336.2 & 75.3\% \\
        PMN J2331$-$2148 & Quadratic & Multiplicative & $\rm{85x10^{-8}}\pm\rm{2x10^{-8}}$ & $\rm{-104x10^{-3}}\pm\rm{3x10^{-3}}$ & 2754.6$\pm$64.7 & 87.7\% \\
        PKS 2345$-$16 & Quadratic & Multiplicative & $\rm{19x10^{-7}}\pm\rm{1x10^{-7}}$ & $\rm{-22x10^{-2}}\pm\rm{1x10^{-2}}$ & 6128.2$\pm$348.7 & 87.1\% \\
	\\
	\enddata
\end{deluxetable*}

%% file: table_3.tex
\begin{table*}
\centering
\caption{Detection rates for time series data exhibiting long-term trends affected by various types of noise. We considered two types of trends, linear and quadratic, and four types of noise: ``white noise'' (WN), ``pink noise'', (PN) ``red noise'' (RN), and ``bending power-law'' (BPL). The detection rates are evaluated for each stage of the analysis pipeline: Lomb-Scargle analysis (LBA), R$^{2}$, and the characterization of trend parameters (CTP). The CTP for linear trends are the parameters ``b’’ and ``c’’, while for quadratic trends, they are ``a'', ``b'' and ``c'' (see Table \ref{tab:trend_properties}). The rates of R$^{2}$ and CTP indicate the detection rates of the LCs previously identified in the preceding stage of the pipeline. Finally, the overall detection rate is summarized in the last column. \label{tab:experiment_1}} 
{%
\begin{tabular}{cccccc}
\hline
\hline
Type of Trend & Type of Noise & LSA (\%) & R$^{2}$ (\%) & CTP (\%) & Total Detection (\%) \\
\hline
\multirow{3}{*}{Linear} & WN & 98.9 & 100 & 91.2 & 90.2 \\ 
                        & PN & 93.3 & 100 & 52.7 & 49.2 \\
                        & RN & 96.5 & 100 & 88.5 & 85.4 \\
                        & BPL & 95.7 & 100 & 86.8 & 83.1 \\
\hline
\multirow{3}{*}{Quadratic} & WN & 100 & 94.8 & 100 & 94.8 \\ 
                        & PN & 98.6 & 82.7 & 100 & 81.5 \\
                        & RN & 98.3 & 94.9 & 100 & 93.3 \\
                        & BPL & 97.6 & 93.9 & 100 & 91.6 \\
\hline
\end{tabular}%
}
\end{table*}

%% file: table_4.tex
\begin{table*}
\centering
\caption{Detection rates for time series data exhibiting long-term trends affected by various types of flares. We consider two types of trends, linear and quadratic. The flares are characterized by amplitude and duration. Two levels of amplitude are defined: A1 (26.5$\times$ 10$^{-8}$ ph cm$^{-2}$ s$^{-1}$) and A2 (70.1$\times$ 10$^{-8}$ ph cm$^{-2}$ s$^{-1}$). Two durations are defined: 2 months and 12 months. The detection rates are evaluated for each stage of the analysis pipeline: Lomb-Scargle analysis (LBA), R$^{2}$, and the characterization of trend parameters (CTP). The CTP for linear trends are the parameters ``b’’ and ``c’’, while for quadratic trends, they are ``a'', ``b'' and ``c'' (see Table \ref{tab:trend_properties}). The rates of R$^{2}$ and CTP indicate the detection rates of the LCs previously identified in the preceding stage of the pipeline. Finally, the overall detection rate is summarized in the last column\label{tab:experiment_2}} 
{%
\begin{tabular}{cccccc}
\hline
\hline
Type of Trend & Type of Flare & LSA (\%) & R$^{2}$ (\%) & CTP (\%) & Total Detection (\%) \\
\hline
\multirow{3}{*}{Linear} & A1 (2 months) & 99.99 & 99.12 & 15.1 & 15.0 \\ 
                        & A1 (12 months) & 99.99 & 70.36 & 0.24 & 0.17 \\
                        & A2 (2 months) & 99.8 & 81.88 & 0.43 & 0.35 \\
                        & A2 (12 months) & 95.7 & 35.18 & 0 & 0 \\
\hline
\multirow{3}{*}{Quadratic} & A1 (2 months) & 99.8 & 58.8 & 100 & 57.7 \\ 
                        & A1 (12 months) & 98.6 & 0 & 0 & 0 \\
                        & A2 (2 months) & 98.8 & 0 & 0 & 0 \\
                        & A2 (12 months) & 99.8 & 0 & 0 & 0 \\
\hline
\end{tabular}%
}
\end{table*}